\documentclass[aps,preprint,groupedaddress]{revtex4}  

\usepackage{amssymb}
\usepackage{amsmath}
\usepackage{amssymb}
\usepackage{graphicx}
\usepackage{subfigure}
\usepackage{color}
\usepackage{mathrsfs}
\usepackage[dvipsnames]{xcolor}

\usepackage[breaklinks=true,colorlinks=true]{hyperref}
\hypersetup{colorlinks=true,citecolor=blue,linkcolor=blue,urlcolor=blue}

\usepackage[utf8]{inputenc}

\begin{document}


\title{Collisions of kinks in deformed $\varphi^4$ and $\varphi^6$ models}

\author{Aliakbar Moradi Marjaneh$^{1}$, Fabiano C. Simas$^{2,3}$ and D. Bazeia$^{4}$}

\email{moradimarjaneh@gmail.com; fc.simas@ufma.br; bazeia@fisica.ufpb.br}

\affiliation{$^{1}$Department of Physics, Quchan Branch, Islamic Azad university, Quchan, Iran\\
$^{2}$Programa de P\'os-Gradua\c c\~ao em F\'isica, Universidade Federal do Maranh\~ao
(UFMA), Campus Universit\'ario do Bacanga, 65085-580, S\~ao Lu\'is, Maranh\~ao, Brazil\\
$^{3}$Centro de Ci\^encias Agr\'arias e Ambientais (CCAA), Universidade Federal do Maranh\~ao (UFMA), 65000-000 Chapadinha, MA, Brazil\\
{$^{4}$Departamento de F\'isica, Universidade Federal da Para\'iba (UFPB), 58051-970 Jo\~ao Pessoa, PB, Brazil}}


\begin{abstract}


Two \emph{hyperbolic-deformed} field theoretic models are discussed. In both of them, due to the effect of specific deformation function on the well known $\varphi^4$ and $\varphi^6$ models, their internal structure may change significantly. Unlike the $\varphi^4$ kinks solutions, which has only one internal mode in addition to its translational mode, the kinks of the hyperbolic-deformed $\varphi^4$ model can have \emph{several} internal modes. Moreover, the deformation on the $\varphi^6$ model has other interesting features, because the kinks of the $\varphi^6$ model have only a zero mode and the deformation may cause the appearance of internal mode for both kink and antikink. The presence of the new internal modes modify the collisions which we explore in the present work. The deformation relies on a real parameter, which controls the number of internal modes, and we also study how the deformation parameter alter the mass of the kinks and the critical velocities involved in the collisions. 
\end{abstract}



\maketitle


\section{Introduction}\label{sec:introduction}


Nonlinear field models have essential role to describe phenomena in several areas, running from subatomic to the cosmological scale. In particular, {\it{solitons}} as nontrivial static solutions of some classical field theory with degenerate minima, have attracted the attention of theoretical and experimental researchers for more than 50 years; see, e.g., Refs. \cite{Bishop.PhysD.1980, Rajaraman.book.1982, Vilenkin.book.2000, Manton.book.2004, Vachaspati.book.2006, Dauxois.book.2006, Kevrekidis.book.2019} and references therein. The topological defects in $(1 + 1)$ dimensional space-time are the simplest form of localized structures, called {\it{kinks}}, and they have been widely considered in the past. For example, we can mention the study of kinks, their scattering and multi-kinks collision in integrable and non-integrable models such as sine-Gordon model \cite{Peyrard.PhysD.1983.msG, Bazeia.EPJC.2011,  Bazeia.EPJC.2013, Moradi.EPJB.2018}, double sine-Gordon model \cite{Campbell.dsG.1986, Peyravi.EPJB.2009,Gani.EPJC.2018, Belendryasova.JPCS.2019, Gani.EPJC.2019}, $\varphi^4$ \cite{Kevrekidis.book.2019, Campbell.PhysD.1983.phi4, anninos, Campbell.PhysD.1986, Dorey.JHEP.2017, Moradi.CNSNS.2017, Askari.CSF.2020, Mohammadi.CNSNS.2020}, $\varphi^6$ \cite{Hoseinmardy.IJMPA.2010, Dorey.PRL.2011, Weigel.AHEP.2017, Moradi.JHEP.2017, Demirkaya.JHEP.2017, Lima.JHEP.2019}, semi-compactness kink \cite{Bazeia.EPJC.2021}, logarithmic models \cite{Ekaterina.PLB.2021} and higher order models \cite{Christov.PRD.2019, Belendryasova.CNSNS.2019, Christov.PRL.2019,Gani.PRD.2020, Christov.CNSNS.2021, Gani.EPJC.2021}, which have many applications to understand, for instance, phase transitions in materials \cite{Gufan.DAN.1978}, dynamics of crowdions \cite{Crowdions0, Crowdions1, Crowdions2, Crowdions3, Crowdions4}, field evolution in the early Universe \cite{Gani.JCAP.2018}, solitonlike solutions for a Higgs scalar field ~\cite{Kudryavtsev.JETPLett.1975}, Q-lump and skyrmion on a domain wall~\cite{blyankinshtein,nitta4,jennings,nitta5,GaLiRaconf} and scalar field models for branes \cite{free,cza,Baz,Baz2,Zhong.PRD.2014,Peyravi.EPJC.2016}. In some specific investigations, we can highlight the scattering of wobbling kinks \cite{Alonso.PRD.2021,Azadeh.JHEP.2021,Azadeh.JHEP.2021.1}, models with two scalar fields \cite{Shnir.PRD.2012,Alonso.Comm.2020,Alonso.Comm.2021} and also, studies on collective coordinate dynamics, which solve a difficulty linked to the collision process and prove that the appearance of the fractal structure is related to energy exchange between the internal modes \cite{Manton.PRL.2021,Manton.PRD.2021,Adam.PRD.2022}.

Although the study of integrable models with localized solutions such as sine-Gordon model provides a lot of information, nevertheless the non-integrable models, that have inner structure \cite{Peyrard.PhysD.1983.msG, Campbell.PhysD.1983.phi4, Panos.JHEP.2017, Dorey.PLB.2018, Zhong.JHEP.2020, Bazeia.JModPhysA.2019} or their solutions are asymmetric \cite{Hoseinmardy.IJMPA.2010, Dorey.PRL.2011, Weigel.AHEP.2017, Moradi.JHEP.2017, Demirkaya.JHEP.2017, Lima.JHEP.2019, Christov.PRD.2019, Belendryasova.CNSNS.2019, Christov.PRL.2019,Gani.PRD.2020}, present more attractive properties. In particular, we can mention the bounce windows in kink scattering in $\varphi^4$ and $\varphi^6$ models. The symmetric kinks of the $\varphi^4$ model can carry energy with internal mode. When the initial velocity is smaller than the critical velocity, bounce windows are created for some initial conditions. This mechanism is explained by the energy exchange between the internal mode and the translation mode \cite{Campbell.PhysD.1983.phi4}. In the $\varphi^6$ model, the kinks are asymmetric, and there is no presence of an internal mode, however, the scattering shows the two-bounce windows. The explanation is provided by the mechanism of resonant energy exchange between the vibrational modes of the antikink-kink pair \cite{Dorey.PRL.2011}. On the other hand, new potentials and their kinks are always used with various applications. For example, hyperbolic models can describe black holes  \cite{QiangWen.PRD.2015}, tachyon matter cosmology \cite{Pourhassan.IJMPD.2017} and quintessential inflation  \cite{Agarwal.PLB.2017}. Another example of the application of kinks can be seen in the study of molecular simulations in trans-polyacetylene \cite{Berna,An}. In Ref.~\cite{Berna}, the study of photogeneration of topological defects  showed a two-bounce behavior, characteristic of nonintegrable models. More recently, in Ref.~\cite{An} it was proposed the study of the relaxation dynamics to form soliton pairs in trans-polyacetilene. The investigation suggested that the excitation of the electron-hole pair generated a kink-antikink pair that collide and reflect with constant velocities. Interestingly, in many cases, several potentials can be obtained by using simple deformation functions on well known models such as  $\varphi^4$ and $\varphi^6$ models \cite{Bazeia.PRD.2002,Bazeia.PRD.2004,Bazeia.PRD.2006,Bazeia:2005hu,Bazeia.EPJC.2018}. The selection of a deformation function can result in the introduction of new physics and the production of new models. In Ref. \cite{Yama.car}, for example, deformation in graphene was addressed, with the aim to understand the positive and negative radiation pressure effects.

In this work, we deal with several issues concerning specific properties of two distinct hyperbolic potentials, described by deformation of the standard $\varphi^4$ and $\varphi^6$ models. To do this, we organize the paper as follows. In section \ref{sec:generalstatement}, we review the well-known $\varphi^4$ and $\varphi^6$ field theories. Then, in the  two sections \ref{sec:TanhDeformedPhi4Model} and \ref{sec:TanhDeformedPhi6Model}, we define a deformation function dependent on real parameter $r$, and introduce tanh-deformed $\varphi^4$ and tanh-deformed $\varphi^6$ models. In these models, we investigate the explicit kinks, their masses, the stability potentials, the number and the values of internal modes and also critical velocity and the corresponding collisions. As we explain in sections \ref{sec:TanhDeformedPhi4Model} and \ref{sec:TanhDeformedPhi6Model}, the deformed models are obtained via a deformation function that depends on a single real parameter, which is used to controls the main features of the models and guide us to investigate the two models. The work in ended in section \ref{sec:conclusion}, where we add our conclusions and comment on issues concerning future directions of new research on the subject.

\section{General considerations}\label{sec:generalstatement}

We consider a field theoretical model in $(1+1)$ spacetime dimensions, with a real scalar field $\varphi(x,t)$. 
The dynamics of the field is described by the Lagrangian density
\begin{eqnarray}\label{eq:lagrangian}
\mathcal{L}=\frac{1}{2}\left(\frac{\partial \varphi}{\partial t}\right)^2-\frac{1}{2}\left(\frac{\partial \varphi}{\partial x}\right)^2-V(\varphi).
\end{eqnarray}
The energy functional corresponding to the Lagrangian Eq.~\eqref{eq:lagrangian} is
\begin{eqnarray}\label{eq:energyfunctional}
E = \int_{-\infty}^{+\infty}\Biggl[\frac{1}{2}\left(\frac{\partial \varphi}{\partial t}\right)^2+\frac{1}{2}\left(\frac{\partial \varphi}{\partial x}\right)^2+V(\varphi)\Biggr]dx,
\end{eqnarray}
where $V(\varphi)$ is a non-negative potential which we suppose engenders a set of minima $\mathcal{V}=\left\{\bar\varphi_1,\bar\varphi_2,\bar\varphi_3,... \right\}.$ The equation of motion can be obtained from the Lagrangian Eq.~\eqref{eq:lagrangian} as
\begin{equation}\label{eq:EOM}
\frac{\partial^2 \varphi}{\partial t^2}-\frac{\partial^2 \varphi}{\partial x^2}+\frac{dV}{d\varphi} = 0,
\end{equation}
and for static field $\varphi=\varphi(x) $ we have
\begin{equation}\label{eq:EOMstatic}
\frac{d^2\varphi}{dx^2} = \frac{dV}{d\varphi}.
\end{equation}
We can write the first order differential equations
\begin{equation}\label{eq:EOMstaticfirstorder}
\frac{d\varphi}{dx}=\pm \sqrt{2V(\varphi)},
\end{equation}
which may also be used to obtain the kink and antikink solutions of the model. By substituting these solutions into Eq.~\eqref{eq:energyfunctional}, one can obtain the (classical) mass of each kink or antikink. In addition, moving kinks with the velocity $v$ along the x-axis can be obtained by the Lorentz boost $\varphi(x,t)=\varphi(\gamma(x-vt))$, where $\gamma=1/\sqrt{(1-v^2)}$. We search for static configurations having finite energy and interpolating between neighboring minima of the model. It means that, for the kink, $\varphi(-\infty)=\lim_{x \to -\infty} \varphi(x)=\bar\varphi_i$ with $\bar\varphi_i \in \mathcal{V}$, and $\varphi(+\infty)=\lim_{x \to +\infty} \varphi(x)=\varphi_{i+1}$ with $\bar\varphi_{i+1} \in \mathcal{V}$.

The next important subject has to deal with internal modes in kink excitation spectra and kink stability analysis because many phenomena, including the interaction of kinks with impurities and the kink scattering, are explained based on such internal modes. To deal with this, one adds a small perturbation $\varphi(x,t)=\varphi(x)+ \eta(x) \cos (\omega t)$ to the static kink $\varphi(x)$. This leads to the Schr\"odinger-like equation

\begin{equation}\label{eq:schrodingerlike}
-\frac{\partial^2 \eta}{\partial \varphi^2}+U\eta=\omega^2\eta,
\end{equation}
where $U=U(x)$ is the kink stability potential
\begin{equation}\label{eq:schrodingerlikepotential}
U=\frac{d^2V}{d\varphi^2}\bigg|_{\varphi(x)}.
\end{equation}
It is straightforward to check from Eq.~\eqref{eq:schrodingerlike} that there is always a zero mode, $\eta_0(x)={d\varphi}/{dx}$ corresponding to $\omega=0$, which is ensured by translational invariance of the system.

Based on the deformation procedure described in \cite{Bazeia.PRD.2002} and further considered in Refs. \cite{Bazeia.PRD.2004,Bazeia.PRD.2006,Bazeia:2005hu,Bazeia.EPJC.2018}, one can take the potential $V(\varphi)$ of a given model and the deformation function $f(\varphi)$ to write the new deformed model with potential $\Tilde{V}(\varphi)$ in the form of
\begin{equation}\label{eq:deformedpotential}
\Tilde{V}(\varphi)=\frac{V[\phi\to f(\varphi)]}{[f^{\prime}(\varphi)]^2},
\end{equation}
such that, in the new model the new kinks are given by 
\begin{equation}\label{eq:deformedkinks}
\Tilde{\varphi}_K(x)=f^{-1}[\varphi_K(x)].
\end{equation}
%
 \begin{figure}
 	\subfigure[]{\includegraphics[width=8cm]{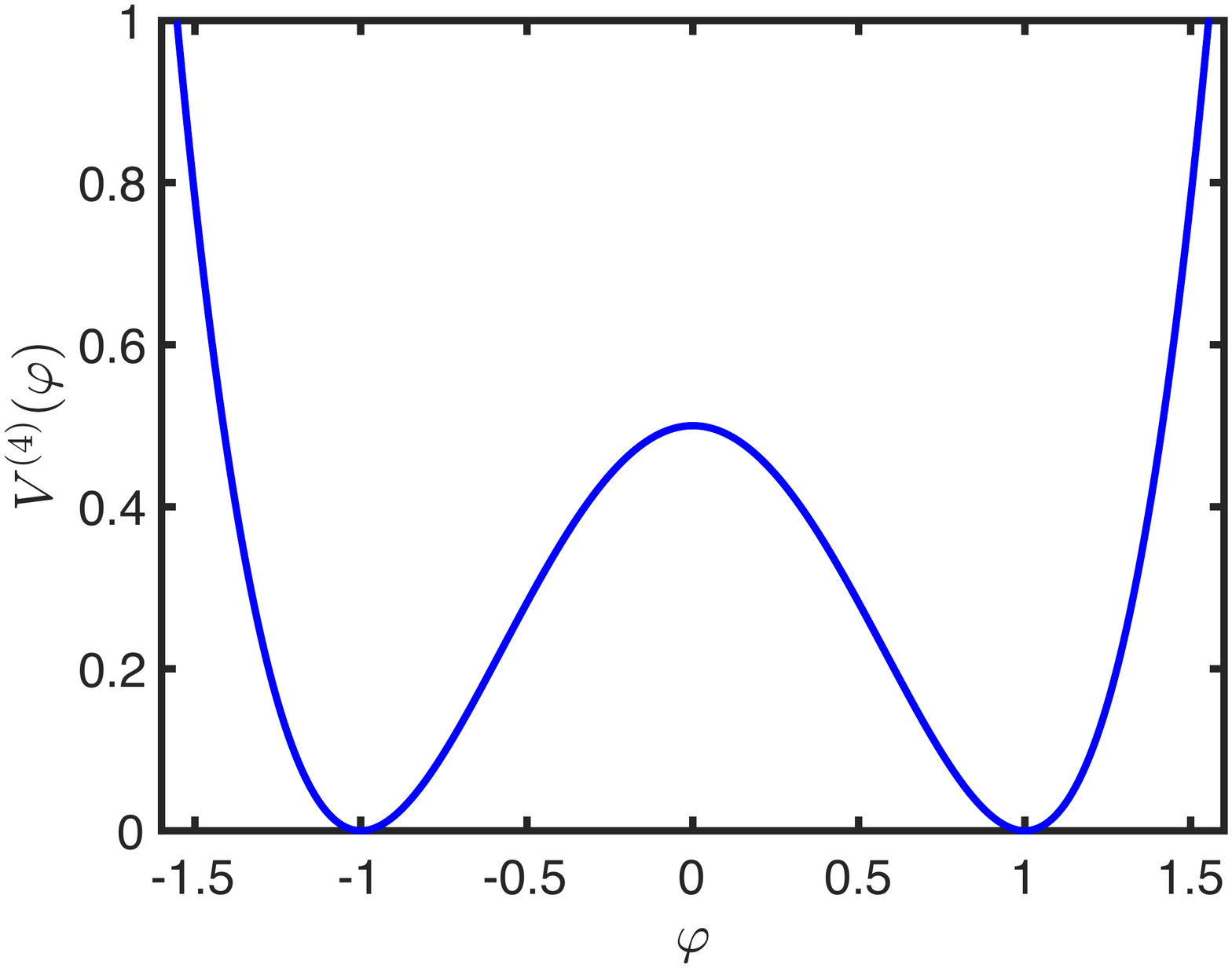}\label{fig:phi4potential}}
 	\subfigure[]{\includegraphics[width=8cm]{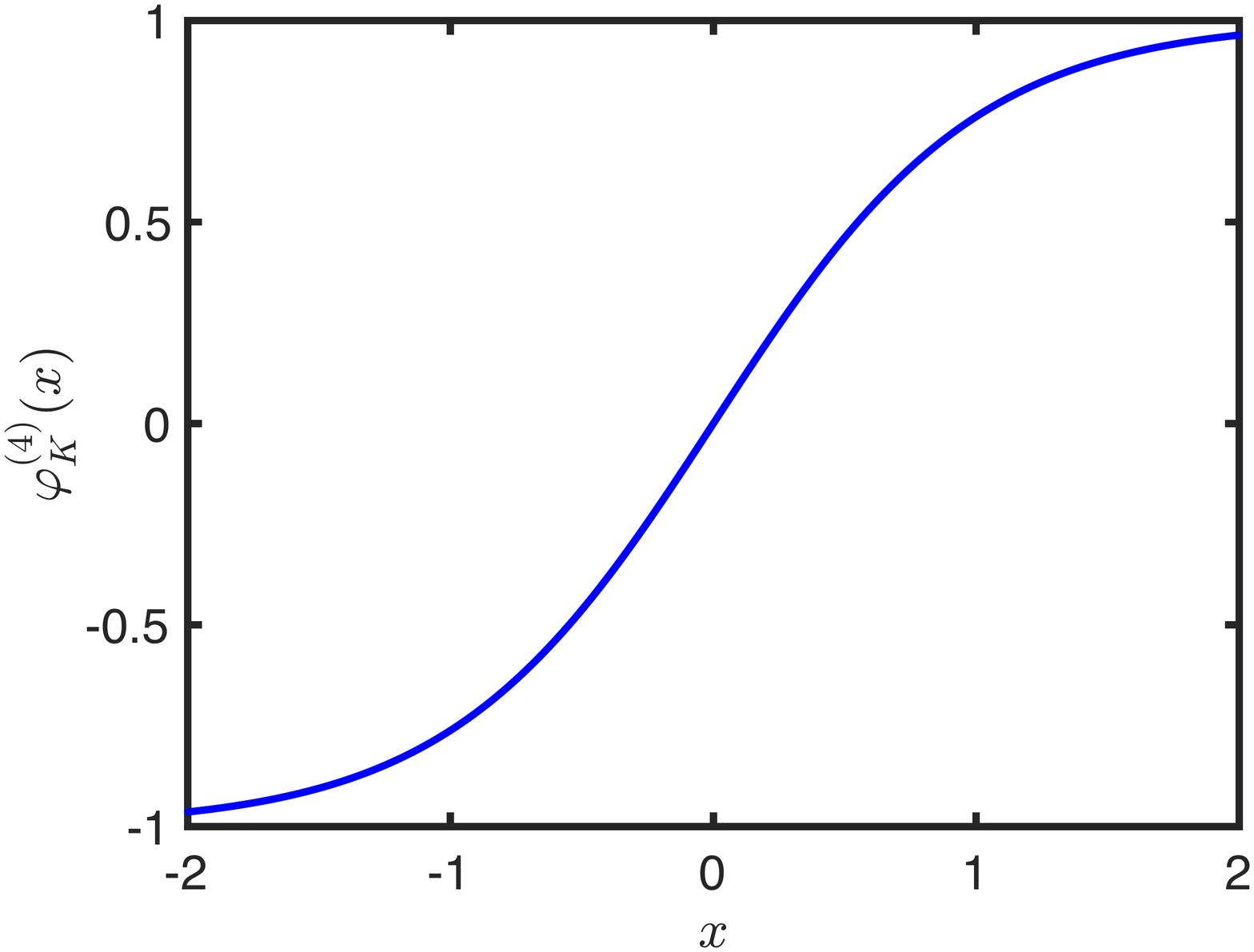}
 	\label{fig:phi4kinks}}
 	\subfigure[]{\includegraphics[width=8cm]{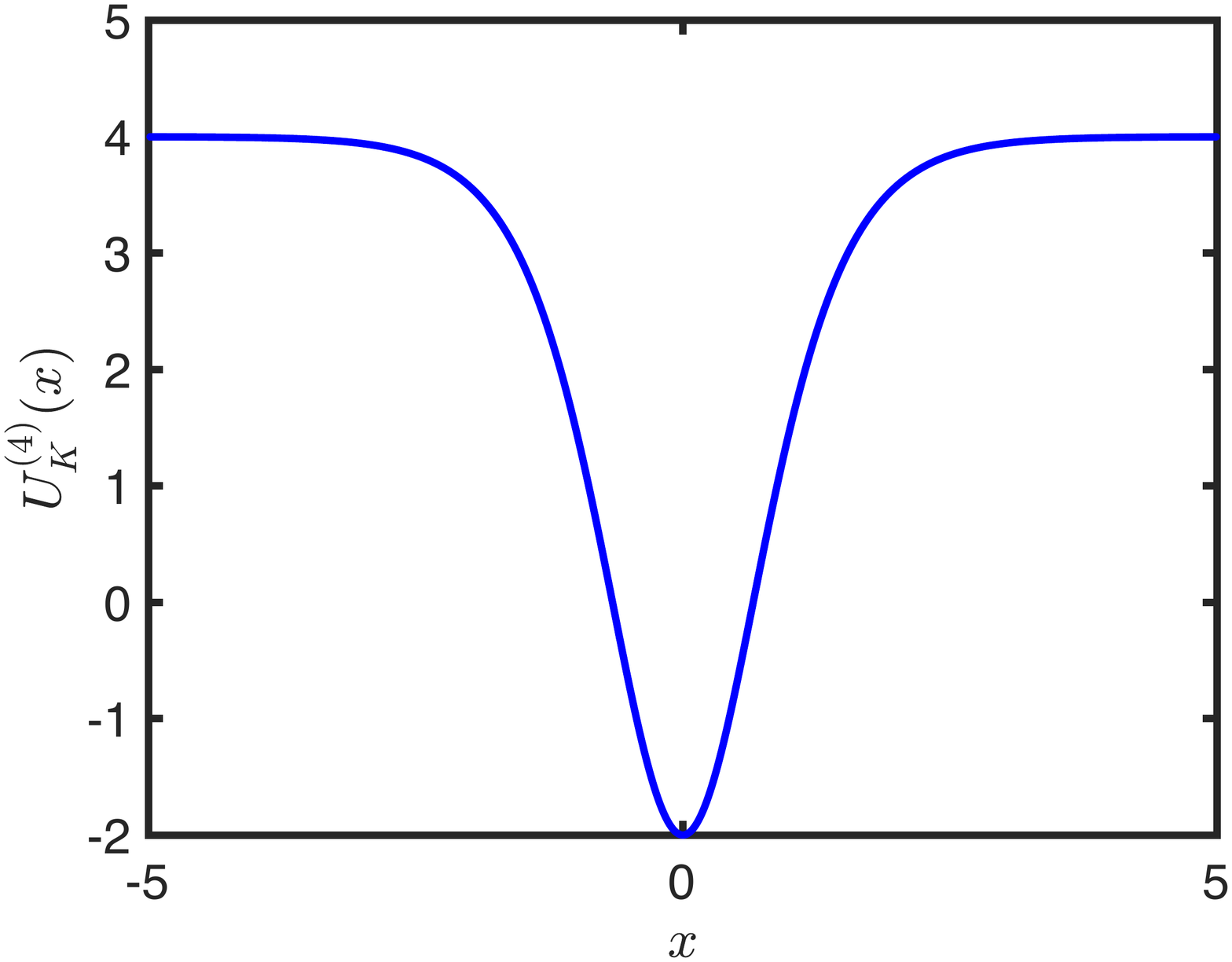}
 	\label{fig:phi4qmp}}
 	\subfigure[]{\includegraphics[width=8cm]{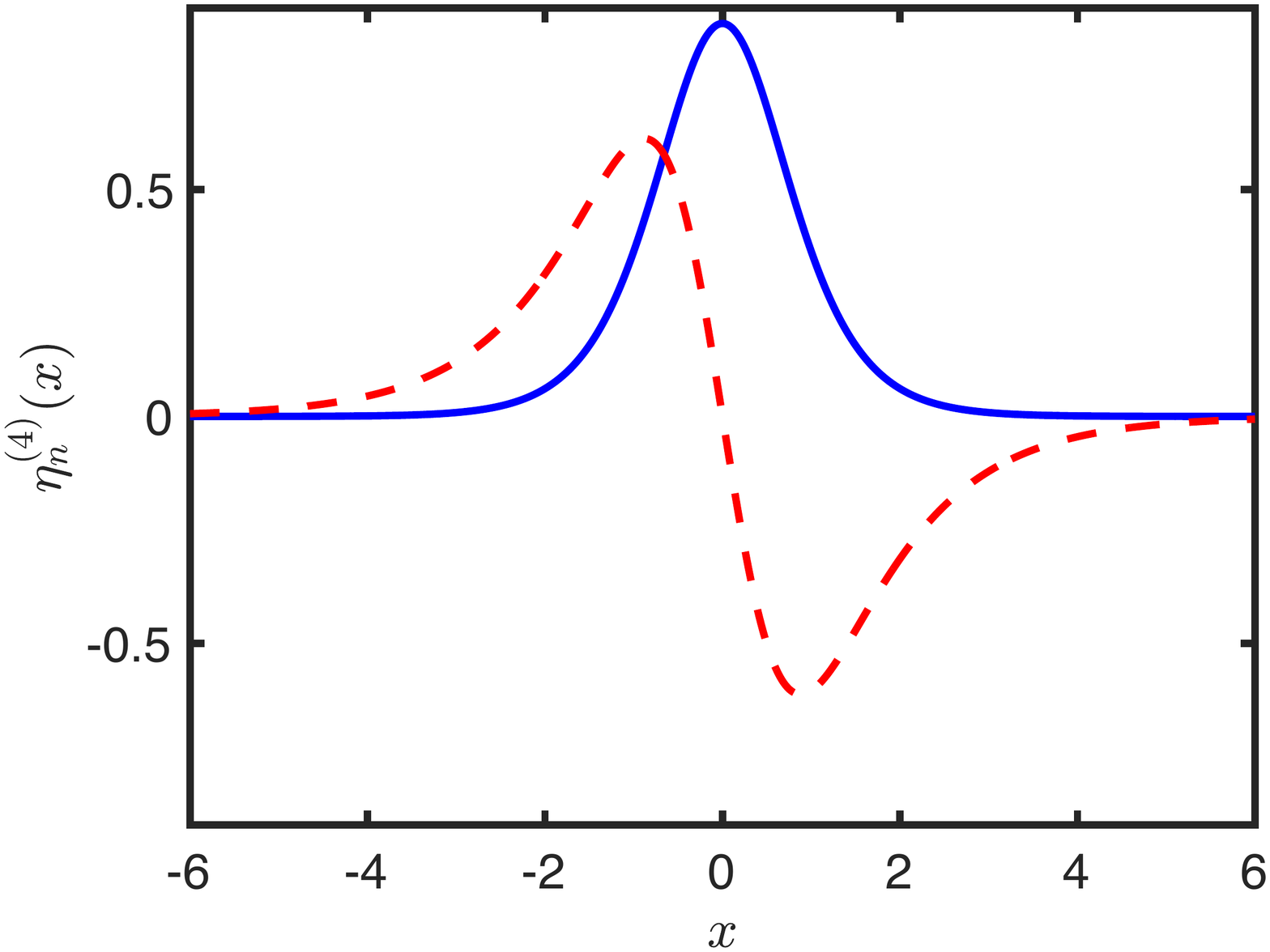}
 	\label{fig:phi4modes}}
 	 \caption{(a) The potential of Eq.~\eqref{eq:phi4potential}, (b) field configuration Eq.~\eqref{eq:phi4kinks}, (c) quantum mechanical potential of Eq.~\eqref{eq:phi4qmp} and (d) the corresponding wave functions of the bound states $\eta_0(x)$ (blue solid) and $\eta_1(x)$ (red dashed) associated to the $\varphi^4$ model.}
 	\label{fig:phi4}
\end{figure}
For example, well known $\varphi^4$ model with the potential 
\begin{equation}\label{eq:phi4potential}
V^{(4)}(\varphi)=\frac{1}{2}(1-\varphi^2)^2,
\end{equation}
has two minima at $\{-1,1\}$ (see Fig.~\ref{fig:phi4potential}) and the kink solution (Fig.~\ref{fig:phi4kinks}) which connect these minima are obtained from Eq.~\eqref{eq:EOMstaticfirstorder}:
\begin{equation}\label{eq:phi4kinks}
\varphi_{K}^{(4)}(x)= \tanh x,
\end{equation}
where the antikink solutions are given by $\varphi_{\bar{K}}^{(4)}(x) = - \varphi_{K}^{(4)}(x)$. The $K$ and $\bar{K}$ denote kink antikink solutions, respectively.
The quantum mechanical potential (Fig.~\ref{fig:phi4qmp}) is 
\begin{equation}\label{eq:phi4qmp}
U^{(4)}_{K}(x)=\frac{d^2V^{(4)}}{d\varphi^2}|_{\varphi\to \varphi^{(4)}_{K}}=4- 6\, {\rm sech}^2(x).
\end{equation}
%
 \begin{figure}
 	\subfigure[]{\includegraphics[width=8cm]{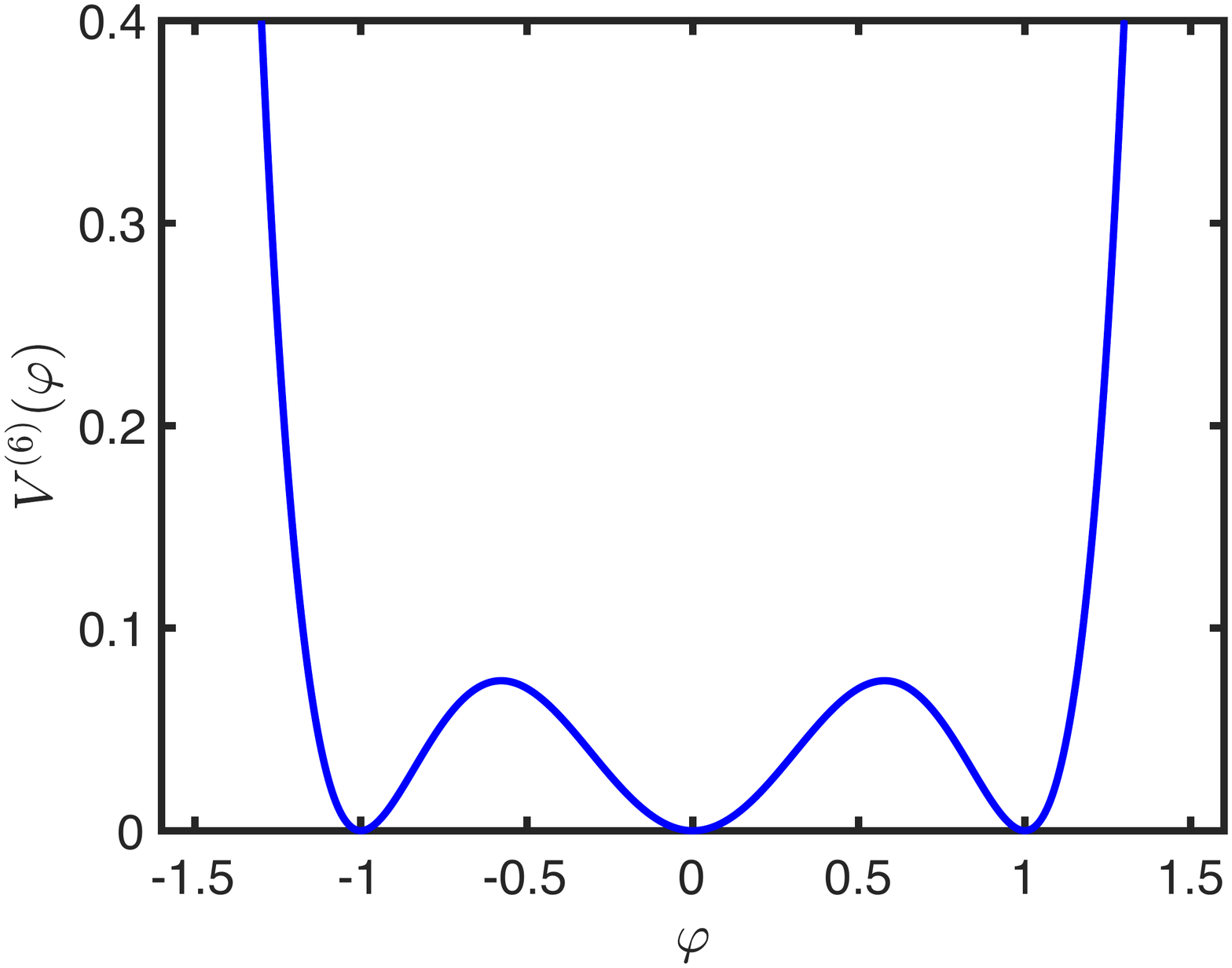}
 	\label{fig:phi6potential}}
 	\subfigure[]{\includegraphics[width=8cm]{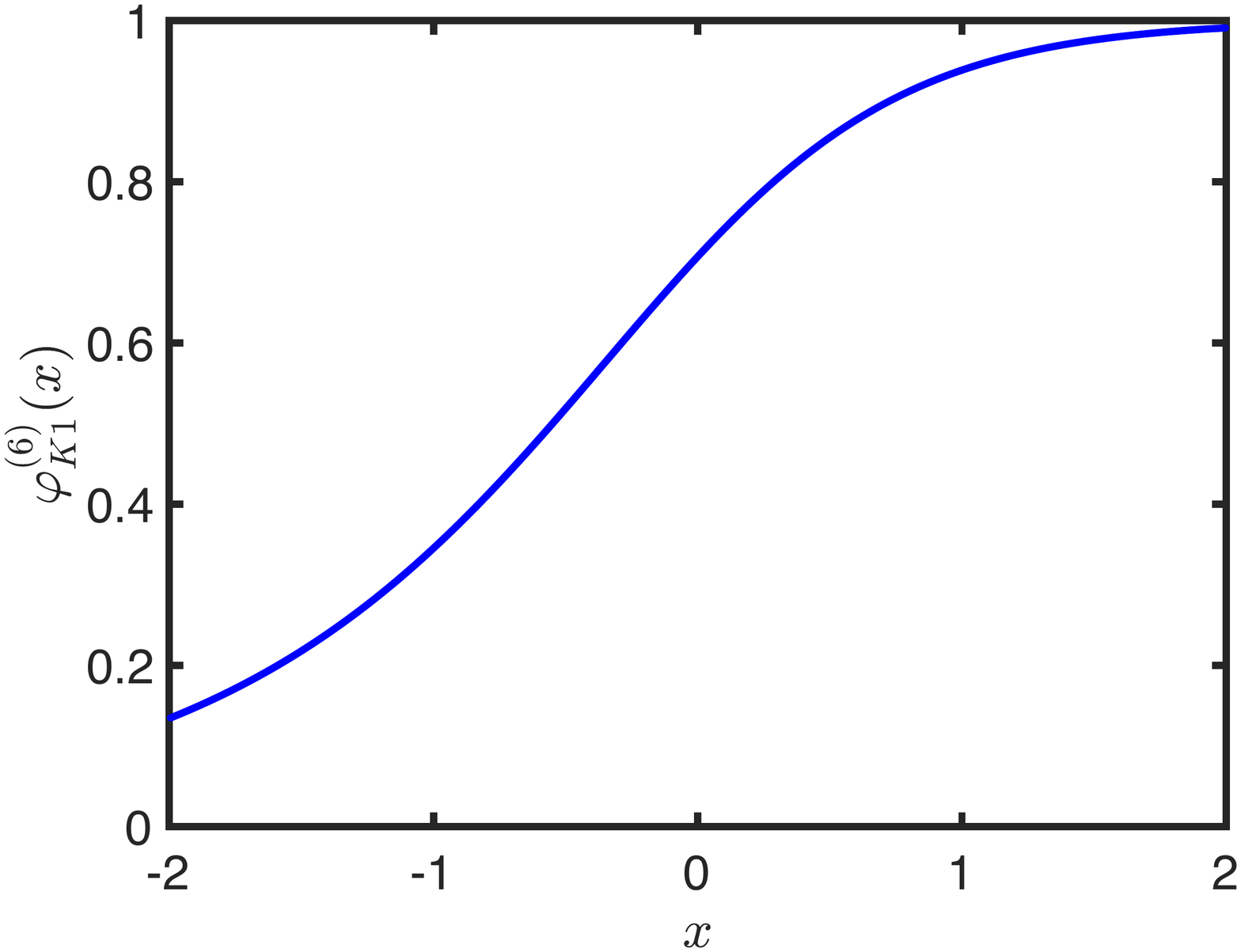}
 	\label{fig:phi6kinks}}
 	\subfigure[]{\includegraphics[width=8cm]{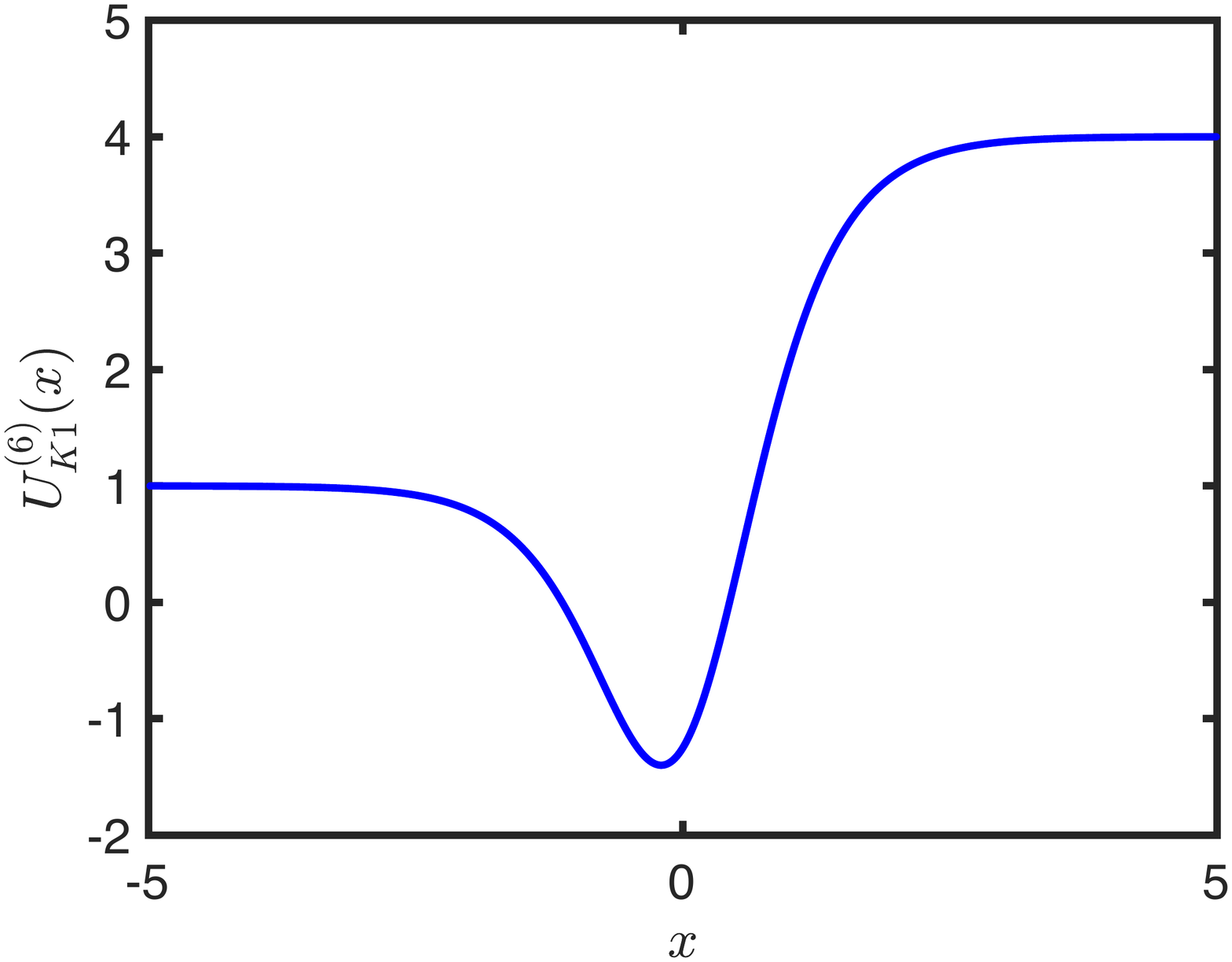}
 	\label{fig:phi6qmp}}
 	\subfigure[]{\includegraphics[width=8cm]{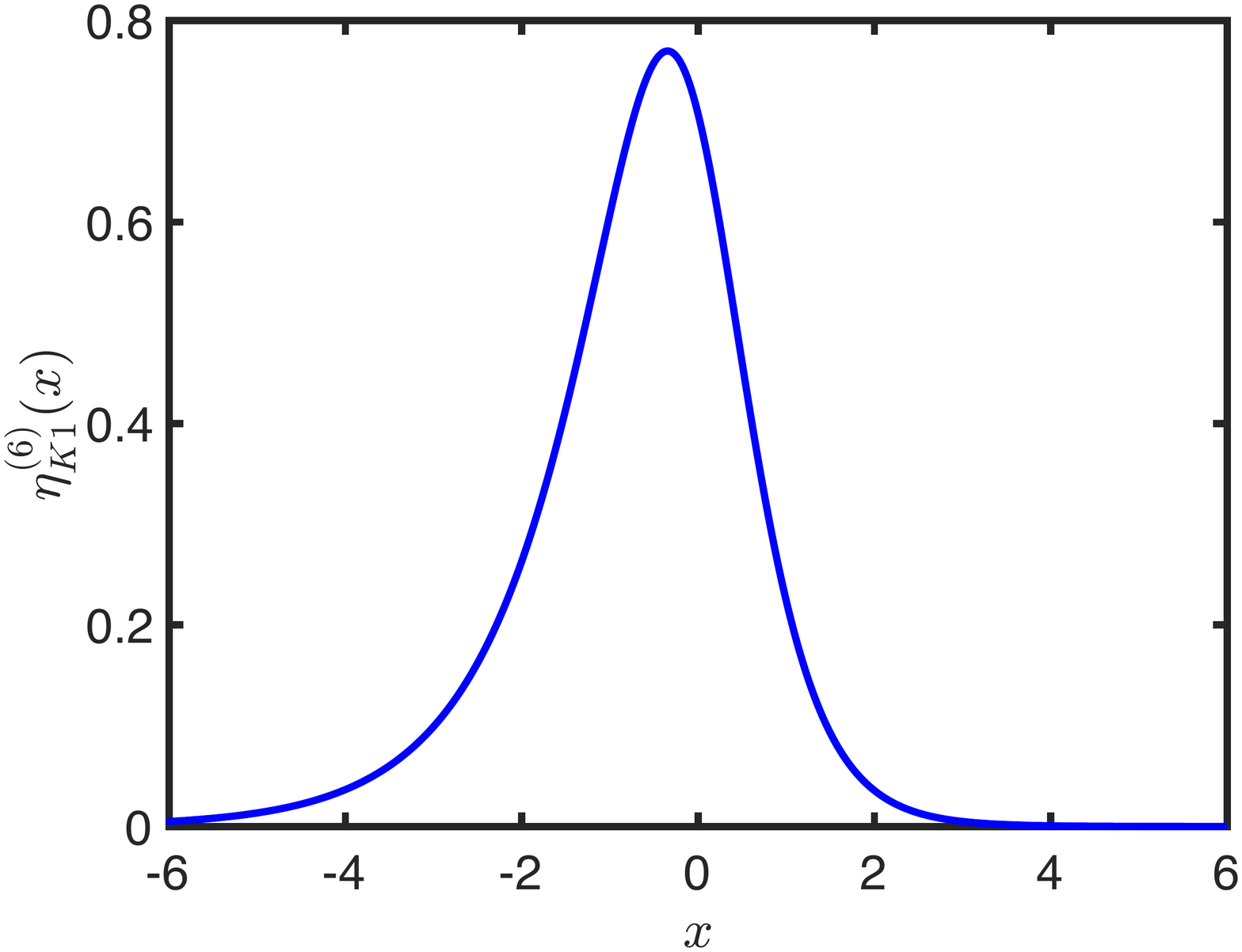}
 	\label{fig:phi6mode}}
 	 \caption{(a) The potential in Eq.~\eqref{eq:phi6potential}, (b) the kinks (Eq.~\eqref{eq:phi6kinks}), (c) quantum mechanical potential (Eq.~\eqref{eq:phi6qmp}) and (d) the corresponding wave function of the $\varphi^6$ model.}
 	\label{fig:phi6}
\end{figure}

The classical mass of the $\varphi^4$ kinks is $M^{(4)}_{K}={3}/{4}$, and they have an internal mode in addition to the zero mode, see Fig.~\ref{fig:phi4modes}.
Another example is the $\varphi^6$ model, given by the following potential
\begin{equation}\label{eq:phi6potential}
V^{(6)}(\varphi)=\frac{1}{2}\varphi^2(1-\varphi^2)^2,
\end{equation}
it has three minima at $\{-1,0,1\}$, as we can see in Fig.~\ref{fig:phi6potential}, therefore it has two kinks and two antikinks. The kink static connecting the minima $\{0,1\}$ is depicted in Fig. \ref{fig:phi6kinks}, given by

\begin{equation}\label{eq:phi6kinks}
\varphi^{(6)}_{K_1}(x)= \sqrt{\frac{1 + \tanh x}{2}}.
\end{equation}
The antikink interpolating between $\{1,0\}$ is $\varphi^{(6)}_{\bar{K}_1}(x) = \varphi^{(6)}_{K_1}(-x)$. The another kink and antikink connecting the other topological sector are given by $\varphi^{(6)}_{K_2}(x) = - \varphi^{(6)}_{K_1}(-x) $ and $\varphi^{(6)}_{\bar{K}_2}(x)= - \varphi^{(6)}_{K_1}(x)$. The quantum mechanical potential of the $\varphi^6$ model is
\begin{equation}\label{eq:phi6qmp}
U^{(6)}=\frac{d^2V^{(6)}}{d\varphi^2}\biggr|_{\varphi\to \varphi^{(6)}},
\end{equation}
that is,
\begin{eqnarray}\label{eq:phi6qmp1}
U^{(6)}_{K_1}(x)= \frac{15\tanh^2(x)}{4} + \frac{3\tanh(x)}{2}  - \frac54.
\end{eqnarray}
In Fig.~\ref{fig:phi6qmp} we depicted the plot of this potential. The $\varphi^6$ kinks have mass $M^{(6)}={1}/{4}$ and they have no internal mode. Moreover, the model has one zero mode, for each topological sector. The Fig.~\ref{fig:phi6mode} show the zero mode for sector $\{0,1\}$.


\section{Tanh-deformed $\varphi^4$ model} \label{sec:TanhDeformedPhi4Model}


Let us now consider the deformed $\varphi^4$ model with a \emph{hyperbolic tangent} deformation function. 
We consider the deformation function $f(\varphi)=r \tanh{\varphi}$, $r$ real, obeying $r>1$, and we use it to deform the $\varphi^4$ model, the first model studied above. The procedure leads us to the new potential
\begin{eqnarray}\label{eq:hypphi4potential}
\Tilde{V}_{r}^{(4)}(\varphi)=\frac{1}{2 r^2}\left(\cosh ^2 \varphi-r^2 \sinh ^2 \varphi \right)^2.
\end{eqnarray}
The Eq.~\eqref{eq:EOM} with this potential yields the equation of motion of the model
\begin{eqnarray}\label{eq:hypphi4EQM}
\frac{\partial^2 \varphi}{\partial t^2}-\frac{\partial^2 \varphi}{\partial x^2}+\frac{\left(r^2-1\right) }{r^2}\sinh 2\varphi \left(r^2 \sinh ^2 \varphi-\cosh ^2\varphi \right)=0.
\end{eqnarray}
The potential Eq.~\eqref{eq:hypphi4potential} has two degenerate minima $\Tilde{\varphi}_{\pm}=\pm \tanh ^{-1}({1}/{r})$. Therefore, the model has only one kink and one antikink. The kink configuration appears according to the Eq.\eqref{eq:deformedkinks}, and it has the form
\begin{eqnarray}\label{eq:hypphi4kinks}
\Tilde{\varphi}^{(4)}_K(x)= \tanh ^{-1}\left(\frac{1}{r}\tanh (x)\right),
\end{eqnarray}
and the antikink is obtained by changing $x\to-x$. 
 \begin{figure}
 	\subfigure[]{\includegraphics[width=8cm]{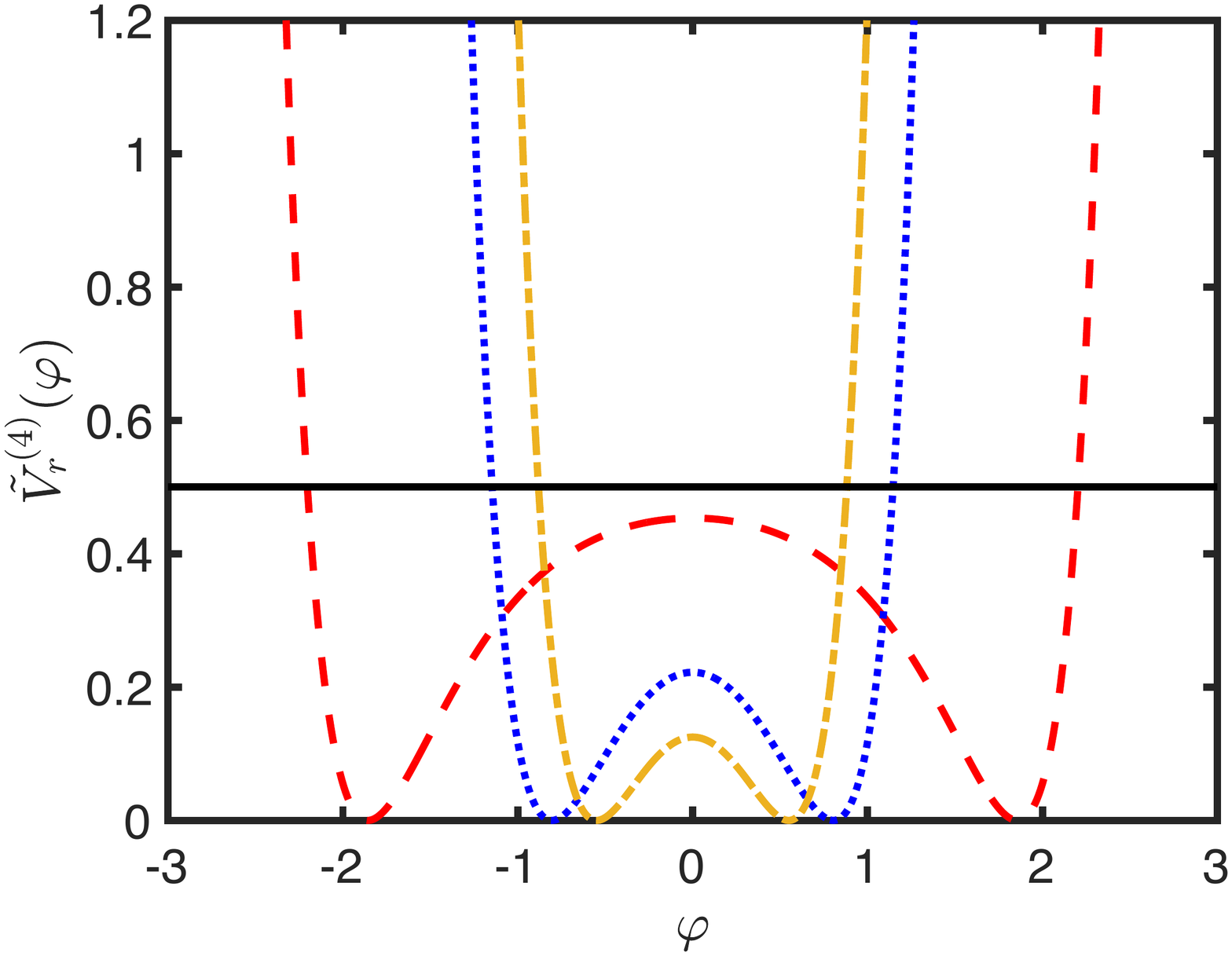}\label{fig:hypphi4potentials}}
 	\subfigure[]{\includegraphics[width=8cm]{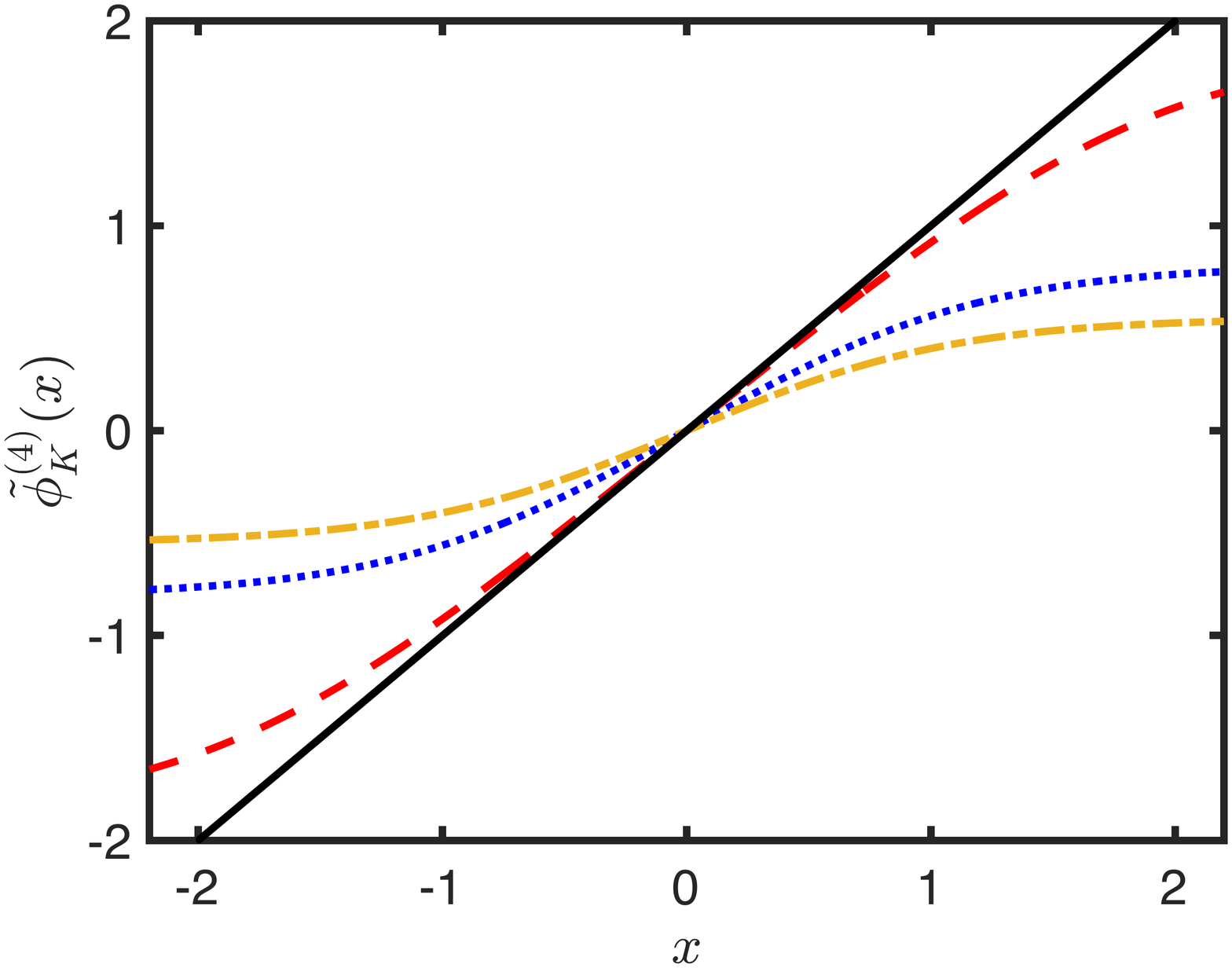}\label{fig:hypphi4kinks}}
 	 \caption{(a) The potential in Eq.~\eqref{eq:hypphi4potential} and (b) the kinks Eq.~\eqref{eq:hypphi4kinks} of the tanh-deformed $\varphi^4$ model for $r=1.0$ (solid black), $r=1.05$ (red dash), $r=1.5$ (blue dot) and $r=2$ (yellow dot-dash).}
 	\label{fig:hypphi4solutionspotentials}
\end{figure}

\begin{figure}
 	\subfigure[]{\includegraphics[width=8cm]{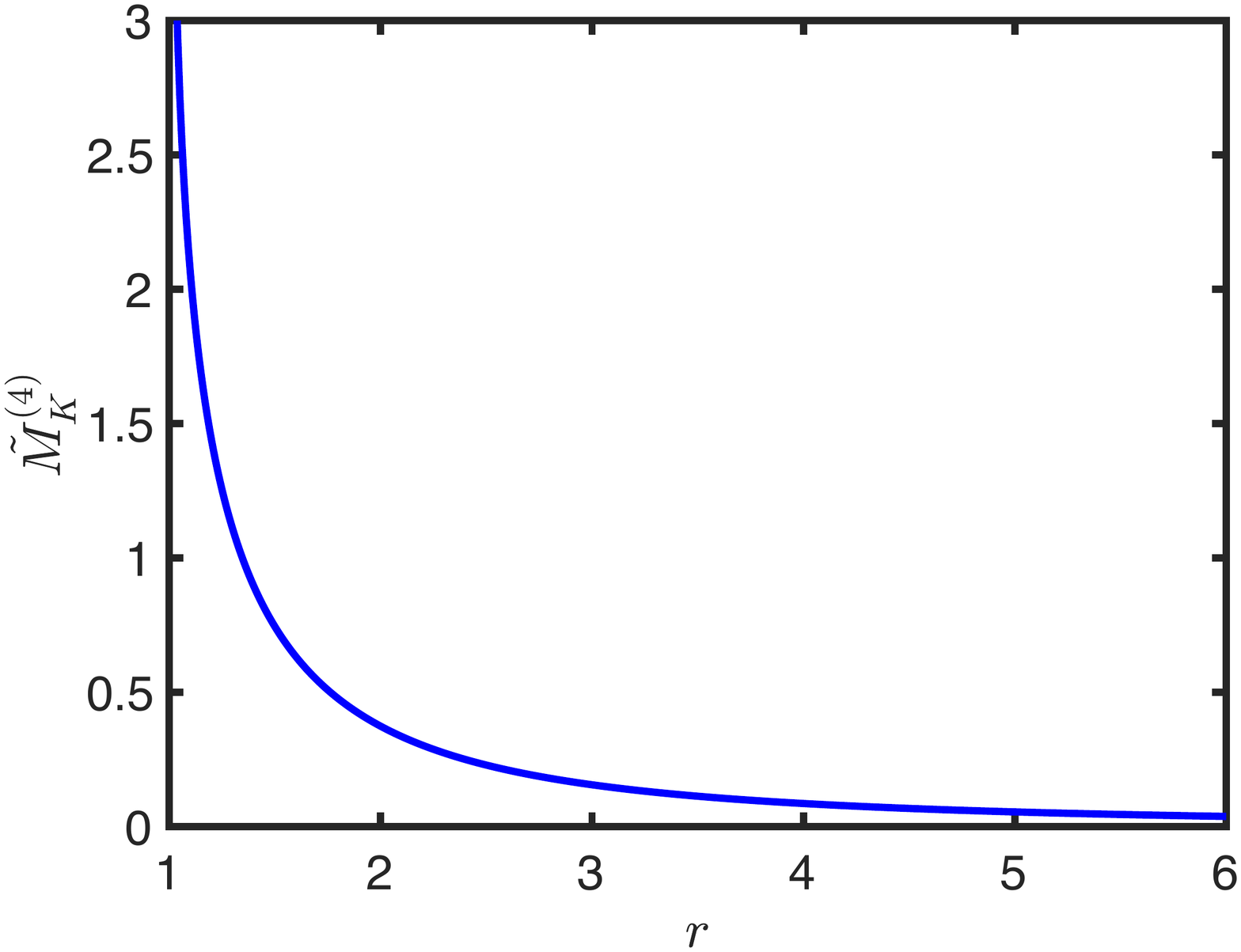}\label{fig:hypphi4mass}}
 	\subfigure[]{\includegraphics[width=8cm]{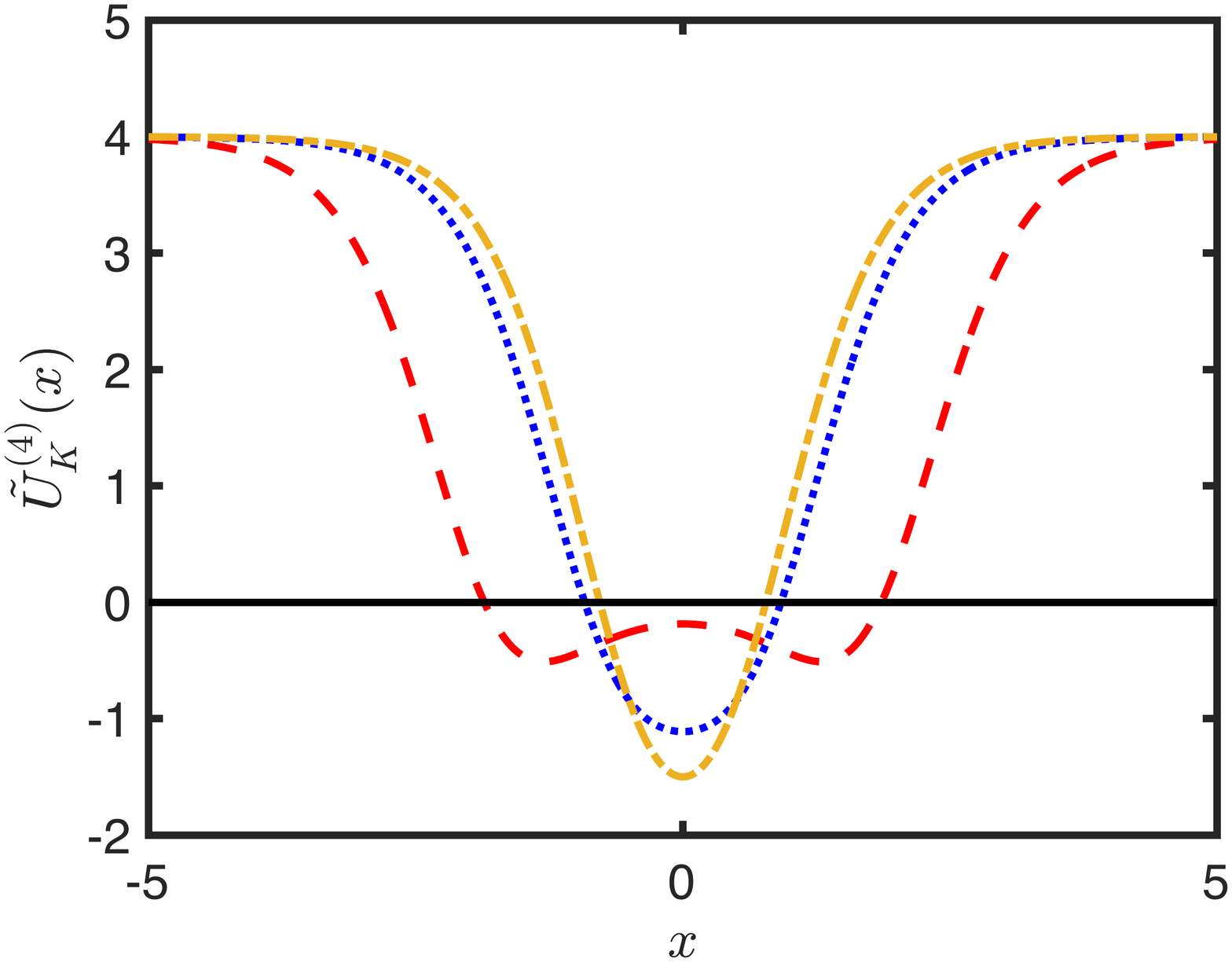}\label{fig:hypphi4QMPs}}
 	 \caption{(a) The masses of kink Eq.~\eqref{eq:tanhdeformedphi4mass} and (b) quantum mechanical potentials Eq.~\eqref{eq:tanhdeformedphi4QMP} of the tanh-deformed $\varphi^4$ model for $r=1.0$ (solid black), $r=1.05$ (red dash), $r=1.5$ (blue dot) and $r=2$ (yellow dot-dash).}
 	\label{fig:hypphi4qmpmass}
\end{figure}

Depending on the parameter $r$, the potential Eq. \eqref{eq:hypphi4potential} and the kinks in Eq. \eqref{eq:hypphi4kinks} look different, as shown in Figs.~\ref{fig:hypphi4potentials} and \ref{fig:hypphi4kinks}. The dependence of the potential center on $r$ is in the form of ${1}/{2r^2}$. It is clear that at $r \to 1$, the center of potential tends towards $1/2$, the minima move away from each other and the kinklike solution tends to a straight line, $\Tilde{\varphi}^{(4)}_{r\to 1}(x) \to x$. At  $r \to \infty$, the center of potential tends to zero, the minima get closer to each other and therefore the potential turns into an infinitely narrow square well. 
\begin{eqnarray}\label{eq:potential2}
\Tilde{V}_r^{(4)}(\varphi=0) \to 
\begin{cases}
\frac{1}{2} \quad for \quad r \to 1 
\\
0 \quad for \quad r \to \infty.
\end{cases}
\end{eqnarray}
The masses of these kinks are
\begin{eqnarray}\label{eq:tanhdeformedphi4mass}
\Tilde{M}_{K}^{(4)}=\frac1r(r^2+1) \coth ^{-1}(r)-1,
\end{eqnarray}
which is plotted as a function of $r$ in Fig.~\ref{fig:hypphi4mass}. In this model, the $r$ dependent Scho\"odinger-like  potential has the following form

\begin{eqnarray}\label{eq:tanhdeformedphi4QMP}
\tilde{U}^{(4)}_K(x) = \frac{(r^2-1)^2}{r^2} \cosh (4{\tilde{\varphi}_{K}^{(4)}(x)}) -\frac{r^4-1}{r^2} \cosh (2 {\tilde{\varphi}_{K}^{(4)}(x)} ).
\end{eqnarray}
In the Fig. ~\ref{fig:hypphi4QMPs} we display this potential for some values of $r$. For small values of $r$, the potential has a global maximum at $x=0$ and appear two minima. This behavior suggests the emergence of internal modes at $r$ very close to unity. As $r$ increases, we notice a single minimum forming around $x=0$, with the potential becoming deeper and narrower.

%

\begin{figure}
 	\subfigure[]{\includegraphics[width=8cm]{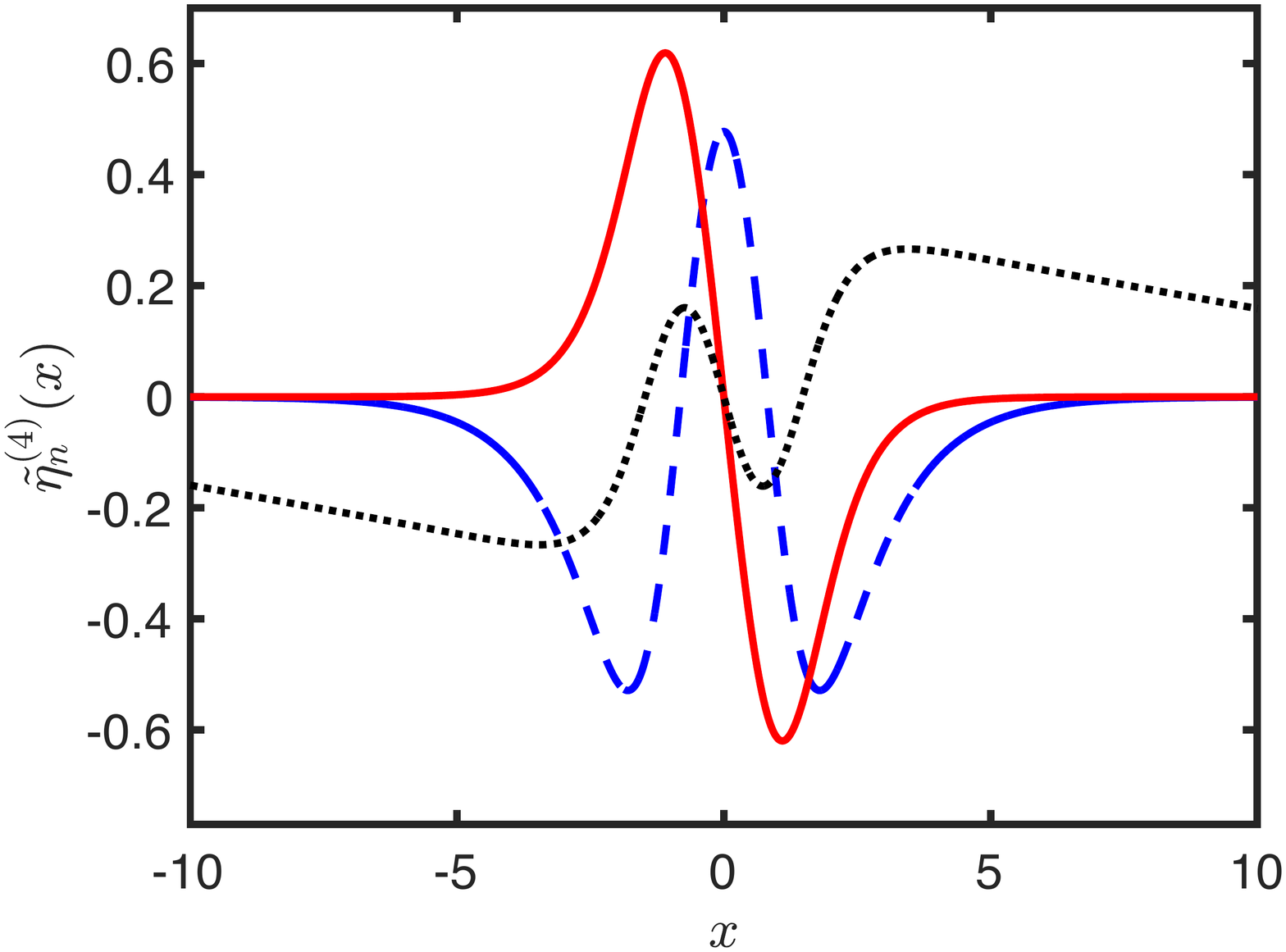}\label{fig:n118modes}}
 	\subfigure[]{\includegraphics[width=8cm]{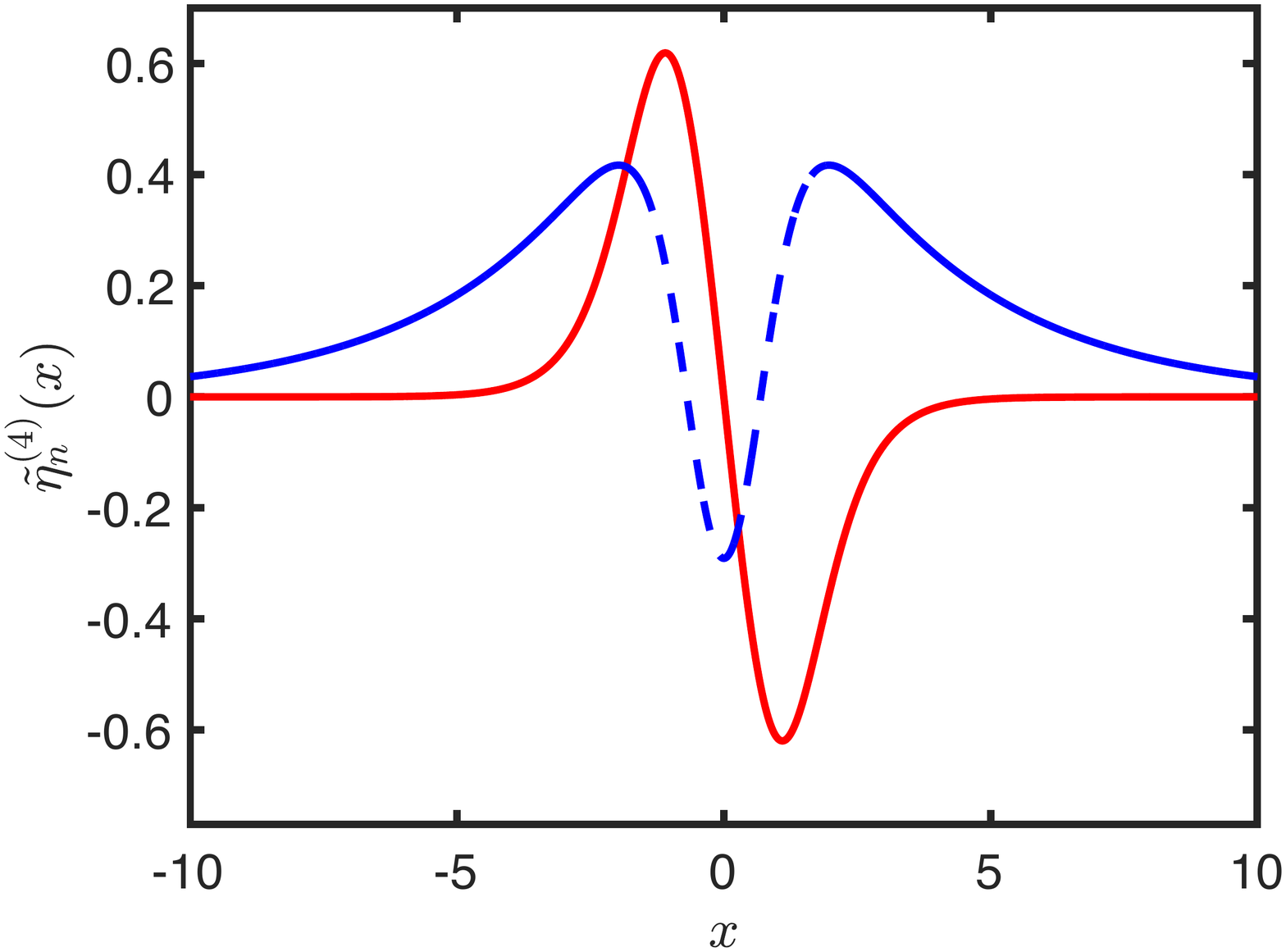}\label{fig:n186modes}}
 	 \caption{Wave functions corresponding for the cases of (a) $r=1.18$ and (b) $r=1.86$.}
\label{fig:tanhphi4examplemodes}
\end{figure}

Using the {\it Shooting method} (see, e.g., \cite{Izaac.book.2018,Belendryasova.CNSNS.2019}) in the $\text{Schr\"odinger-like}$ equation, Eq.~\eqref{eq:schrodingerlike}, the energy levels of the discrete spectrum in the potential well, Eq.~\eqref{eq:tanhdeformedphi4QMP}, can be obtained for different values of parameter $r$. The Fig.~\ref{fig:tanhphi4examplemodes} shows the wave function solutions for the cases of $r=1.18$ and $r=1.86$. The first case has three internal modes in addition to the zero mode, $\omega_1^2=1.5054795$, $\omega_2^2=3.160485$ and $\omega_3^2=3.9957262$, while the second case has two modes, $\omega_1^2=2.4915203$ and $\omega_2^2=3.8953012$.

\begin{figure}
 	\includegraphics[width=9cm]{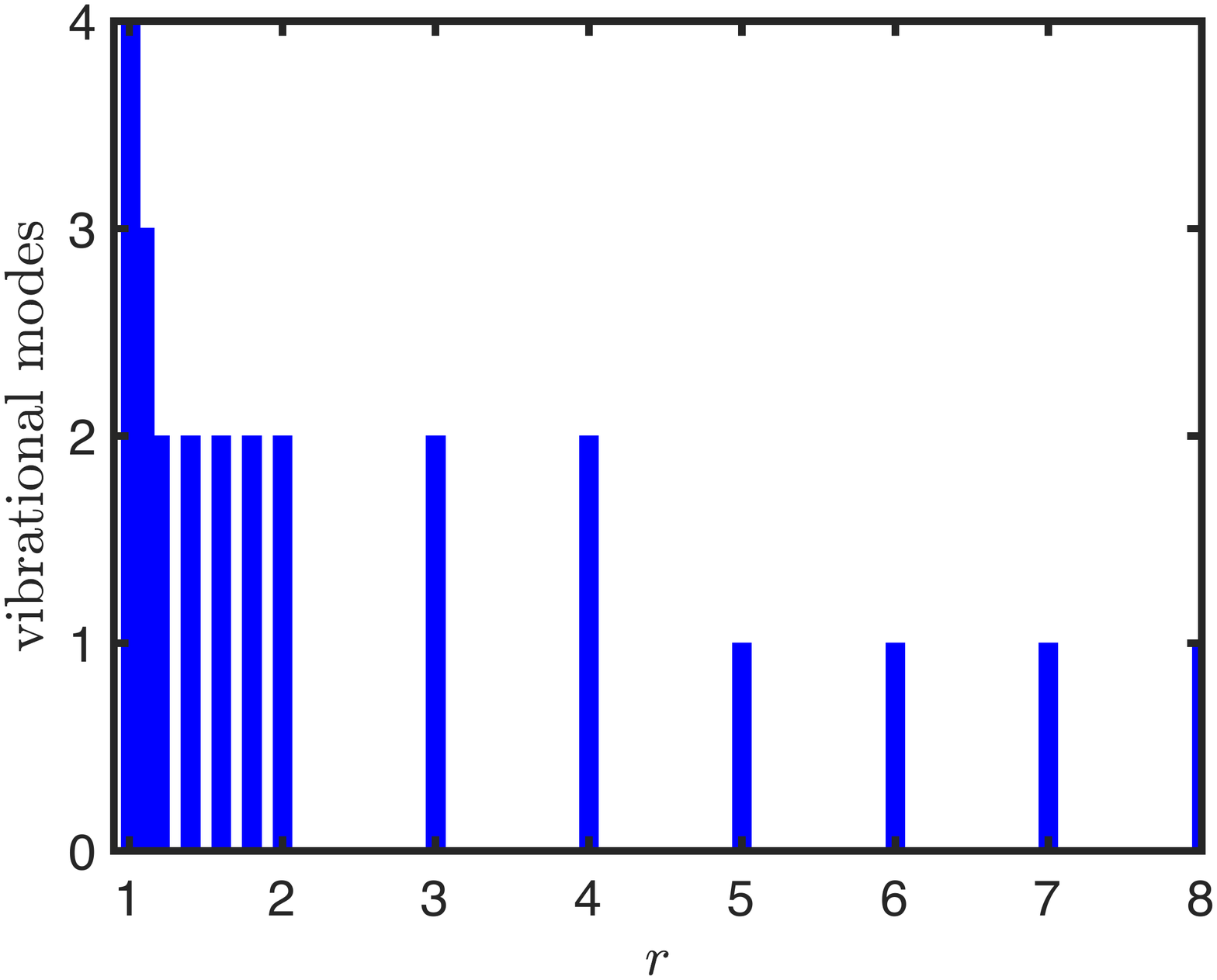}
 	 \caption{The number of vibrational modes for different values of parameter $r$ of the tanh-deformed $\varphi^4$ model.}
 	\label{fig:hypphi4modes}
\end{figure}

The parameter $r$ directly affects the depth and width of the potential well, which is a factor in changing the number of internal modes of the potential. In the Figs. \ref{fig:hypphi4modes} we show our main result for the number of vibrational modes for different values of parameter $r$. The closer the $r$ parameter is to one, the greater the number of internal modes.

\subsection{Numerical results} 


To investigate another effect of parameter $r$, we consider the kink scattering. Due to the fact that the kinks of the tanh-deformed $\varphi^4$ model are symmetric, the kink-antikink and antikink-kink collisions always give the same results, as it also happens in the $\varphi^4$ model. So we only investigate the kink-antikink collision.

For the numerical solutions, we solved the equation of motion with the following initial conditions
\begin{eqnarray}\label{eq:conditionphi4}
\varphi(x,0) & = & \tilde{\varphi}^{(4)}_{K}(x+x_0,v,0) + \tilde{\varphi}^{(4)}_{\bar{K}}(x-x_0,-v,0) - \tanh^{-1}\bigg(\frac1r\bigg),\\
\dot{\varphi}(x,0) & = & \dot{\tilde{\varphi}}^{(4)}_{K}(x+x_0,v,0) + \dot{\tilde{\varphi}}^{(4)}_{\bar{K}}(x-x_0,-v,0).
\end{eqnarray}
We fixed $x_0=10$ for the initial symmetric position of the pair and set the grid boundary at $x_{max}=\pm200$. We use a $4^{th}$ order finite-difference method with the spatial step $\delta x = 0.05$ and $6^{th}$ order sympletic integrator with time step $\delta t = 0.02$. It is important to note that, $\varphi(x,t) = \varphi(\gamma(x-vt))$, means boost of Lorentz for static kink and $\gamma = 1/\sqrt{1-v^2}$.

\begin{figure}
 	\includegraphics[width=9cm]{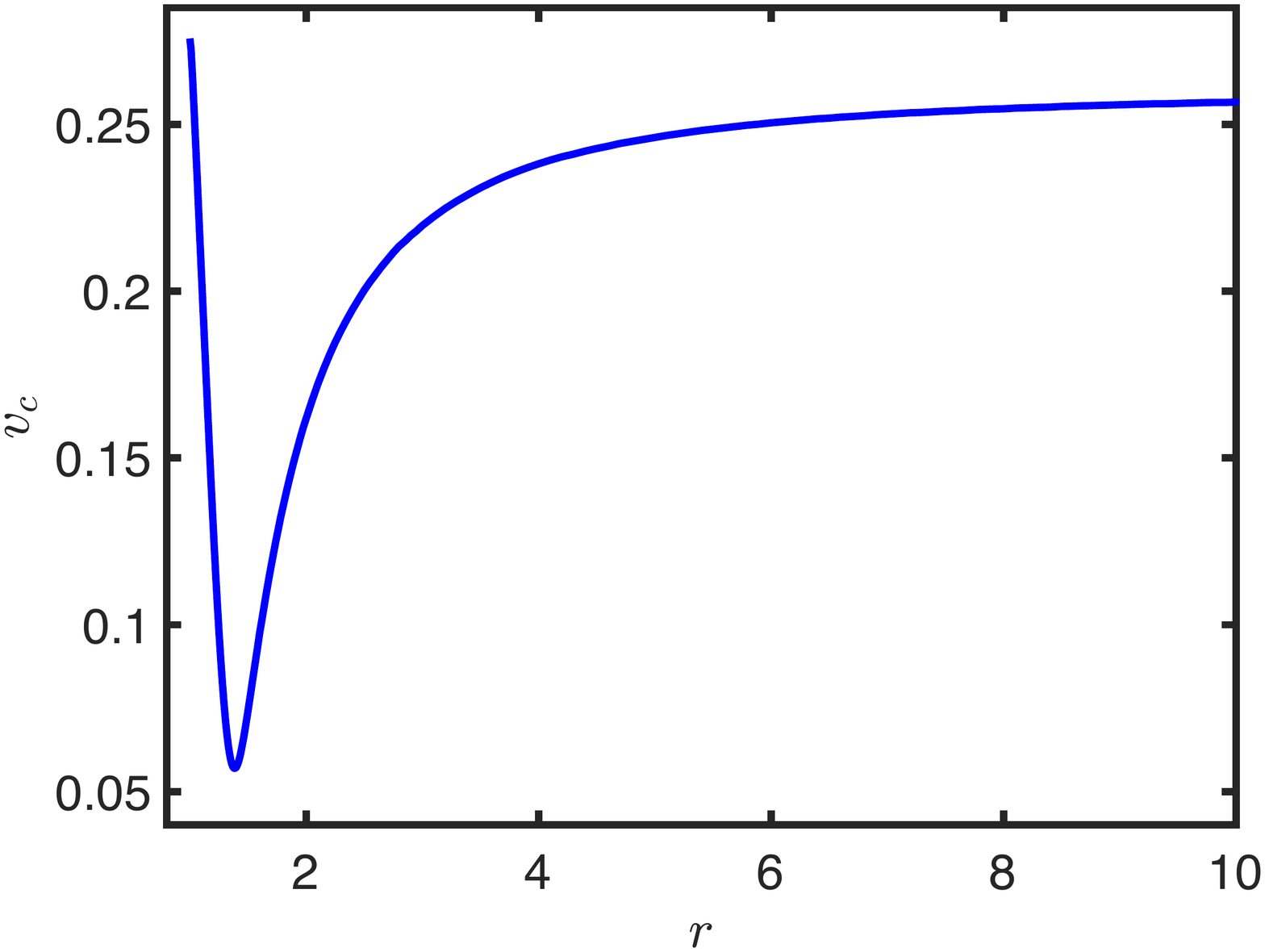}
 	 \caption{Critical velocity in the kink-antikink collision as function of $r$ of the tanh-deformed $\varphi^4$ model.}
 	\label{fig:hypphi4criticalvelocity}
\end{figure}

We illustrate in the Fig. \ref{fig:hypphi4criticalvelocity} the critical velocity $v_c$ as function of the parameter $r$. The critical velocity is the boundary between two different regions in terms of the final products of the collision. If the initial velocity $v$ is larger than  $v_c$ ($v>v_c$), the kink and the antikink escape to infinity after one collision. For the initial velocity less than the critical velocity ($v<v_c$), the kink and the antikink become trapped after the collision and create a bound state. In this region, they can escape to infinity after some bounces. 

\begin{figure}
 	\subfigure[]{\includegraphics[width=8cm]{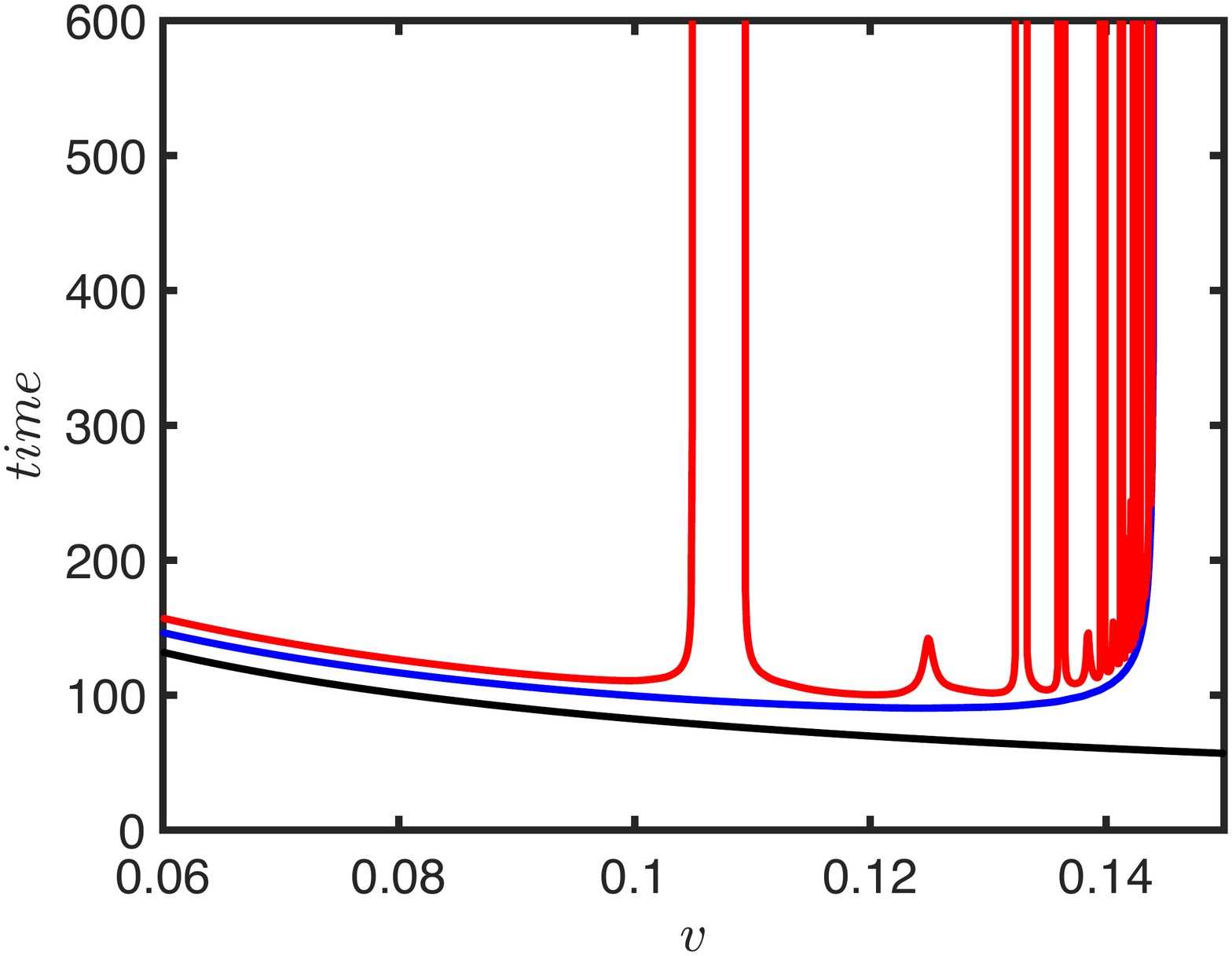}\label{fig:scattn118}}
 	\subfigure[]{\includegraphics[width=8cm]{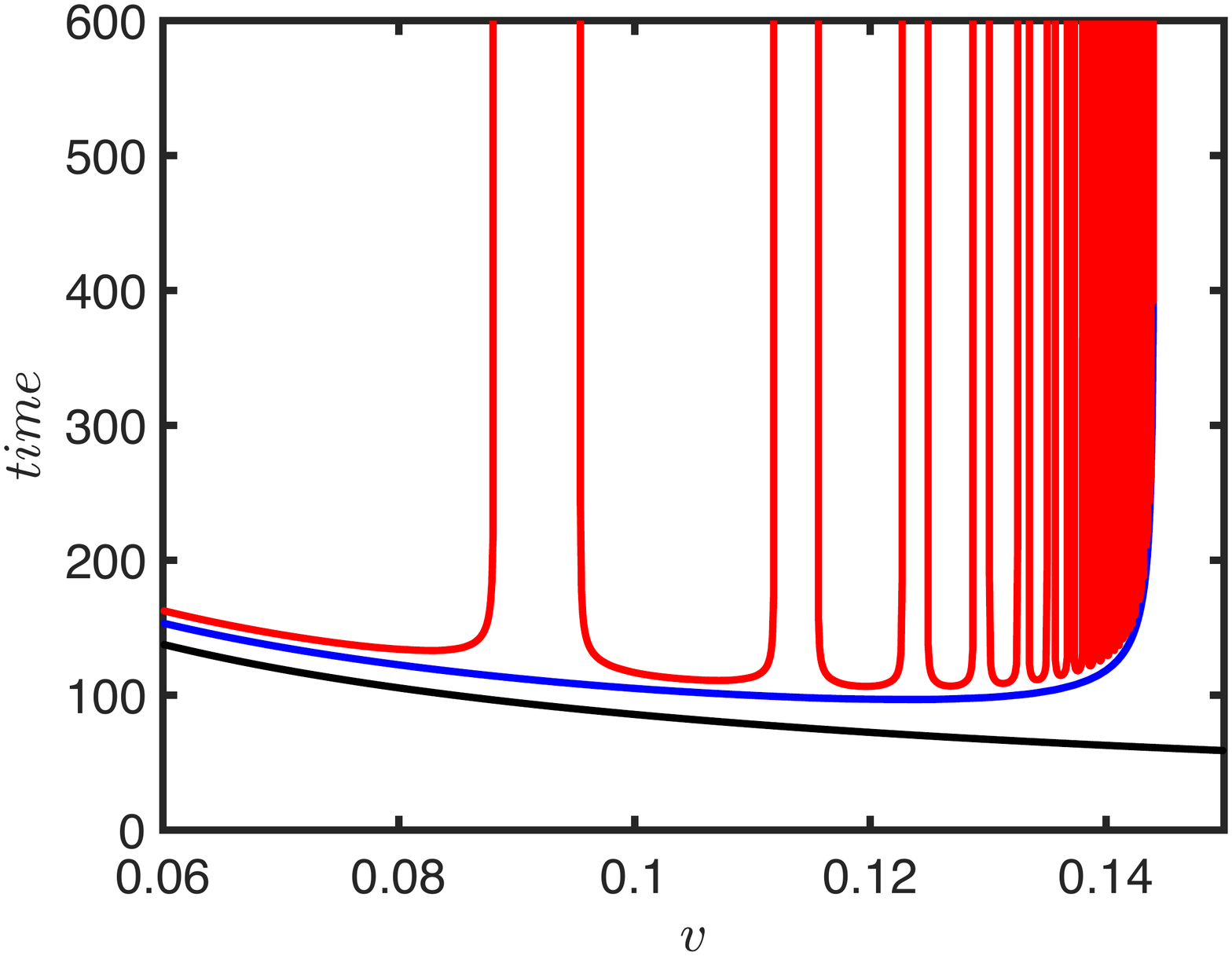}\label{fig:scattn186}}
 	 \caption{Time to the first (black), second (blue) and third (red) kink-antikink collisions for tanh-deformed $\varphi^4$ model as a function of $v$ for (a) $r=1.18$ and (b) $r=1.86$ }
 	\label{fig:scatt}
\end{figure}

Based on the correlation between the number and value of the bound states, we can understand the behavior of  critical velocity as follows. The larger the $\omega$, the smaller the time interval between bounces, which means that the energy transfer from translational to vibrational mode becomes more difficult to realize. Particularly, for small values of $r$, the extra $\omega_3$ bound mode has a larger value. The presence of this mode contributes to the suppression of the two-bounce windows - see Fig.~\ref{fig:scattn118} - and its high value hinders the resonant energy exchange, enlarging the one-bounce region. We notice that the critical velocity $v_c$ reaches a minimum at $r \approx 1.385$, as illustrated in Fig. \ref{fig:hypphi4criticalvelocity}. This region corresponds to the transition from three to two internal modes, where we detected a change in the critical velocity. This figure depicts behavior that is quite similar to Refs. \cite{Bazeia.EPJC.2021,SimasJHEP}. Furthermore, for large values of $r$, the tanh-deformed $\varphi^4$ model has only one vibrational state. For instance, our numerical investigation revealed that for $r=10$, there is only one internal mode with $\omega_1^2=2.9839491$. This value closely resembles the $\varphi^4$ model, where we have $\omega^2=3$. As a result, the kink-antikink scattering process for the tanh-deformed $\varphi^4$ model demonstrates that the critical velocity increases proportionally to $r$. It is worth noting that the critical velocity for large values of $r$ approaches the value of the $\varphi^4$ model, where it is $v_c=0.2598$.

\begin{figure}
    \subfigure[\quad three-bounce for $v=0.1342$]{\includegraphics[width=8cm]{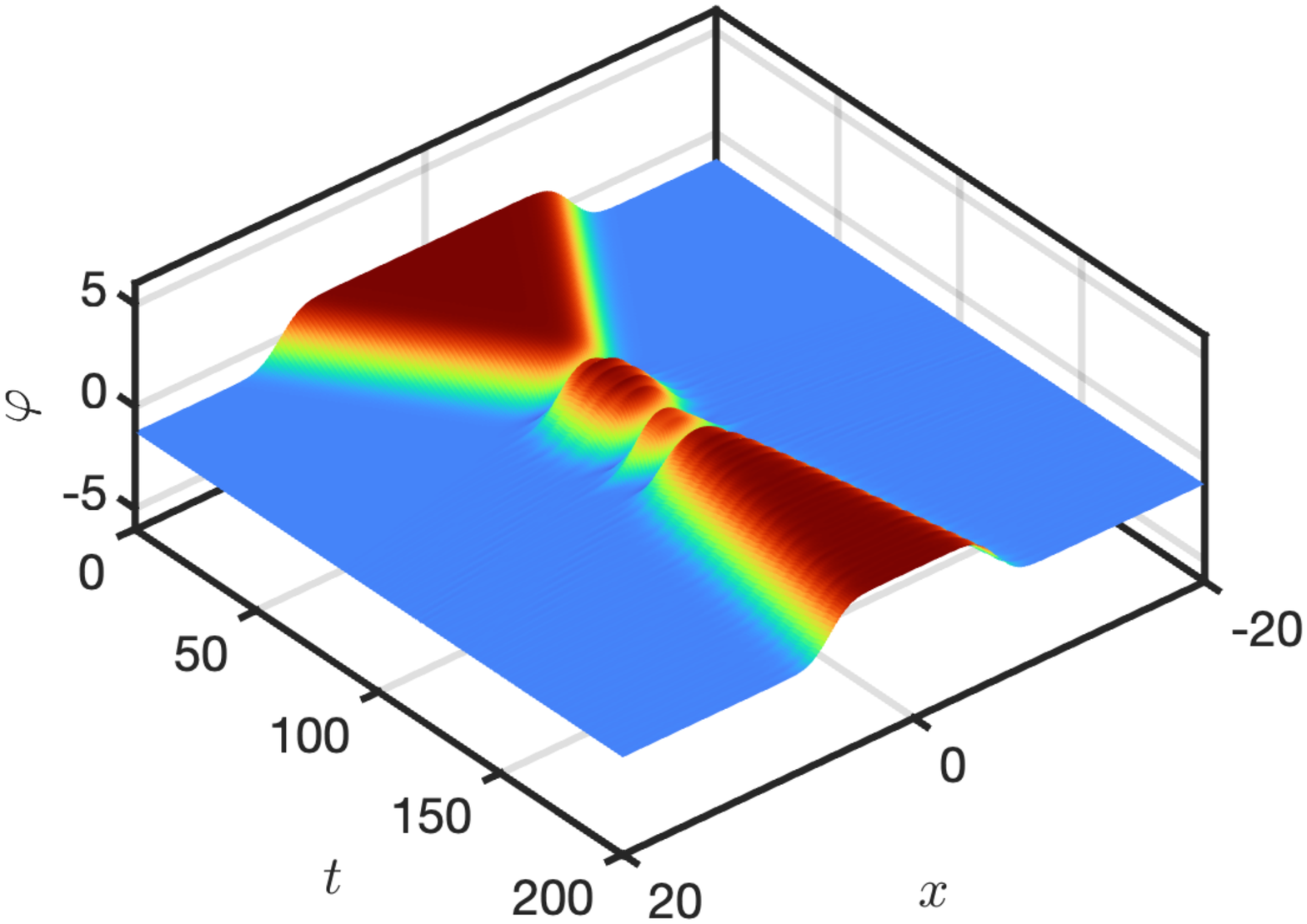}\label{fig:n118v01342}}
 	\subfigure[\quad five-bounce for $v=0.1342$]{\includegraphics[width=8cm]{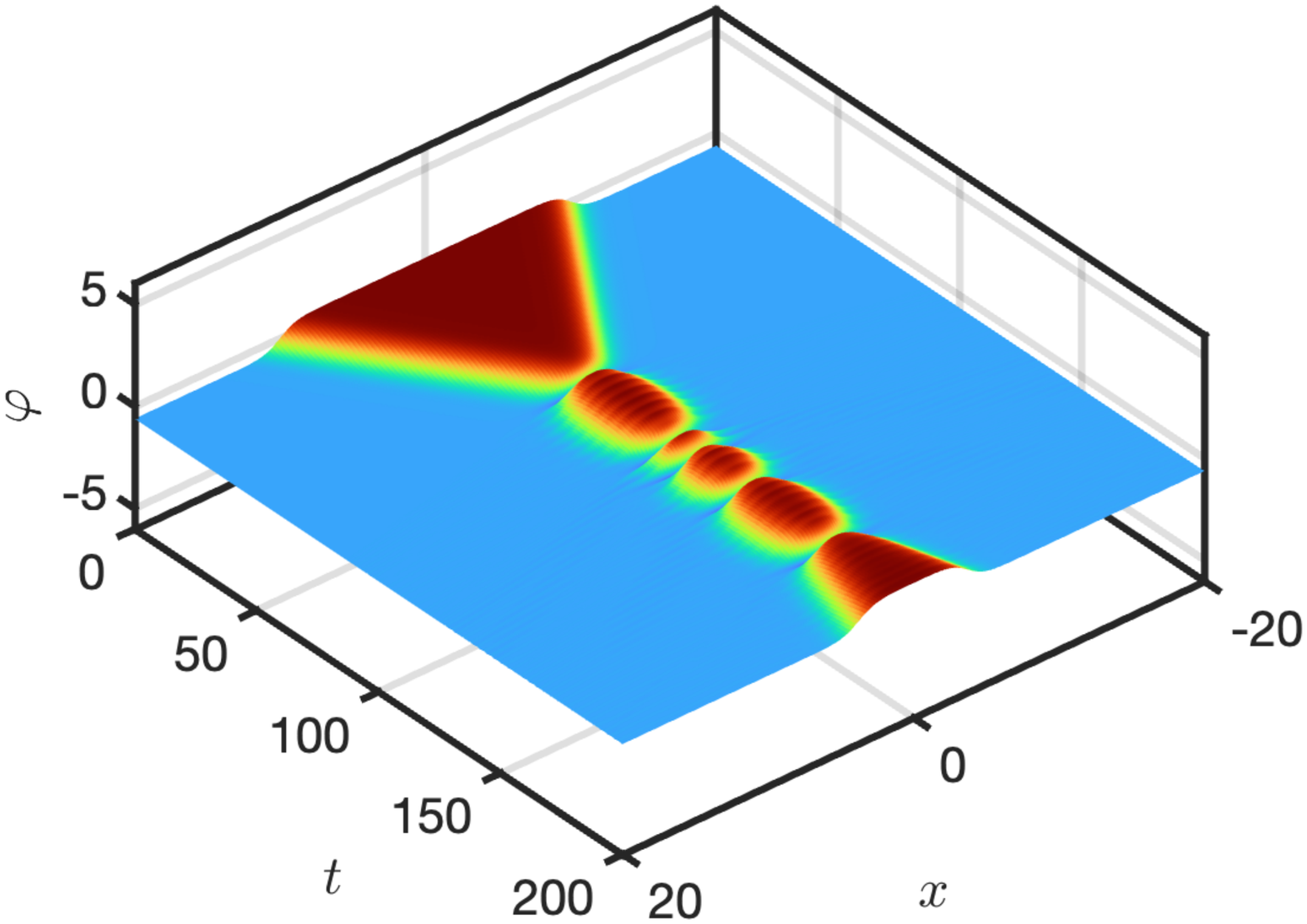}\label{fig:n186v01342}}
 	\\
 	\subfigure[\quad $n$-bounce for $v=0.1358$]{\includegraphics[width=8cm]{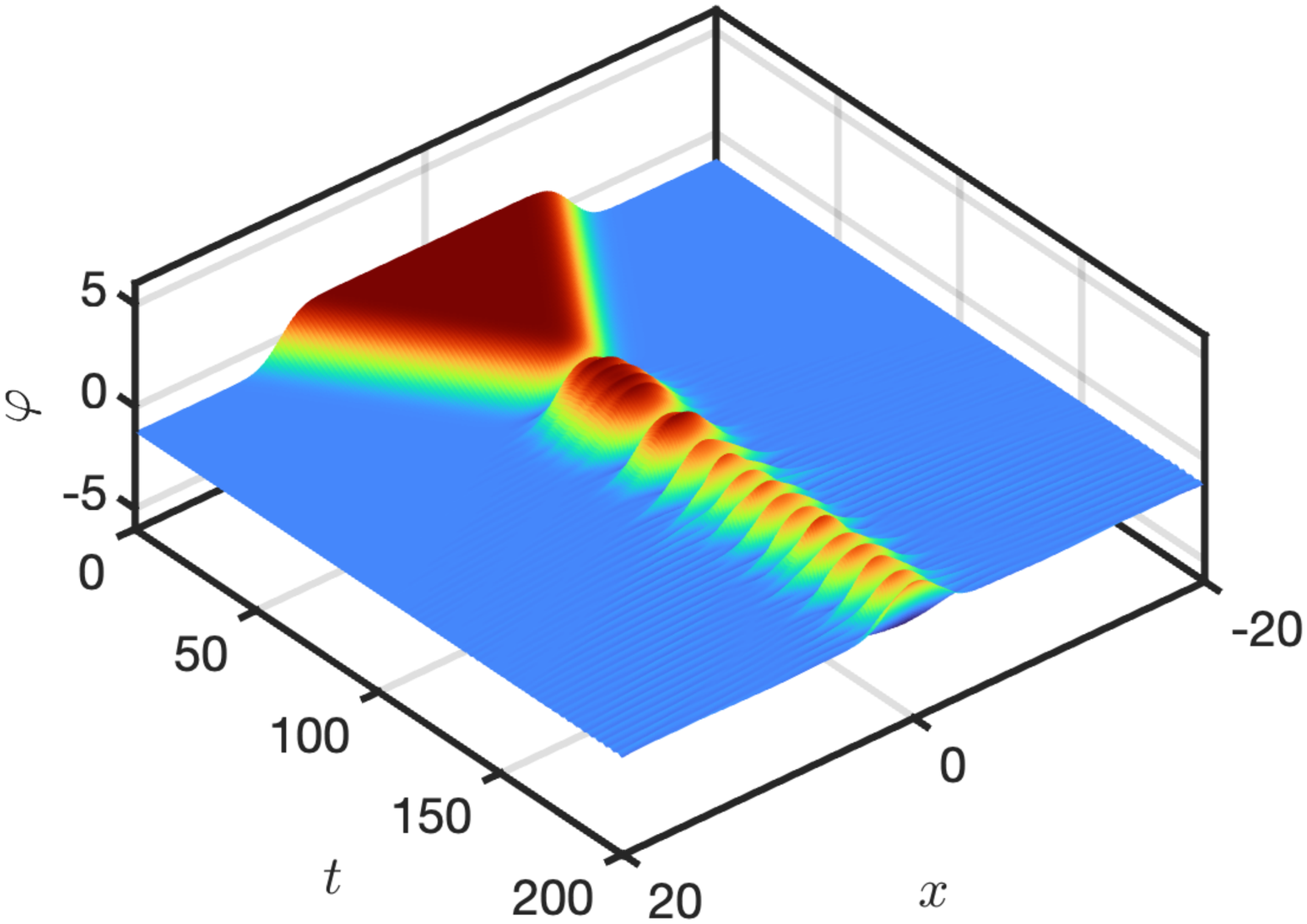}\label{fig:n118v01358}}
 	\subfigure[\quad three-bounce for $v=0.1358$]{\includegraphics[width=8cm]{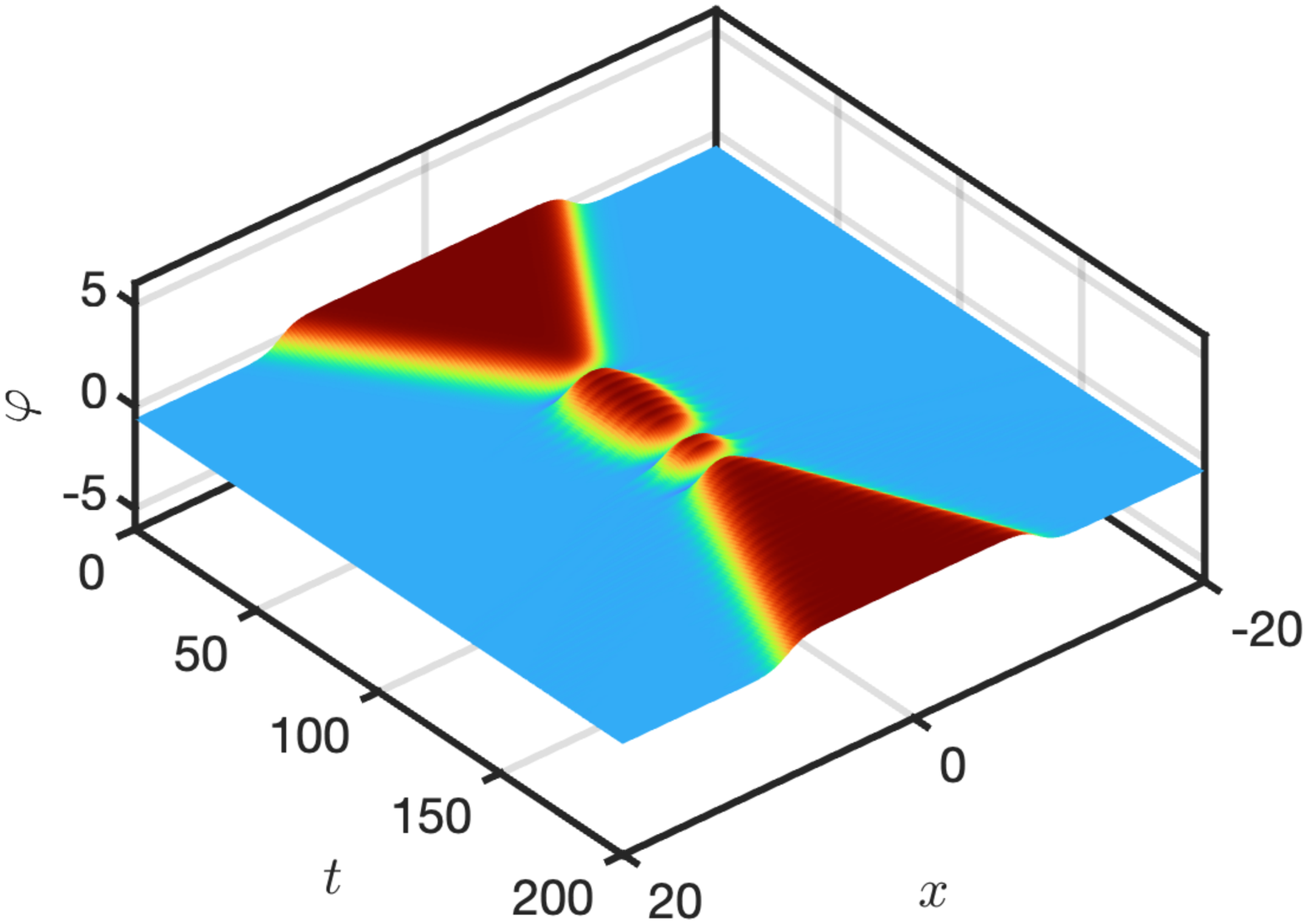}\label{fig:n186v01358}}
 	\\
 	\subfigure[\quad two-bounce for $v=0.1362$]{\includegraphics[width=8cm]{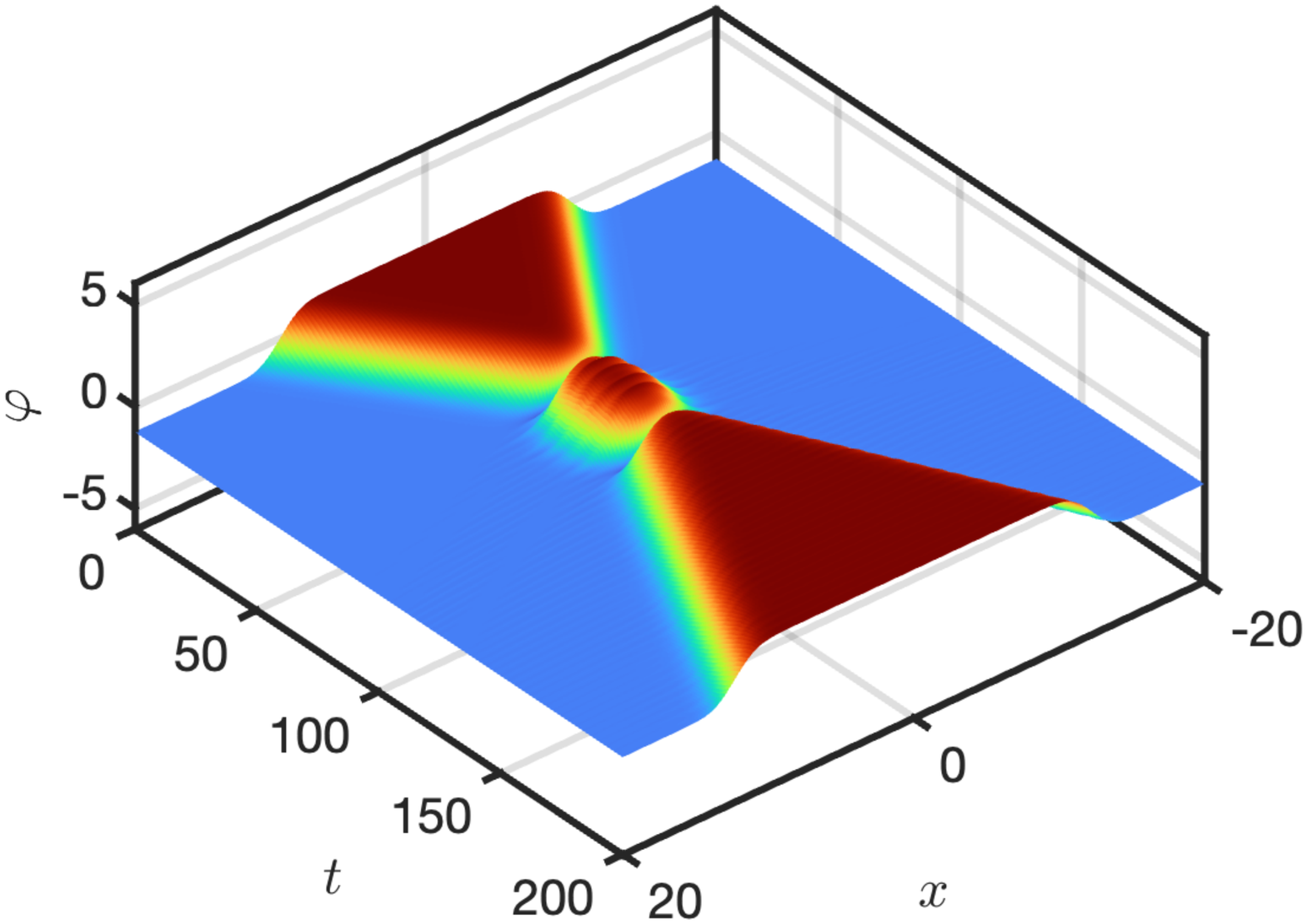}\label{fig:n118v01362}}
 	\subfigure[\quad four-bounce for $v=0.1362$]{\includegraphics[width=8cm]{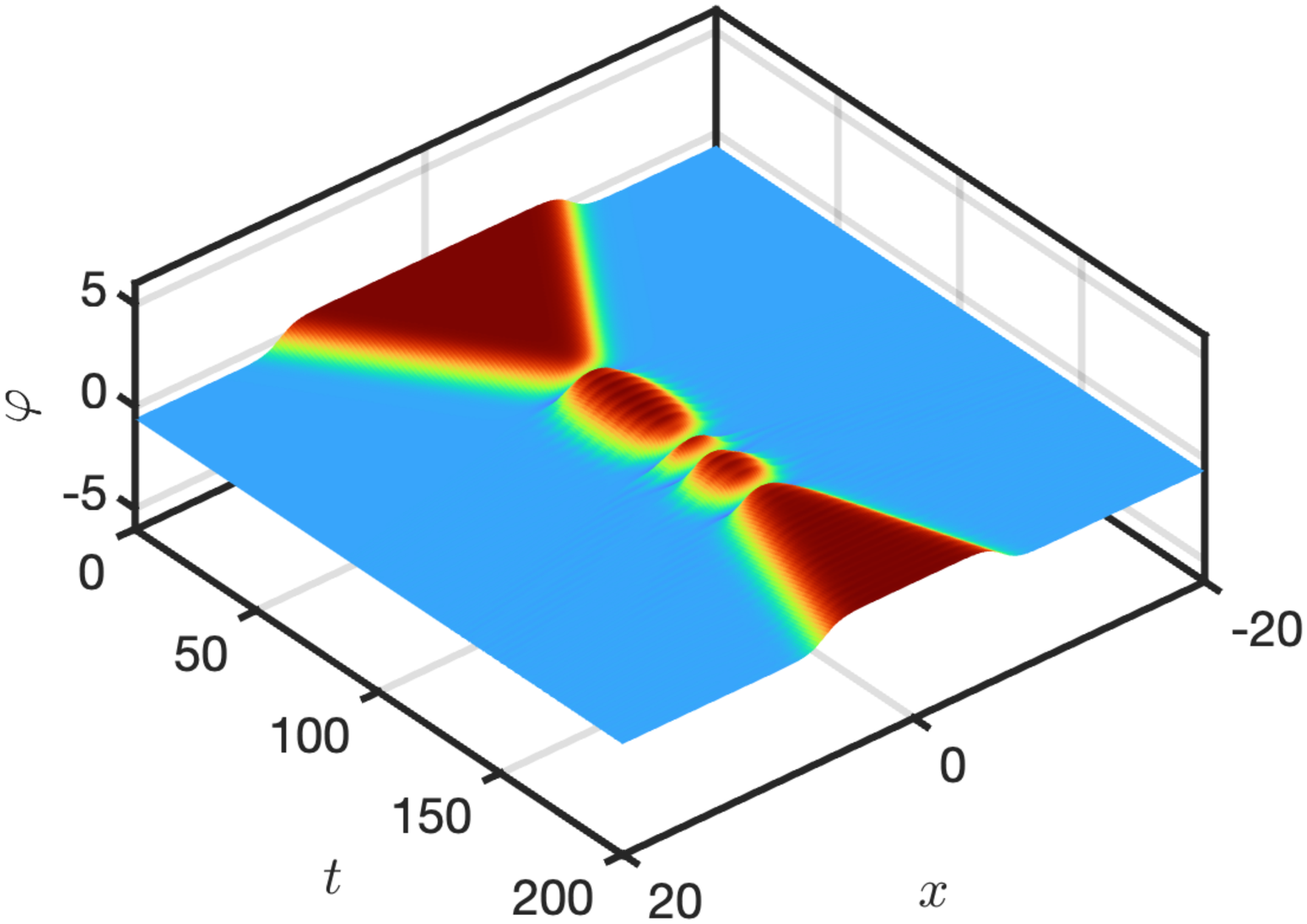}\label{fig:n186v01362}}
 	 \caption{Two kink collisions in the tanh-deformed $\varphi^4$ model for different initial velocities $v$. $r=1.18$ in (a), (c) and (e) panels, and $r=1.86$ in (b), (d) and (f) panels. We used the initial separation between centers of kink and antikink as $2x_0=20$.}
\label{fig:twokinkscollision}
\end{figure}

The effect of $r$ and therefore the effect of internal modes are given in Fig.~\ref{fig:twokinkscollision} with a few examples. The initial position of kink and antikink is $x_0=10$ for all sub-figures of Fig.~\ref{fig:twokinkscollision}. The sub-figures on the left hand side have $r = 1.18$ and the sub-figures on the right hand side have $r = 1.86$. We chose the same initial velocities for each pair of sub-figures. So, the initial velocity is the same in Figs.~\ref{fig:n118v01342} and \ref{fig:n186v01342}, $v=0.1342$, but the final states are different. Fig.~\ref{fig:n118v01342} shows $3$-bounce collision while Fig.~\ref{fig:n186v01342} shows $5$-bounce collision. These differences are seen in pairs of Figs.~\ref{fig:n118v01358} and \ref{fig:n186v01358} as well as in pairs of Figs.~\ref{fig:n118v01362} and \ref{fig:n186v01362}.
In Figs.~\ref{fig:n118v01358} and \ref{fig:n186v01358}, the initial velocity is $v=0.1358$ and the final states are $n$-bounce collision for $r=1.18$ and $3$-bounce collision for $r=1.86$. The initial velocity is $v=0.1362$ for Figs.~\ref{fig:n118v01362} and \ref{fig:n186v01362}, and the final states are $2$-bounce collisions and $4$-bounce collisions respectively.

The Fig.~\ref{fig:scatt} depicts the entire structure of the appearance of the two-bounce windows for $r=1.18$ and $r=1.86$. In this figure, we observe the time of the first, second and third kink-antikink collision in terms of the initial velocity $v$. The black and blue curves depict the first and second collisions, respectively. The appearance of the two-bounce windows is shown by the divergence in the red curve. In the Fig.~\ref{fig:scattn118}, for $v<v_c=0.144$, we note the presence of some two-bounce windows and the appearance of smooth peaks, known as false two-bounce windows. In the case of $v>v_c$, we only see an inelastic collision. Importantly, there are three vibrational states for $r=1.18$ and, as a result, the amount of extra states hinders the resonant energy exchange process, shattering the structure of the two-bounce windows. In comparison to Fig.~\ref{fig:scattn118}, we notice more two-bounce windows and absence of false two-bounce windows in Fig.~\ref{fig:scattn186}. As we already know, only two vibrational modes are present for the case $r=1.86$. Moreover, since of the number of bound states influences the process of resonant window suppression, the fewer the additional states, the more the number of two-bounce windows.

\begin{figure}
 	\subfigure[]{\includegraphics[width=8cm]{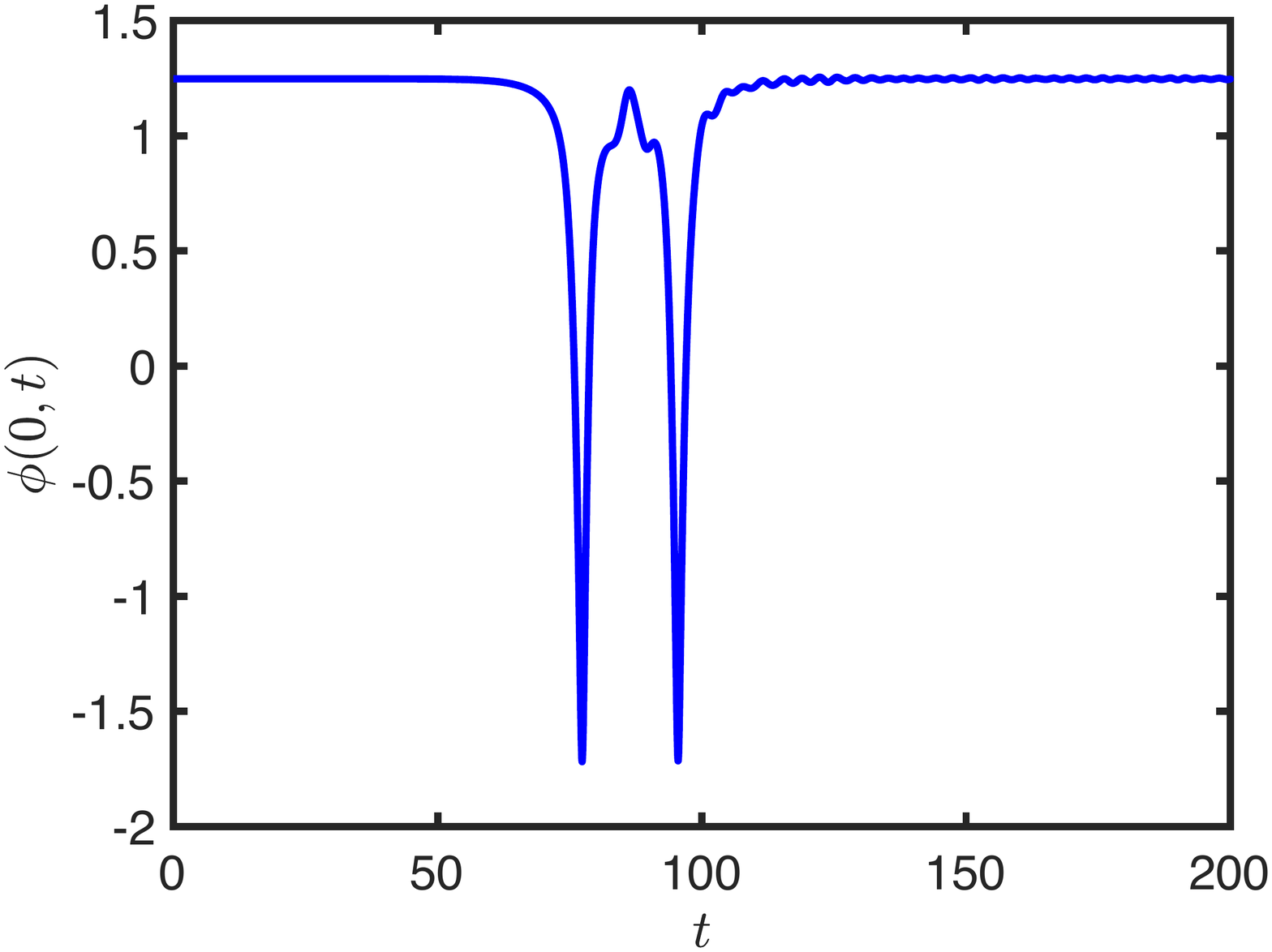}\label{fig:2Dhypphi4r118}}
 	\subfigure[]{\includegraphics[width=8cm]{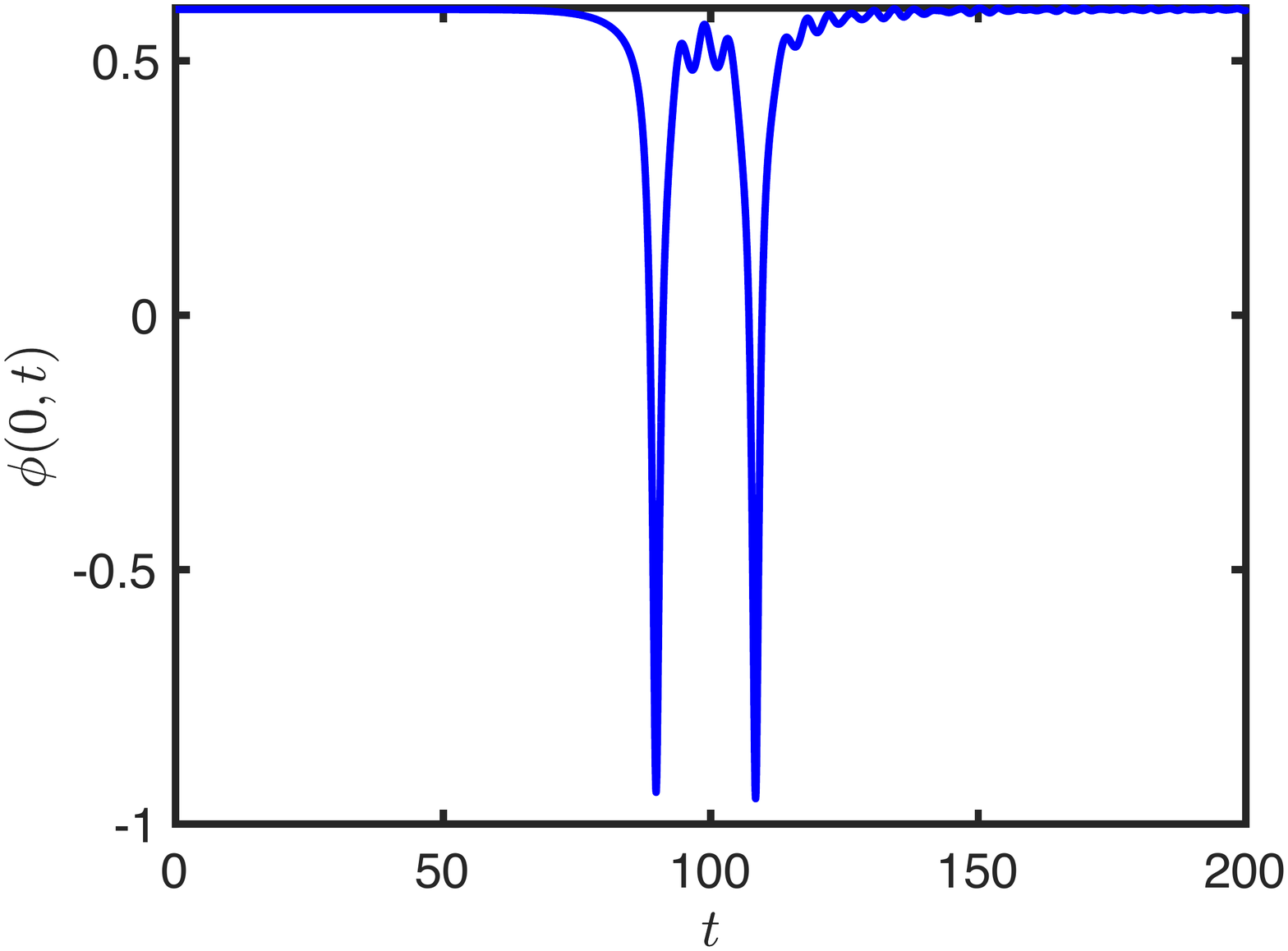}\label{fig:2Dhypphi4r186}}
 	 \caption{Scalar field at the center of mass $\phi(x=0,t)$ as function of time for (a) $v=0.107$ and $r=1.18$ and (b) $v=0.09$ and $r=1.86$.}
 	\label{fig:2dhypphi4}
\end{figure}

The number of vibrational modes also influences how the two-bounce oscillations are deformed. The behavior of the scalar field at the center of mass as a function of time is depicted in Fig. \ref{fig:2Dhypphi4r118} and Fig. \ref{fig:2Dhypphi4r186}, for the case $r=1.18$ and $r=1.86$, respectively. In Fig. \ref{fig:2Dhypphi4r186}, we note that $\phi(x=0,t)$ oscillates four times $(n=4)$ around the vacuum, corresponding to the expected second two-bounce windows $(m=2)$. This collision corresponds to a first two-bounce windows visible in the Fig. \ref{fig:scattn186}. The Fig. \ref{fig:2Dhypphi4r118}, on the other hand, depicts a distinct oscillatory pattern between the two-bounce. In particular, the scalar field oscillates three times around the vacuum, corresponding to $n=3$, labeled by $m=1$. As result, this collision corresponds to the first two-bounce windows in the Fig. \ref{fig:scattn118}.

The information from the two-bounce windows is presented in Tables \ref{tab1} and \ref{tab2}, for $r=1.18$ and $r=1.86$, respectively. The tables are characterized by the order $m$, initial velocity of the start $v_1$ and end $v_2$ of each window, the thickness $\Delta v = v_2-v_1$ and the scaling parameter $\beta$ \cite{anninos}. We notice that the thickness of the resonance windows decreases with increasing window order, which can be seen in Fig. \ref{fig:scatt}. This agrees with a scaling relation, where the windows width and the number of oscillations to some power are inversely related $\Delta v \propto M^{-\beta}$ \cite{anninos}. Table \ref{tab2} shows a scaling relation with $\beta \sim 3.16 \pm 0.05$. However, Table \ref{tab1} demonstrates that the additional modes have an impact on the scaling relation, increasing the uncertainty of the $\beta \sim 3.29 \pm 0.15$. Notice the absence of order $m=2$ and $m=5$ windows in Table \ref{tab1}, as well as in Fig. \ref{fig:scattn118}.

\begin{figure}
 	\subfigure[]{\includegraphics[width=8cm]{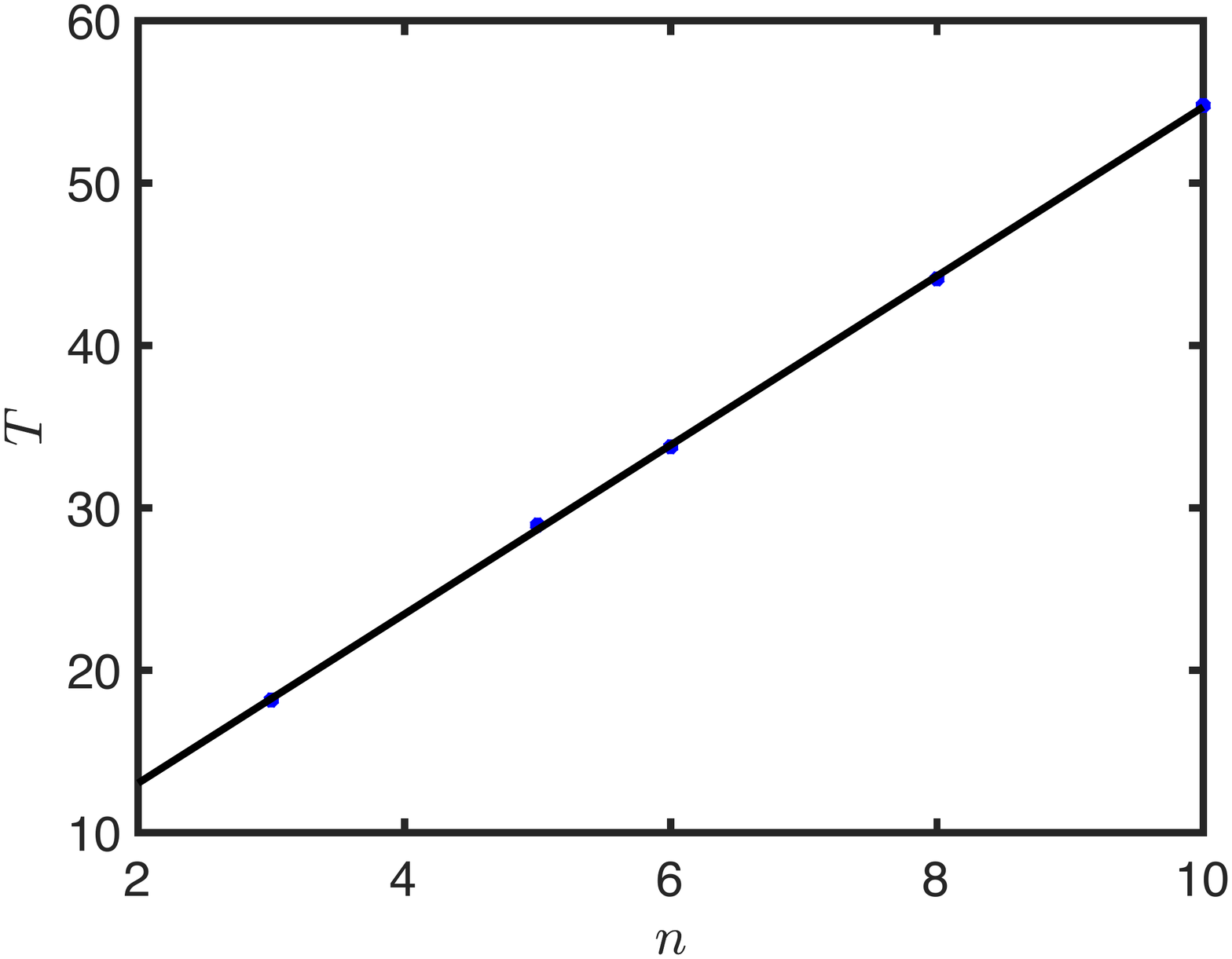}}
 	\subfigure[]{\includegraphics[width=8cm]{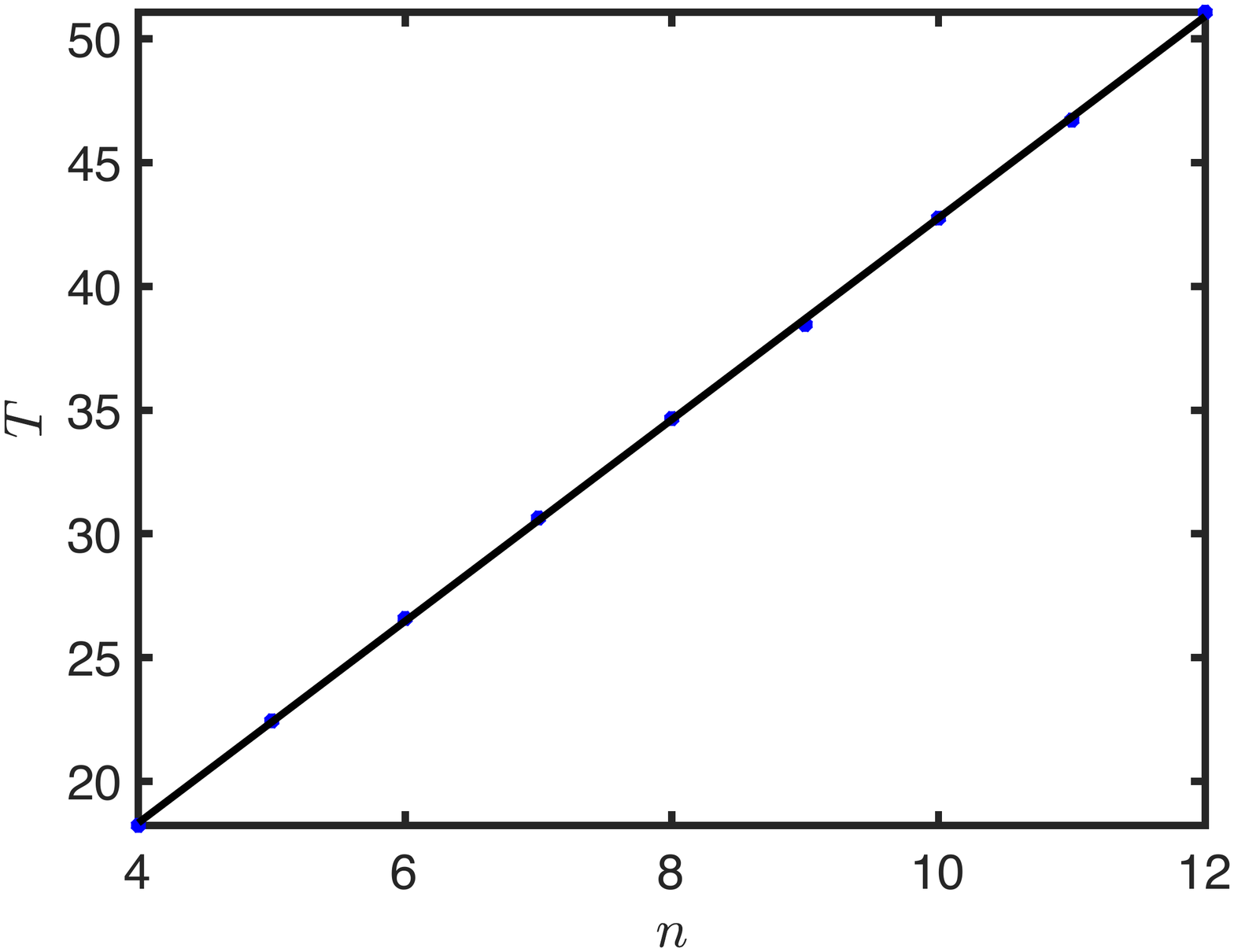}}
 	 \caption{Time between bounces as a function of $n=m+2$ for (a) $r=1.18$ and (b) $r=1.86$.}
 	\label{Txn_hyp_phi4}
\end{figure}

As illustrated in Fig. \ref{Txn_hyp_phi4}, we plot the time between bounces as a function of $n=m+2$, where $m$ corresponds to the windows number. According to \cite{Campbell.PhysD.1983.phi4}, the presence of two-bounce windows obeys the relation $\omega_1T=2\pi m+\delta$, where $\omega_1$ is the frequency of the shape mode, $T$ is the time between bounces and $\delta$ is the phase shift. We can see from the figure that a straight line can accurately approximate the numerical points. We notice that the indicated dots reflect numerical values, whereas the line represents a least squares fitting. The slope of the line for $r=1.18$ is $5.20$, equivalent to the value $2\pi/\omega_1 \approx 5.12$. For $r=1.86$, the slope is $4.07$, and in this case, we have $2\pi/\omega_1 \approx 3.98$. The relative errors are, respectively, $1.5\%$ and $2.2\%$.

\begin{table}[ht]
\centering
\begin{tabular}[t]{cccccc}
\hline
$m$&\quad$v_1$&\quad$v_2$&\quad$\Delta v$&\quad$\beta$\\
\hline
1&\quad0.1050&\quad0.1093&\quad0.0043&\quad-\\
3&\quad0.1324&\quad0.1332&\quad0.0008&\quad3.29\\
4&\quad0.1360&\quad0.1364&\quad0.0004&\quad3.43\\
6&\quad0.1396&\quad0.1398&\quad0.0002&\quad3.13\\
\hline
\end{tabular}
\caption{The order $m$ for two-bounce windows, initial velocity $v_1$ and $v_2$, width $\Delta v$ and scaling $\beta$ for $r=1.18$.}
\label{tab1}
\end{table}

\begin{table}[ht]
\centering
\begin{tabular}[t]{cccccc}
\hline
$m$&\quad$v_1$&\quad$v_2$&\quad$\Delta v$&\quad$\beta$\\
\hline
2&\quad0.0881&\quad0.0953&\quad0.0072&\quad-\\
3&\quad0.1119&\quad0.1155&\quad0.0036&\quad3.11\\
4&\quad0.1228&\quad0.1248&\quad0.0020&\quad3.16\\
5&\quad0.1288&\quad0.13&\quad0.0012&\quad3.20\\
6&\quad0.1326&\quad0.1334&\quad0.0008&\quad3.17\\
\hline
\end{tabular}
\caption{The order $m$ for two-bounce windows, initial velocity $v_1$ and $v_2$, width $\Delta v$ and scaling $\beta$ for $r=1.86$.}
\label{tab2}
\end{table}

\section{Tanh-deformed $\varphi^6$ model} 
\label{sec:TanhDeformedPhi6Model}

\begin{figure}
 	\subfigure[]{\includegraphics[width=8cm]{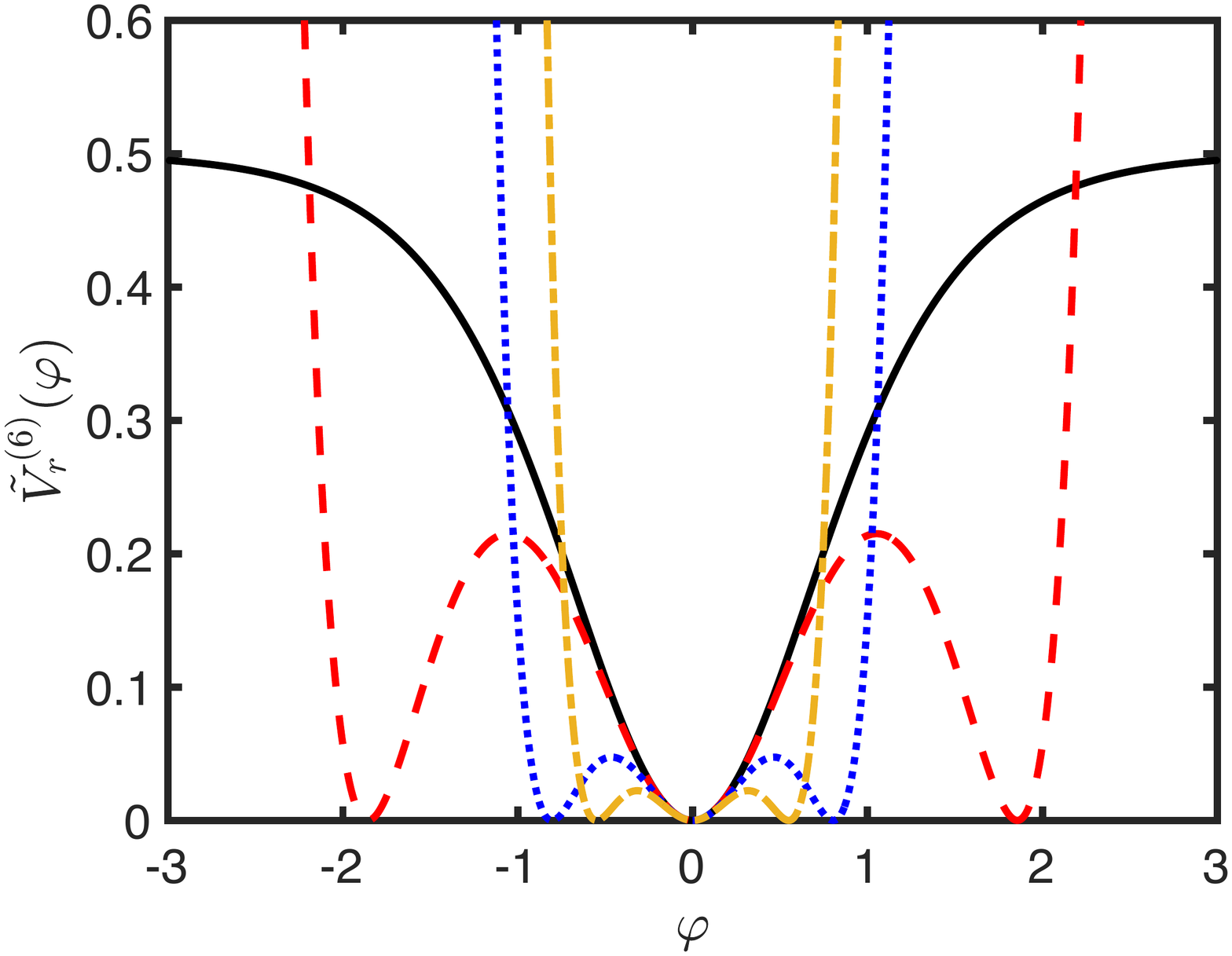}\label{fig:hypphi6potentials}}
 	\subfigure[]{\includegraphics[width=8cm]{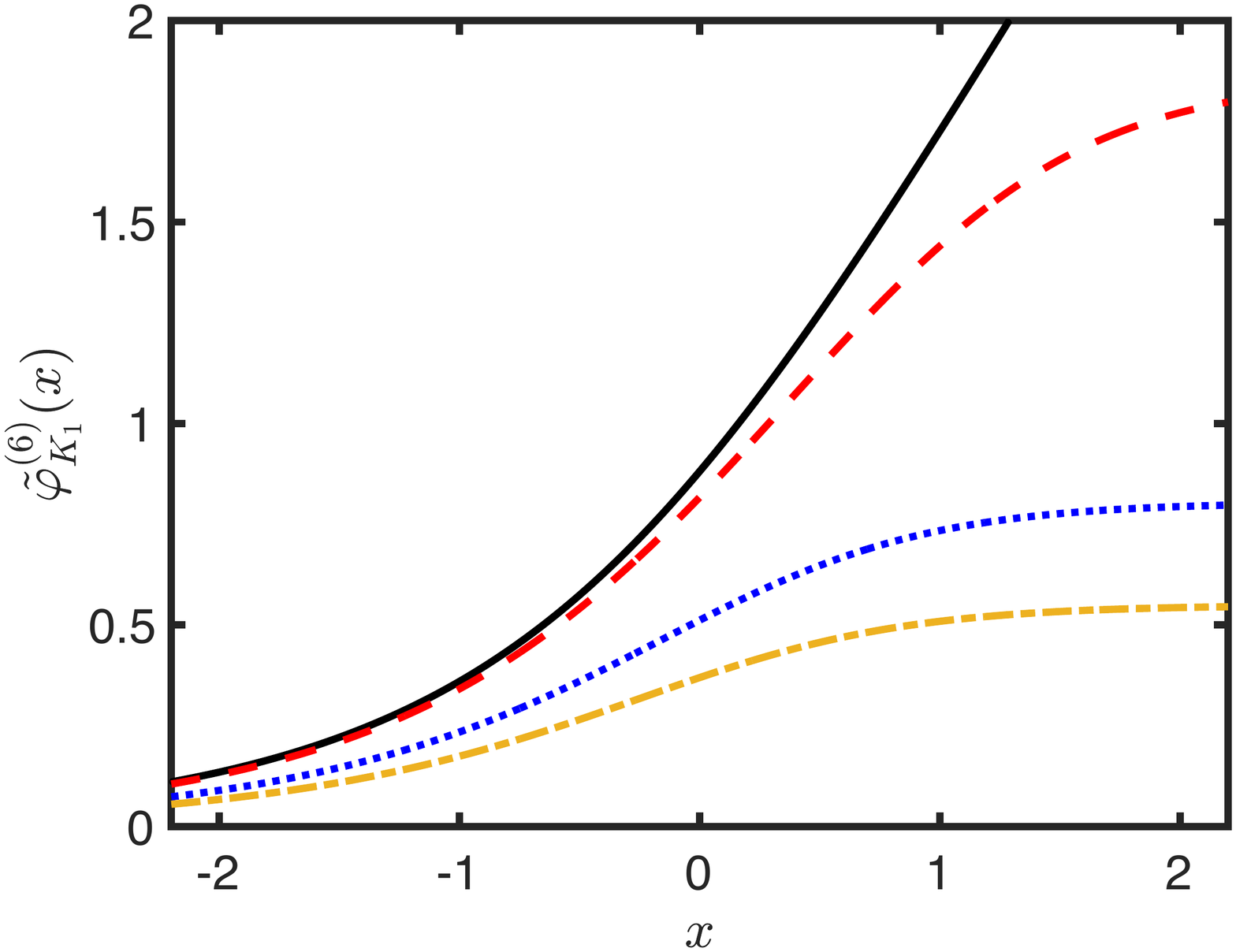}\label{fig:hypphi6kinks}}
 	 \caption{(a) The potentials in Eq.~\eqref{eq:hypphi6potential} and (b) the kinks Eq.~\eqref{eq:hypphi6kinks} of the tanh-deformed $\varphi^6$ model for (a) $r=1.0$ (solid black), (b) $r=1.05$ (red dash), (c) $r=1.5$ (blue dot) and (d) $n=2.0$ (yellow dot-dash).}
\label{fig:hypphi6solutionspotentials}
\end{figure}

Let us now investigate the tanh-deformed $\varphi^6$ model. Under the same deformation used above, the potential is
\begin{eqnarray}\label{eq:hypphi6potential}
\Tilde{V}_{r}^{(6)}(\varphi)=\frac{1}{2} \tanh ^2(\varphi )\!\! \left(\cosh ^2(\varphi )-r^2 \sinh ^2(\varphi )\right)^2.
\end{eqnarray}
Again, $r$ is a real parameter, obeying $r>1$. The Fig.~\ref{fig:hypphi6potentials} shows the potential for some values of $r$, where we realize the presence of two topological symmetric sectors for $r>1$. The potential has three degenerate minima, at  $0,$ and at $\pm \tanh^{-1}({1}/{r})$. The increase of parameter $r$, reduces the height of the maxima of the potential and the minima get closer to each other. The Eq.~\eqref{eq:EOM} with this potential yields the equation of motion of the model, given by
\begin{eqnarray}\label{eq:hypphi6EQM}
\frac{\partial^2 \varphi}{\partial t^2}-\frac{\partial^2 \varphi}{\partial x^2}+r^2 \tanh (\varphi ) \left(1\!-\!\left(r^2-1\right) \sinh ^2(\varphi )\right)\! \left(\text{sech}^2(\varphi )-\frac{\left(r^2-1\right) }{r^2}\cosh (2 \varphi )\!\right)=0.\;
\end{eqnarray}

The static solution is given by
\begin{eqnarray}\label{eq:hypphi6kinks}
{\tilde{\varphi}}^{(6)}_{K_1}(x)= \tanh ^{-1}\left(\frac{1}{r} \sqrt{\frac{1 + \tanh{x}}{2}}\right),
\end{eqnarray}
where this kink interpolates between the minima $\{0,\tanh^{-1}({1}/{r})\}$. Similar to the $\varphi^6$ model Eq.~\eqref{eq:phi6potential}, the antikink connecting the minima $\{\tanh^{-1}({1}/{r}),0\}$ is $\tilde{\varphi}^{(6)}_{\bar{K}_1}(x) = \tilde{\varphi}^{(6)}_{K_1}(-x)$. The kink and antikink connecting the other topological sector are given by $\tilde{\varphi}^{(6)}_{K_2}(x) = - \tilde{\varphi}^{(6)}_{K_1}(-x) $ and $\tilde{\varphi}^{(6)}_{\bar{K}_2}(x)= - \tilde{\varphi}^{(6)}_{K_1}(x)$. Therefore, kinks solutions have the same sign, whereas antikinks solutions have the opposite sign.

%
 \begin{figure}
 	\subfigure[]{\includegraphics[width=8cm]{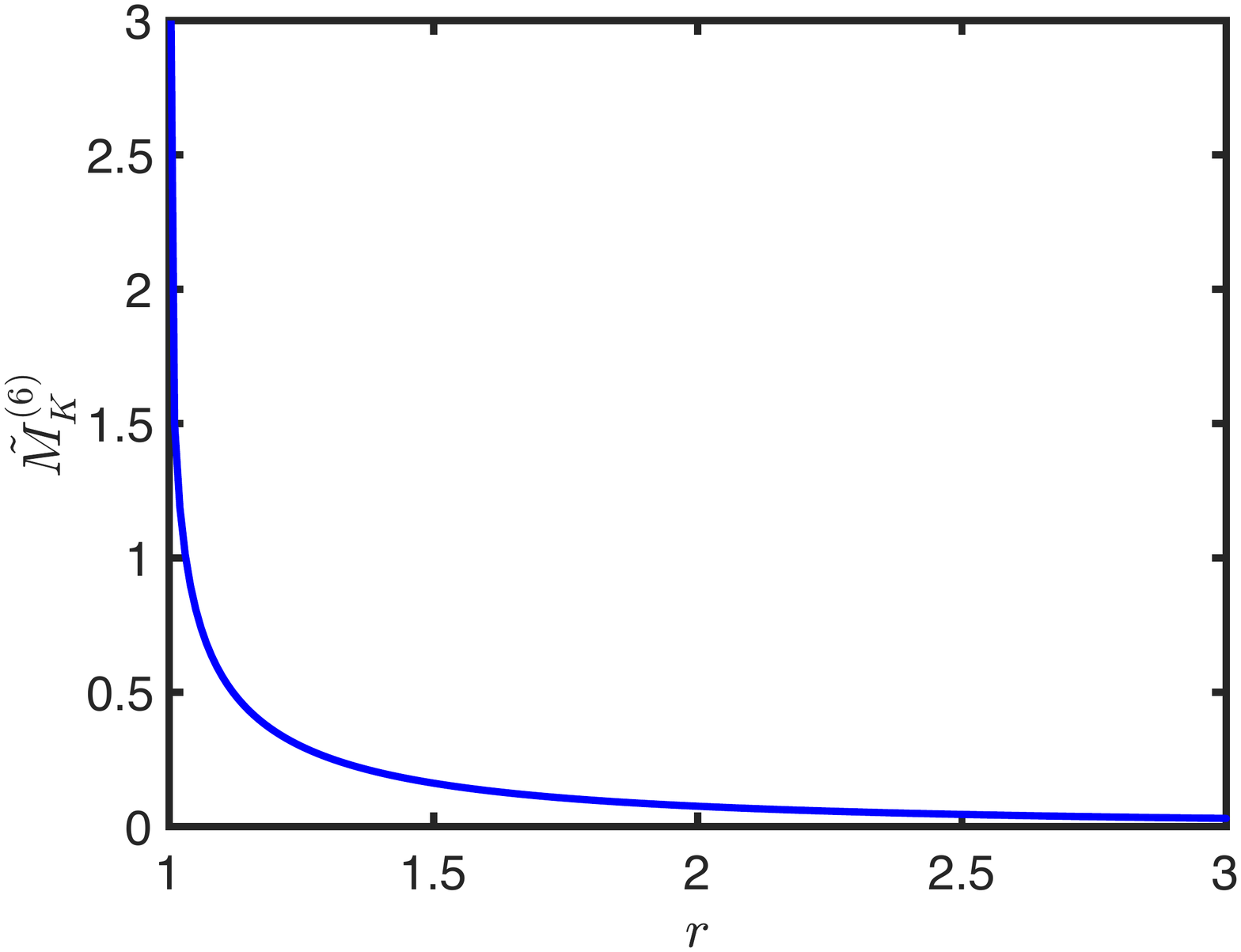}\label{fig:hypphi6mass}}
 	\subfigure[]{\includegraphics[width=8cm]{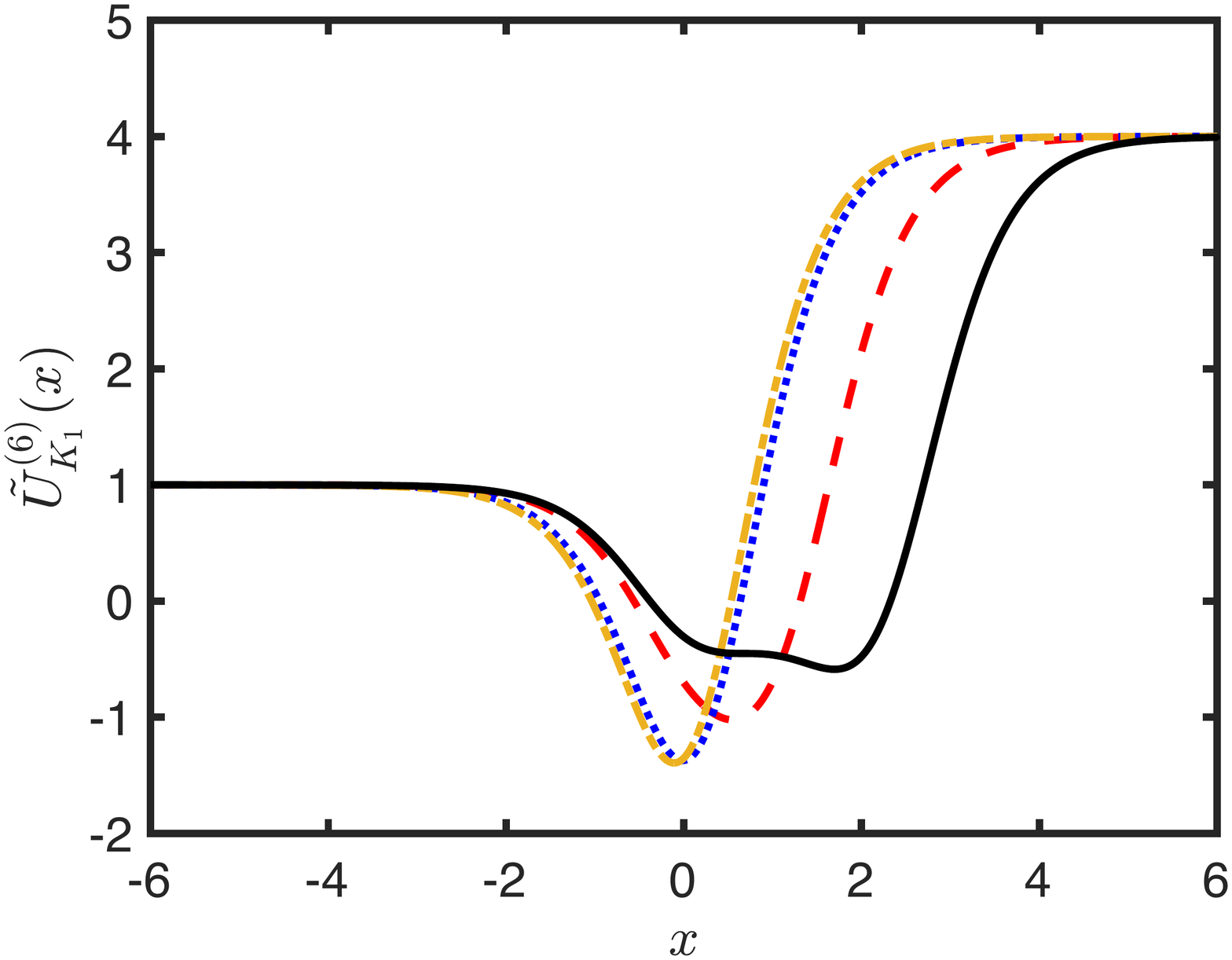}\label{fig:hypphi6QMPs}}
 	 \caption{ (a) The masses of kink Eq.~\eqref{eq:tanhdeformedphi6mass} and (b) quantum mechanical potential in Eq.~\eqref{eq:tanhdeformedphi6QMP} of the tanh-deformed $\varphi^6$ model for (a) $r=1.005$ (solid black), (b) $r=1.05$ (red dash), (c) $r=1.5$ (blue dot) and (d) $r=2$ (yellow dot-dash).}
\label{fig:hypphi6qmpmass}
\end{figure}

We can calculate the mass of these kinks. Since they are the energy of the solutions, we can write
\begin{eqnarray}\label{eq:tanhdeformedphi6mass}
\Tilde{M}_K^{(6)}=\frac{r^2}{2} \log \left(\frac{r^2}{r^2-1}\right)-\frac12.
\end{eqnarray}
Fig.~\ref{fig:hypphi6mass} depicts the mass of the kink, and we observe that the mass diminishes as $r$ increases.

The stability analysis leads to the following quantum mechanical potential
\begin{eqnarray}\label{eq:tanhdeformedphi6QMP}
\tilde{U}^{(6)}_{K}(x)=-r^4 (\cosh (2 \tilde{\varphi}^{(6)})\!-\!2) \text{sech}^4(\tilde{\varphi}^{(6)} )\!-\!2 r^2 \left(r^2\!-\!1\right)\! \cosh (2 \tilde{\varphi}^{(6)} )\!+\!\left(r^2\!-\!1\right)^2 \!\!\cosh (4 \tilde{\varphi}^{(6)} )\!.\;
\end{eqnarray}
The Schr\"odinger-like potential for ${\tilde{\varphi}}^{(6)}_{K_1}(x)$ is displayed in Fig. \ref{fig:hypphi6QMPs}. When $r \rightarrow 1$, the depth of the potential is reduced, resulting in the development of two minima. This behavior facilitates the appearance of a vibrational mode. In particular, bound states were analyzed numerically at this potential for various values of $r$. We can see that in addition to the zero mode, there is a vibrational mode for $r<1.007$. As we know, the appearance of an internal mode is a crucial part of the kink collision process.


\subsection{Numerical results}


In this section we present our results of kink scattering in tanh-deformed $\varphi^6$ model. The kinks of this model are asymmetric and therefore we investigate the kink-antikink and antikink-kink collisions. For the numerical solutions, we solved the equation of motion with the following initial conditions
\begin{eqnarray}\label{eq:condition}
\varphi(x,0) & = & \tilde{\varphi}^{(6)}_{K_1}(x+x_0,v,0) + \tilde{\varphi}^{(6)}_{\bar{K_1}}(x-x_0,-v,0) - \tanh^{-1}\bigg(\frac1r\bigg),\\
\dot{\varphi}(x,0) & = & \dot{\tilde{\varphi}}^{(6)}_{K_1}(x+x_0,v,0) + \dot{\tilde{\varphi}}^{(6)}_{\bar{K_1}}(x-x_0,-v,0).
\end{eqnarray}
We fixed $x_0=10$ for the initial symmetric position of the pair and set the grid boundary at $x_{max}=\pm200$. We use a $4^{th}$ order finite-difference method with the spatial step $\delta x = 0.05$ and $6^{th}$ order sympletic integrator with time step $\delta t = 0.02$. It is important to note that, $\varphi(x,t) = \varphi(\gamma(x-vt))$, means boost of Lorentz for static kink and $\gamma = 1/\sqrt{1-v^2}$.

\begin{figure}
 	\subfigure[\quad n-bounces for $v=0.17$]{\includegraphics[width=8cm]{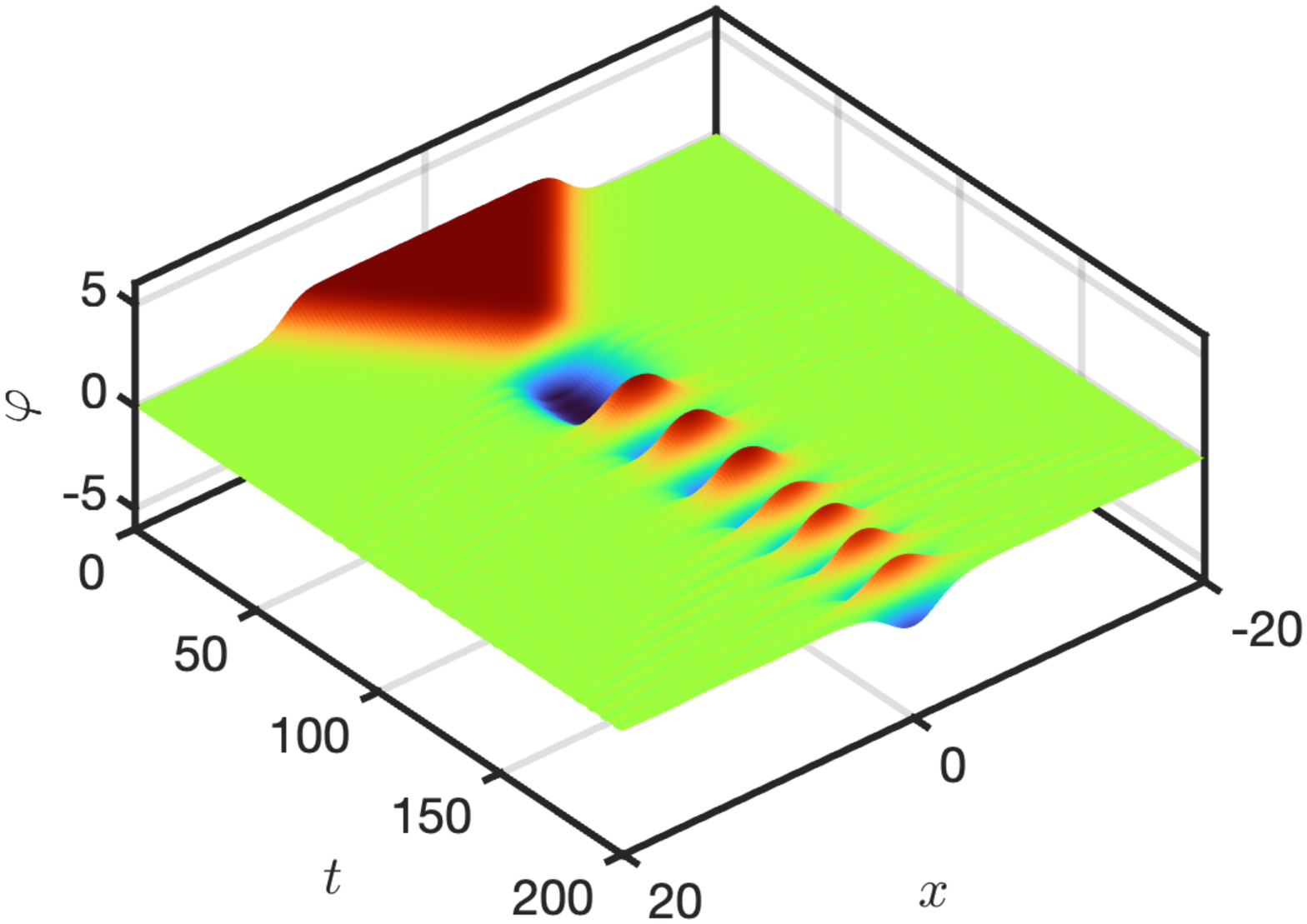}\label{fig:hypphi6v017n105}}
 	\subfigure[\quad one-bounce for $v=0.20$]{\includegraphics[width=8cm]{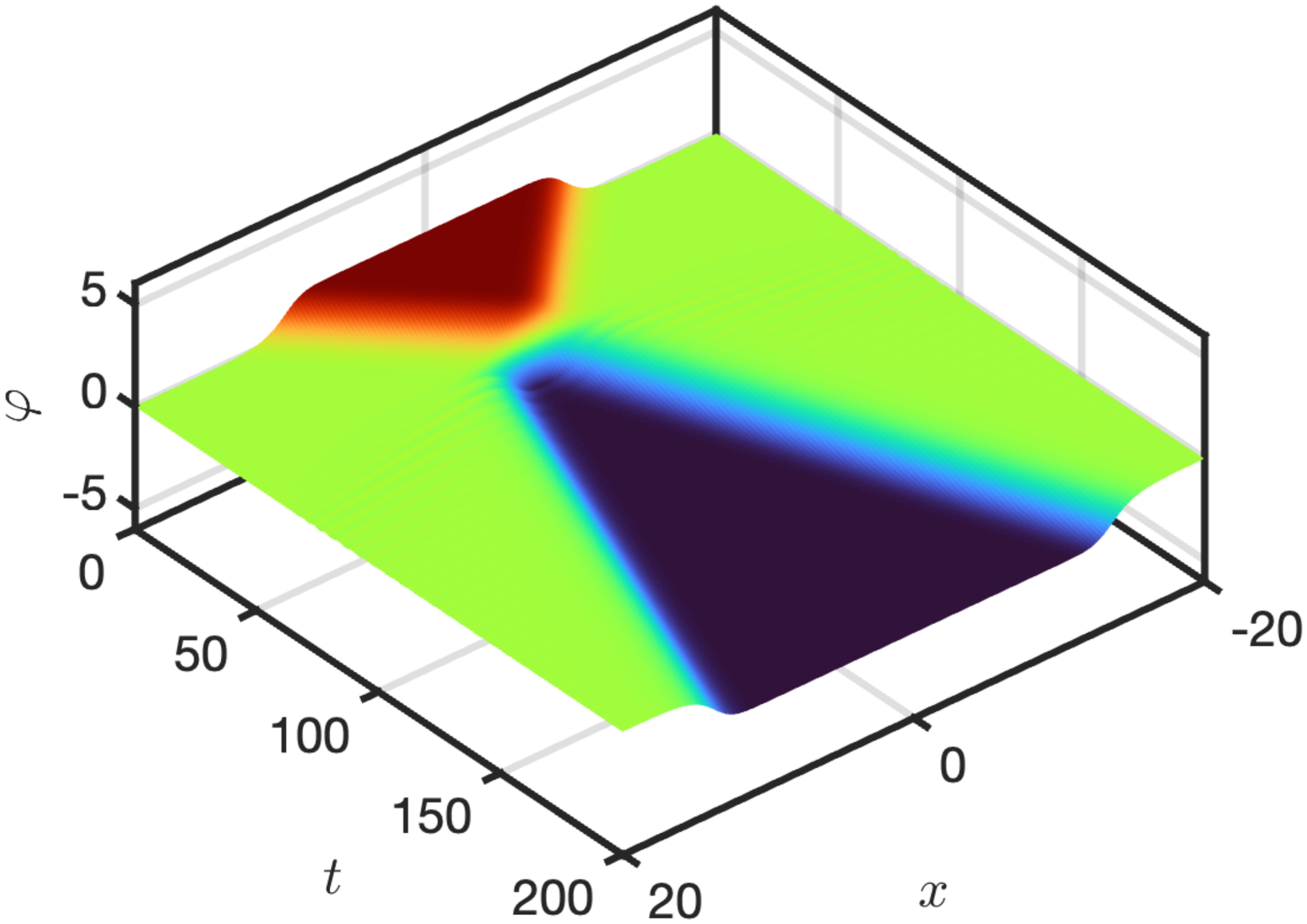}\label{fig:hypphi6v02n105}}
 	 \caption{Kink-antikink collision in the tanh-deformed $\varphi^6$ model for different initial velocities $v$. $r=1.05$. We used the initial separation between centers of kink and antikink as $2x_0=20$.}
    \label{fig:hypphi6KA1}
\end{figure}

It is worthwhile to mention that, unlike in the $\varphi^6$ model \cite{Dorey.PRL.2011}, altering the $r$ parameter results in the appearance of vibrational mode for a single kink or antikink. For instance, we analyzed the kink-antikink scattering for $r=1.05$; this case has only the zero mode. The Fig.~\ref{fig:hypphi6KA1} depicts final states in kink-antikink collision for case $r=1.05$. Fig.~\ref{fig:hypphi6v017n105} shows the scattering of oscillatory waves at $x=0$ after the first impact, followed by a small radiation. This behavior occurs for $v\leq v_c=0.1801$. However, for $v>v_c$, the output is an inelastic collision, with the scalar field at the center of mass scattering into the other vacuum of the model, as we can see in Fig.~\ref{fig:hypphi6v02n105}. The absence of the vibrational mode in this case results in the disappearance of two-bounce windows, as expected. 

%

\begin{figure}
    \subfigure[\quad oscillating pulses for   $v=0.19$]{\includegraphics[width=8cm]{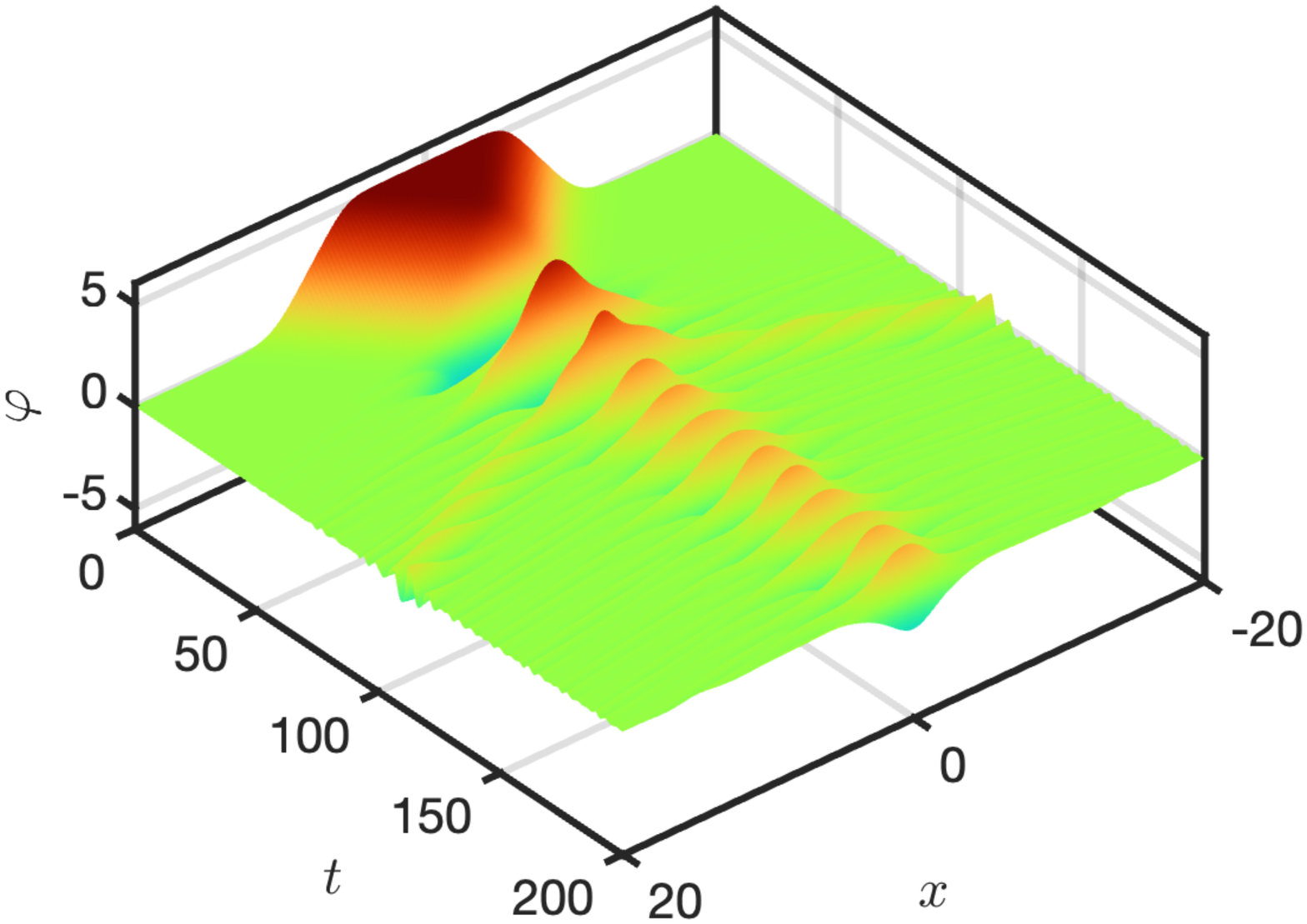}\label{fig:n100005v019000}}
 	\subfigure[\quad $n$-bounce for $v=0.2036$]{\includegraphics[width=8cm]{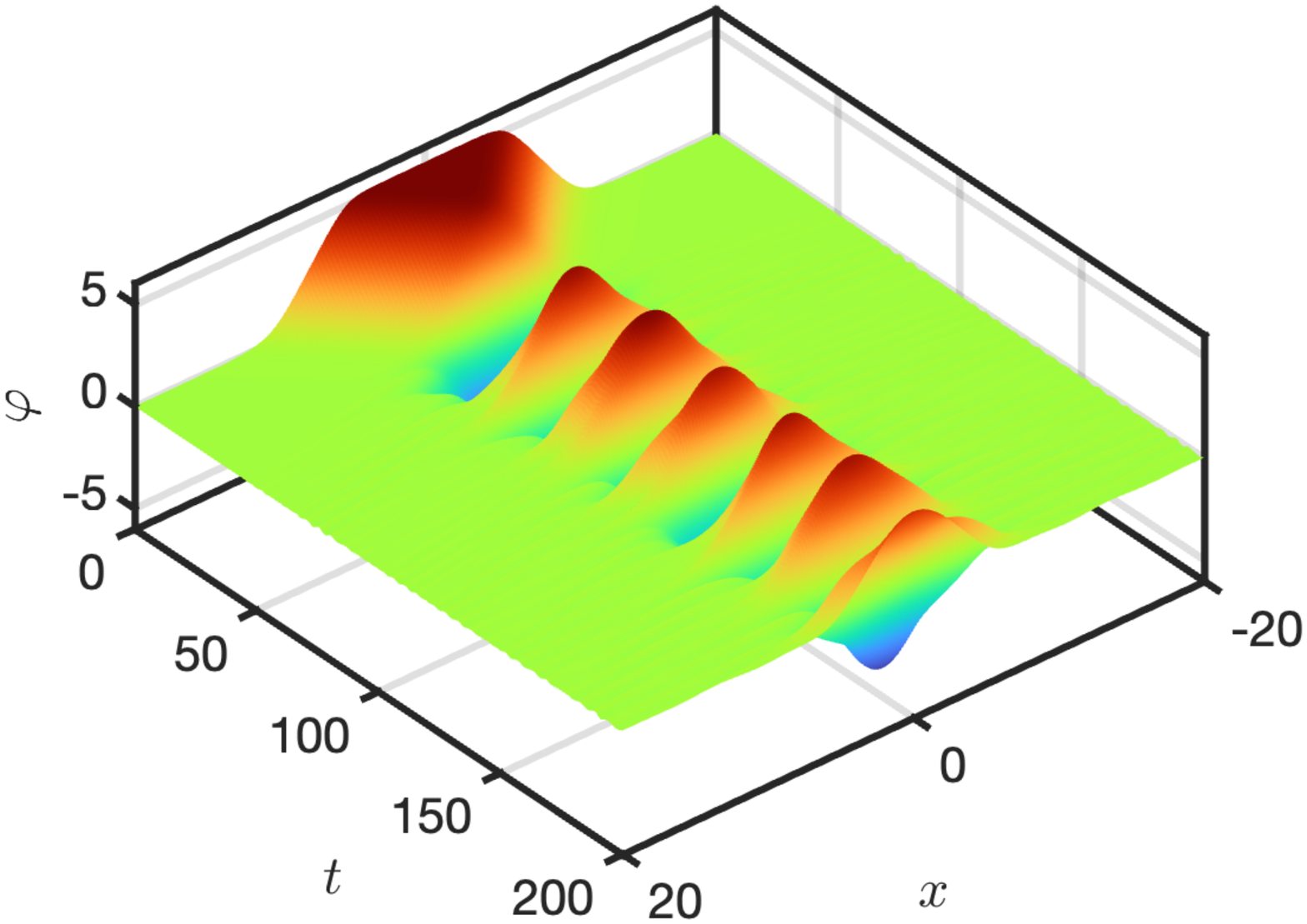}\label{fig:n100005v020360}}
 	\\
 	\subfigure[\quad one-bounce for  $v=0.202$]{\includegraphics[width=8cm]{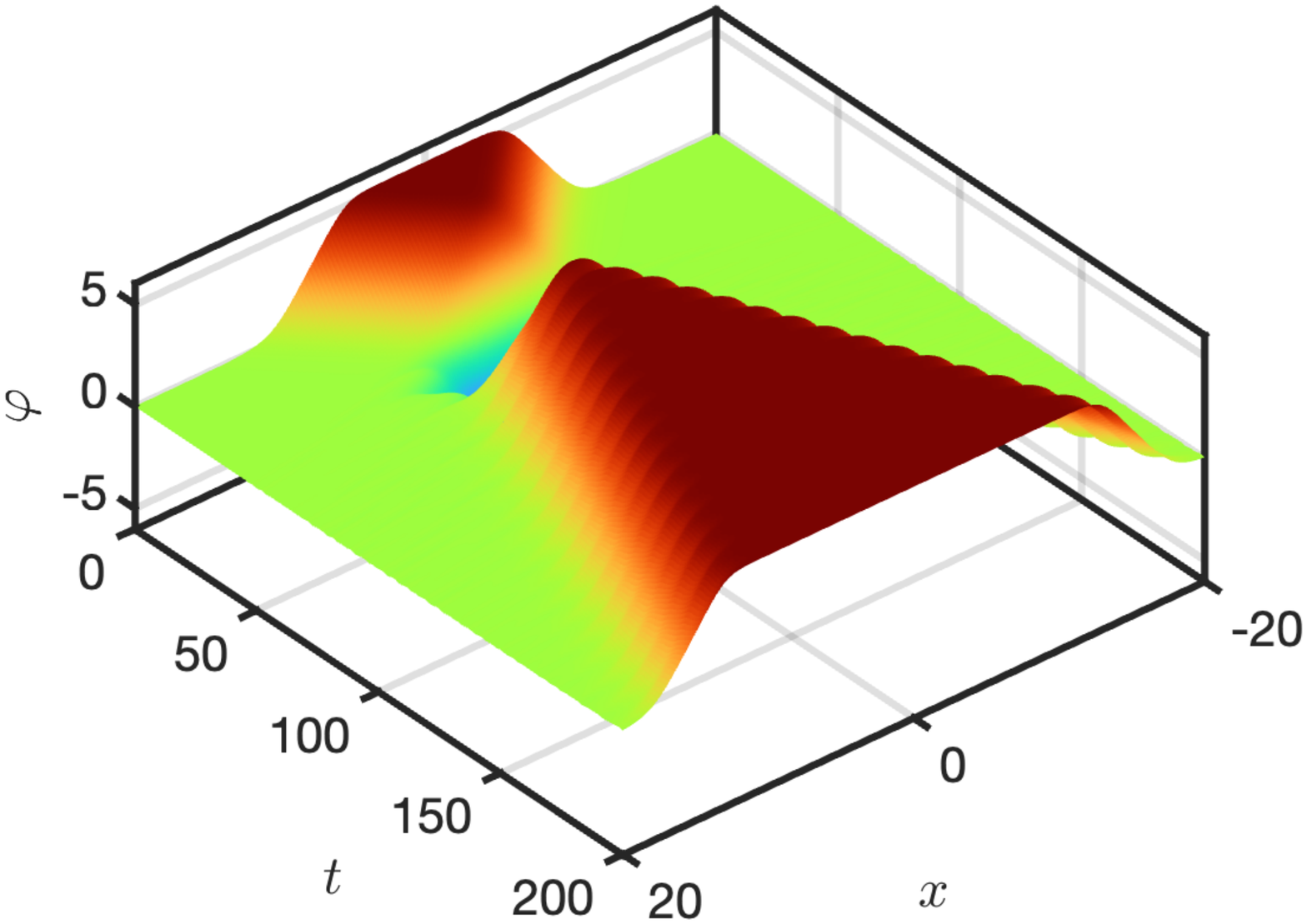}\label{fig:n100005v020200}}
 	\subfigure[\quad two-bounce for $v=0.2038$]{\includegraphics[width=8cm]{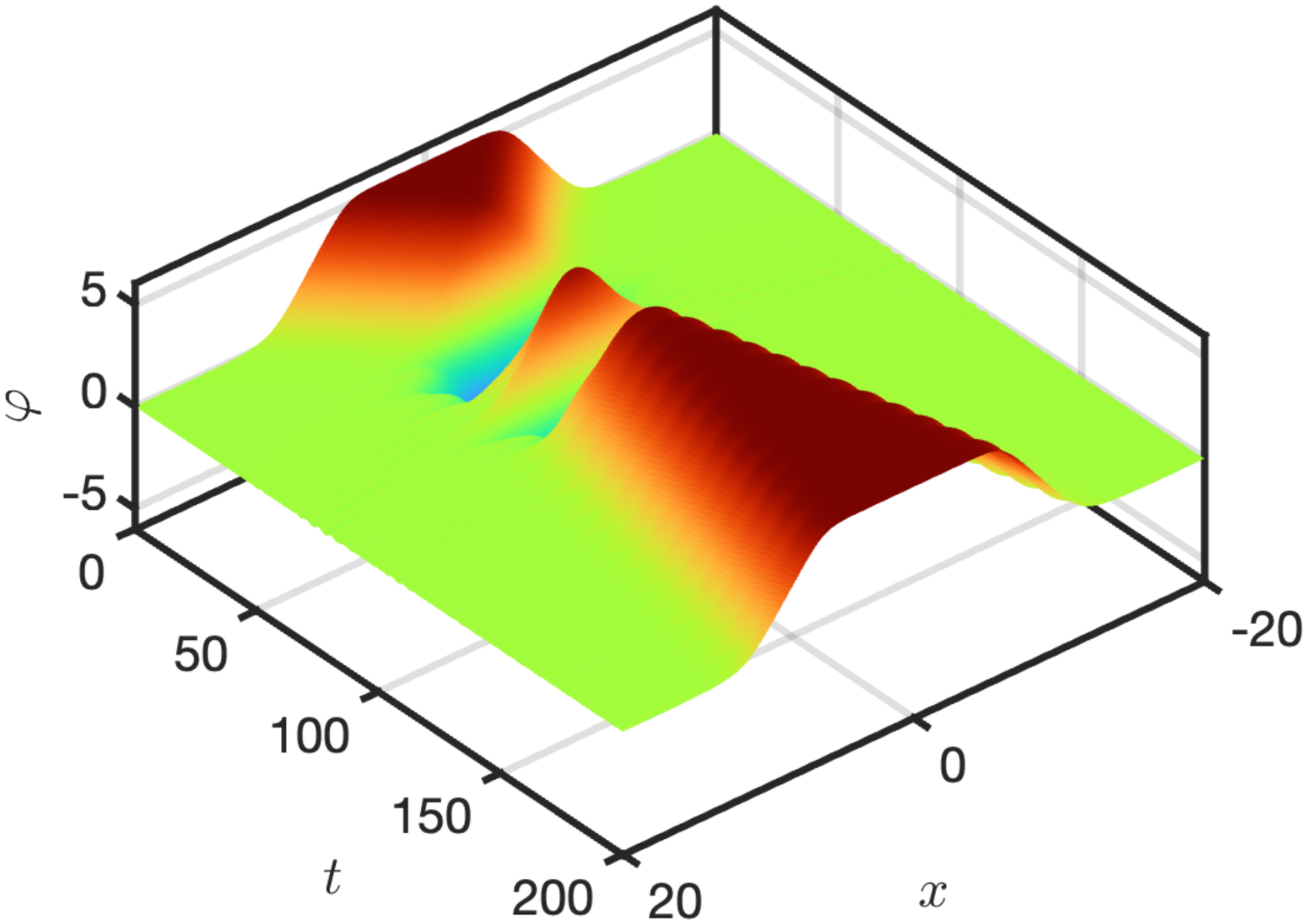}\label{fig:n100005v020380}}
 	\\
 	\subfigure[\quad two-bounce for $v=0.2031$]{\includegraphics[width=8cm]{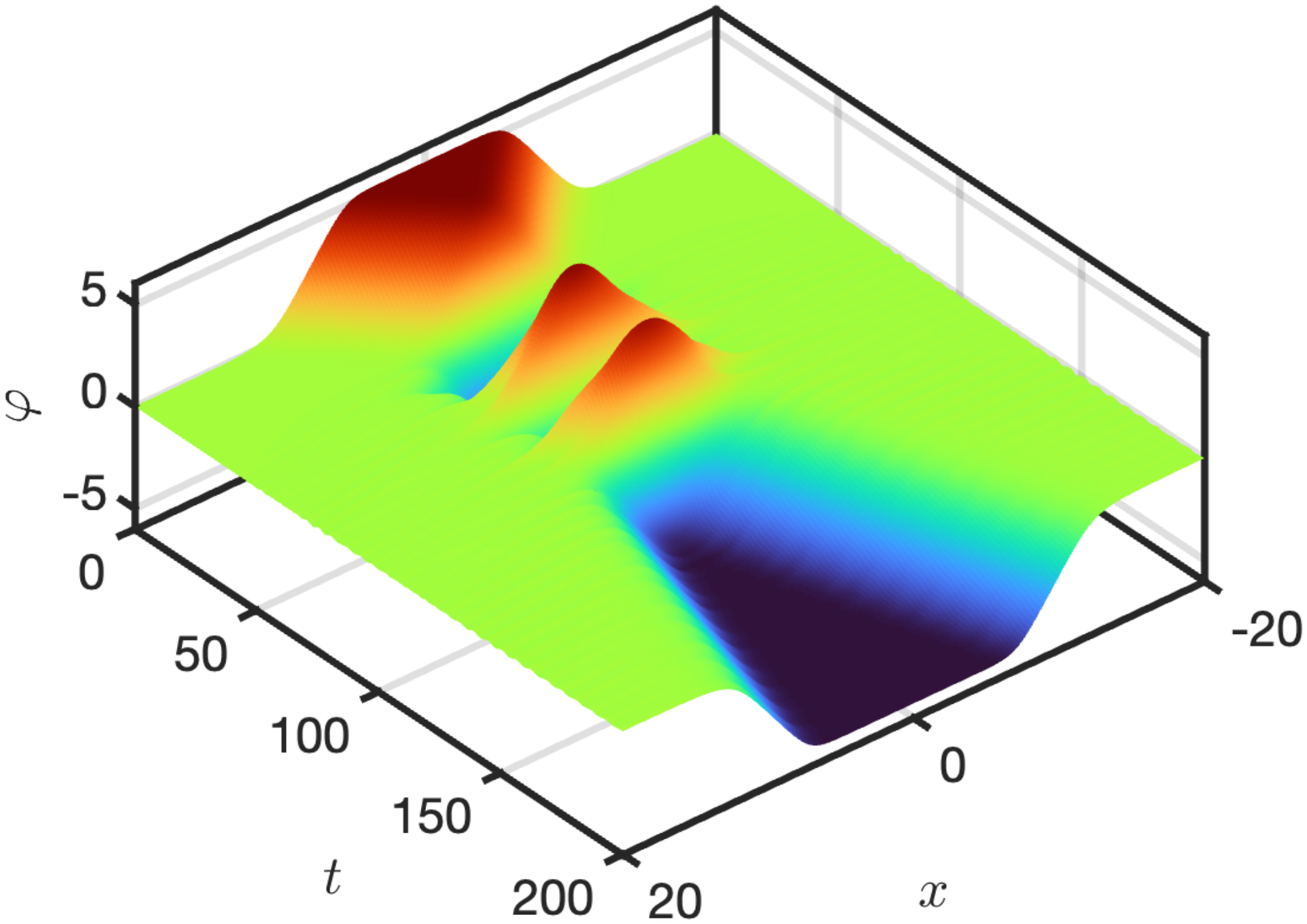}\label{fig:n100005v020310}}
 	\subfigure[\quad one-bounce for $v=0.2095$]{\includegraphics[width=8cm]{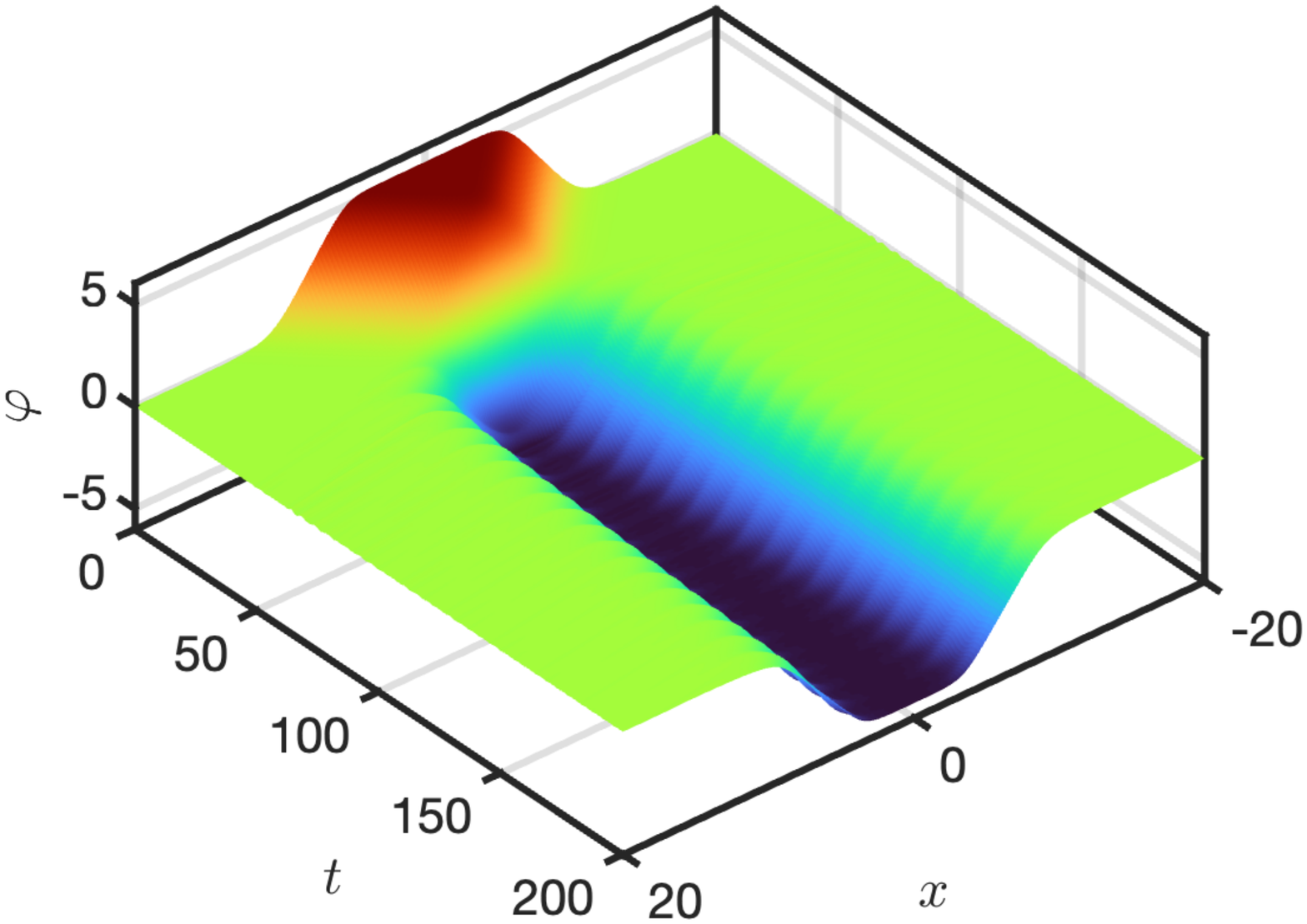}\label{fig:n100005v020950}}
 	 \caption{Kink-antikink collision in the tanh-deformed $\varphi^6$ model for different initial velocities $v$. $r=1.00005$. We used the initial separation between centers of kink and antikink as $2x_0=20$.}
    \label{fig:kinkantikinkcollision}
\end{figure}

We perform the collision for other values of $r$, for instance, $r=1.00005$. Numerical analysis of the Schr\"odinger potential revealed the appearance of an internal state ($\omega^2 = 0.4607$). The output of the kink-antikink process for this case can be seen in Fig.~\ref{fig:kinkantikinkcollision} and shows different final states for some initial velocities. For instance, there are creation of oscillating pulses  - Fig.~\ref{fig:n100005v019000}, as well as the kink-antikink pair trapping - Fig.~\ref{fig:n100005v020360}.

An interesting behavior is shown in Fig.~\ref{fig:n100005v020200}. The kink-antikink pair approaches, collides, oscillates in vacuum $\tilde{\varphi}_{-}=-\tanh^{-1}(1/r)$, and moves away in the initial vacuum. In this scenario, the scalar field begins in the vacuum $\tilde{\varphi}_{+}$, collides, and oscillates in the other topological sector $\tilde{\varphi}_{-}$. This is referred to as a one-bounce window with two oscillations. This result was already reported in Refs.~\cite{Peyrard.PhysD.1983.msG,Bazeia.PLB.2019}. We can notice the appearance of two-bounce behavior for $v=0.2038$, as we can see in Fig. \ref{fig:n100005v020380}. However, the oscillation happens in vacuum $\tilde{\varphi}_{-}$, which differs from the standard two-bounce. After the collisions, we note the scattering of the pair in the vacuum $\tilde{\varphi}_{+}$. More activity is observed as the initial velocity is varied. For instance, with $v=0.2031$, the kink-antikink pair collides twice and scatters to infinity with a phase shift - see Fig.~\ref{fig:n100005v020310}. For large initial velocities ($v>v_c=0.2094$), emerge only a collision with a phase shift; after interaction, the kink-antikink pair produces an antikink-kink pair, as we can see in Fig.~\ref{fig:n100005v020950}.

 \begin{figure}
    \subfigure[\quad bion for $v=0.19$]{\includegraphics[width=8cm]{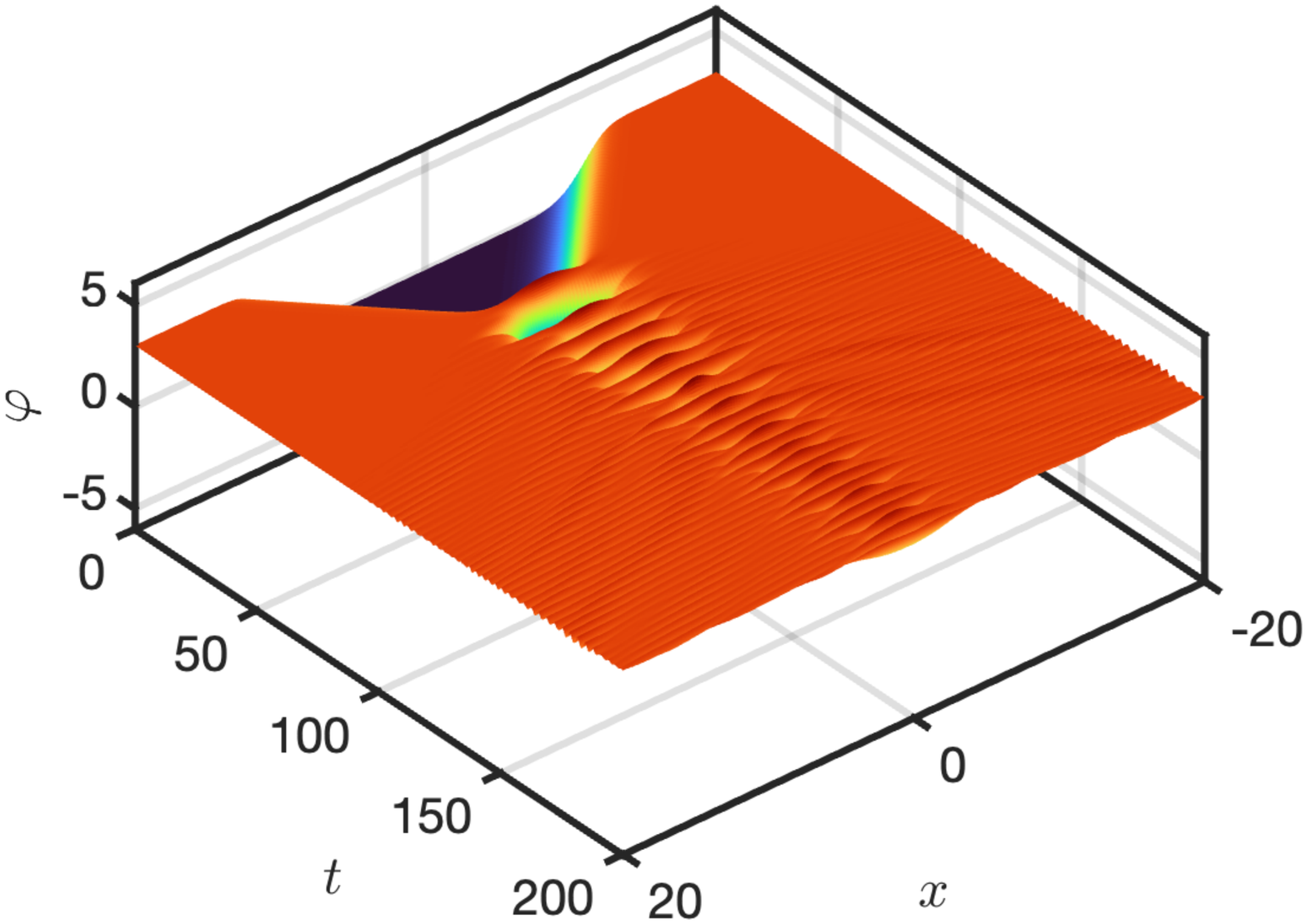}\label{fig:hypphi6akv019n1005}}
 	\subfigure[\quad two-bounce for $v=0.21$]{\includegraphics[width=8cm]{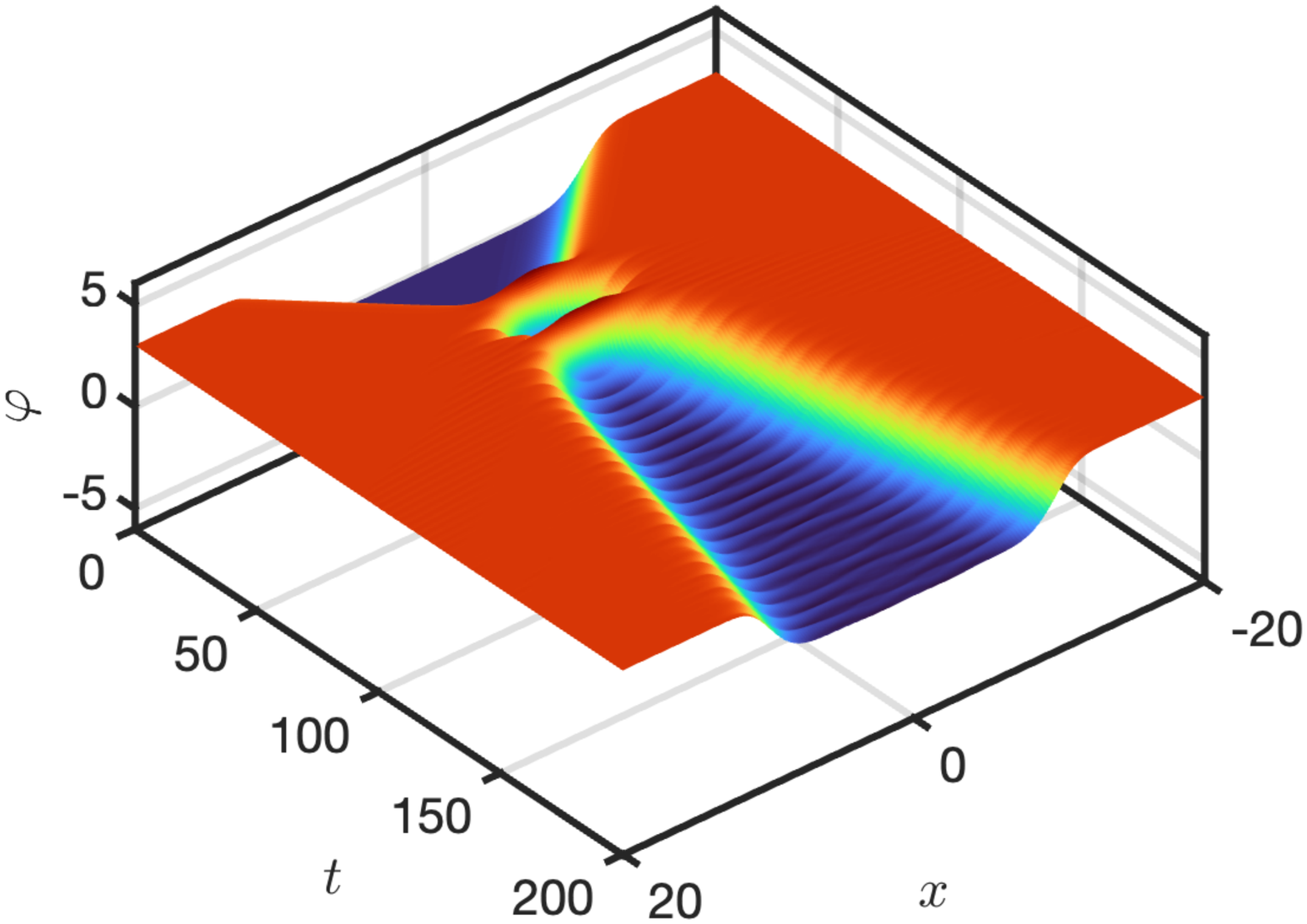}\label{fig:hypphi6akv021r1005}}
 	\\
 	\subfigure[\quad three-bounce for $v=0.215$]{\includegraphics[width=8cm]{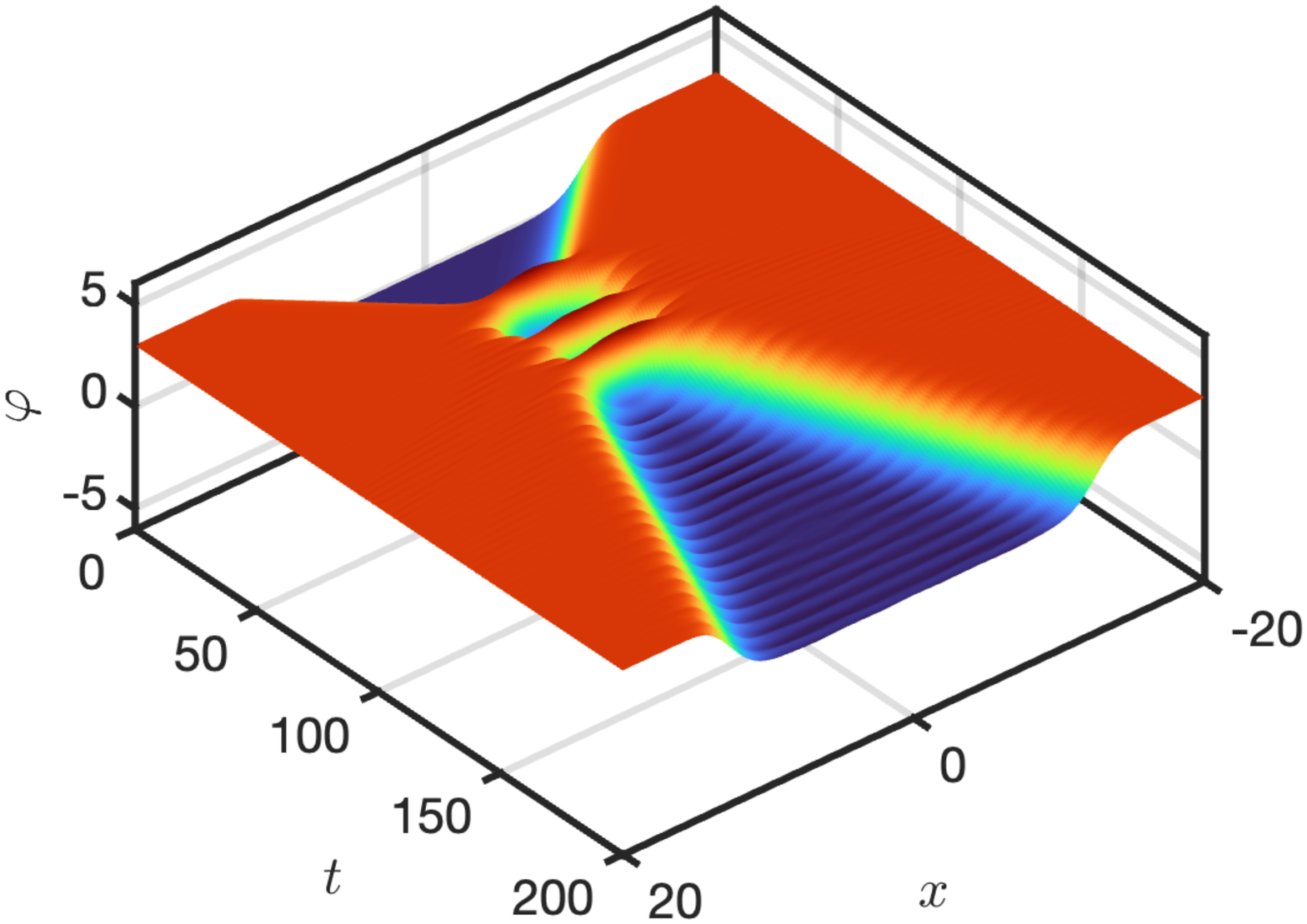}\label{fig:hypphi6akv0215r1005}}
 	\subfigure[\quad one-bounce for $v=0.26$]{\includegraphics[width=8cm]{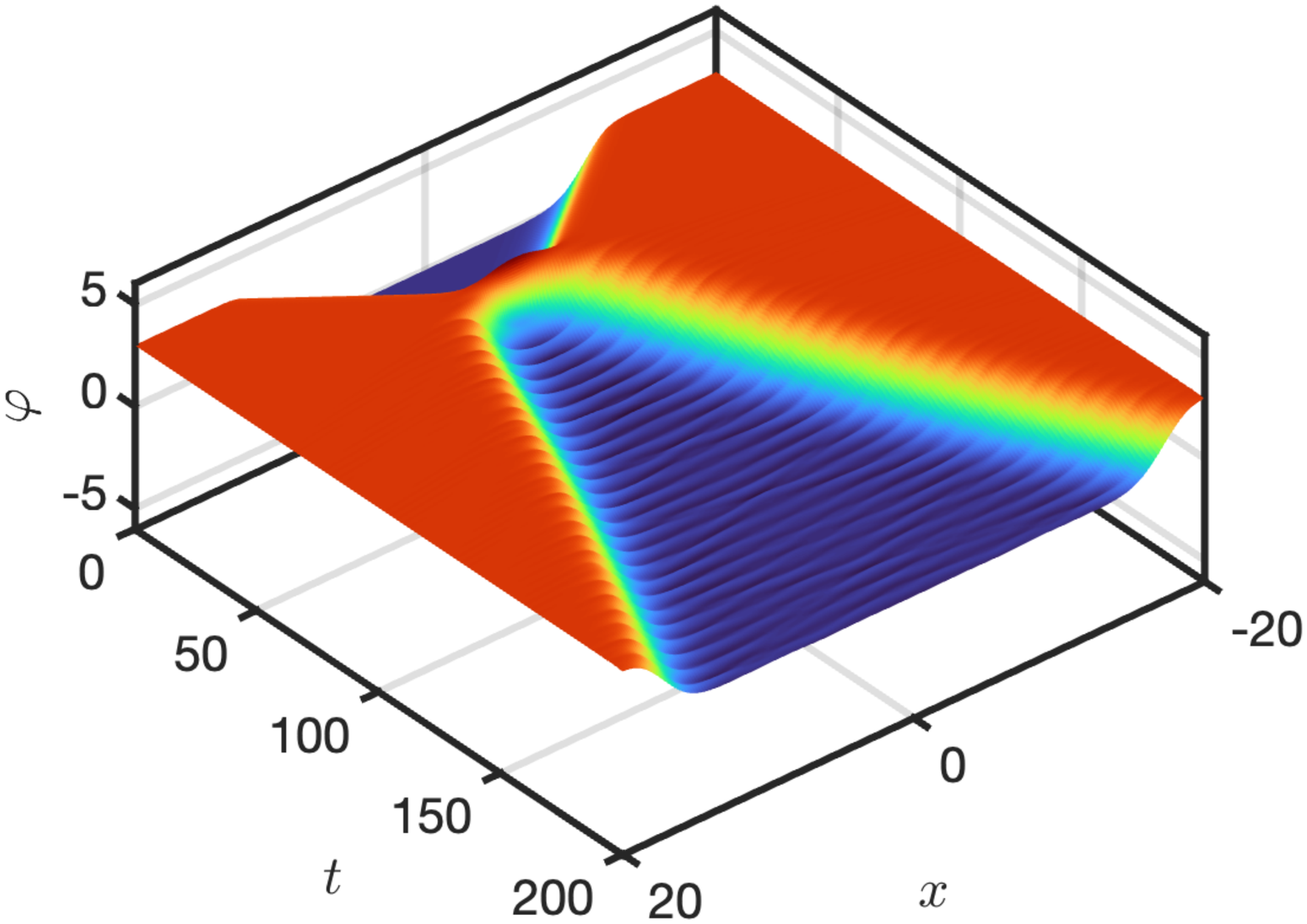}\label{fig:hypphi6akv026n1005}}
 	 \caption{Antikink-kink collision in the tanh-deformed $\varphi^6$ model for different initial velocities $v$. $r=1.005$. We used the initial separation between centers of kink and antikink as $2X_0=20$.}
    \label{fig:hypphi6AK1}
\end{figure}

Finally we consider antikink-kink collision for $r=1.005$. This case has the critical velocity of $v_c=0.2508$ and there is one shape mode. We see the emergence of bion behaviors for $v\leq v_c$. The antikink-kink pair oscillates around the vacuum $\tilde{\varphi}_{+}$ in this case, emitting radiation until the pair is completely annihilated; see Fig. \ref{fig:hypphi6akv019n1005}. Furthermore, we observe the usual fractal behavior of the bounce windows in particular initial velocities ranges. The Figs.~\ref{fig:hypphi6akv021r1005} and \ref{fig:hypphi6akv0215r1005}, for $v=0.21$ and $v=0.215$, show the outcomes for two- and three-bounces, respectively. In contrast, for $v> v_c$, only an inelastic collision is observed. The antikink-kink pair interacts and scatters to infinity, producing some radiation in the process. Notice that in this configuration, the pair does not visit the other topological sector. However, in the kink-antikink scenario, the scalar field reach the other topological sector during and after the collision.

\begin{figure}
 	\includegraphics[width=9cm]{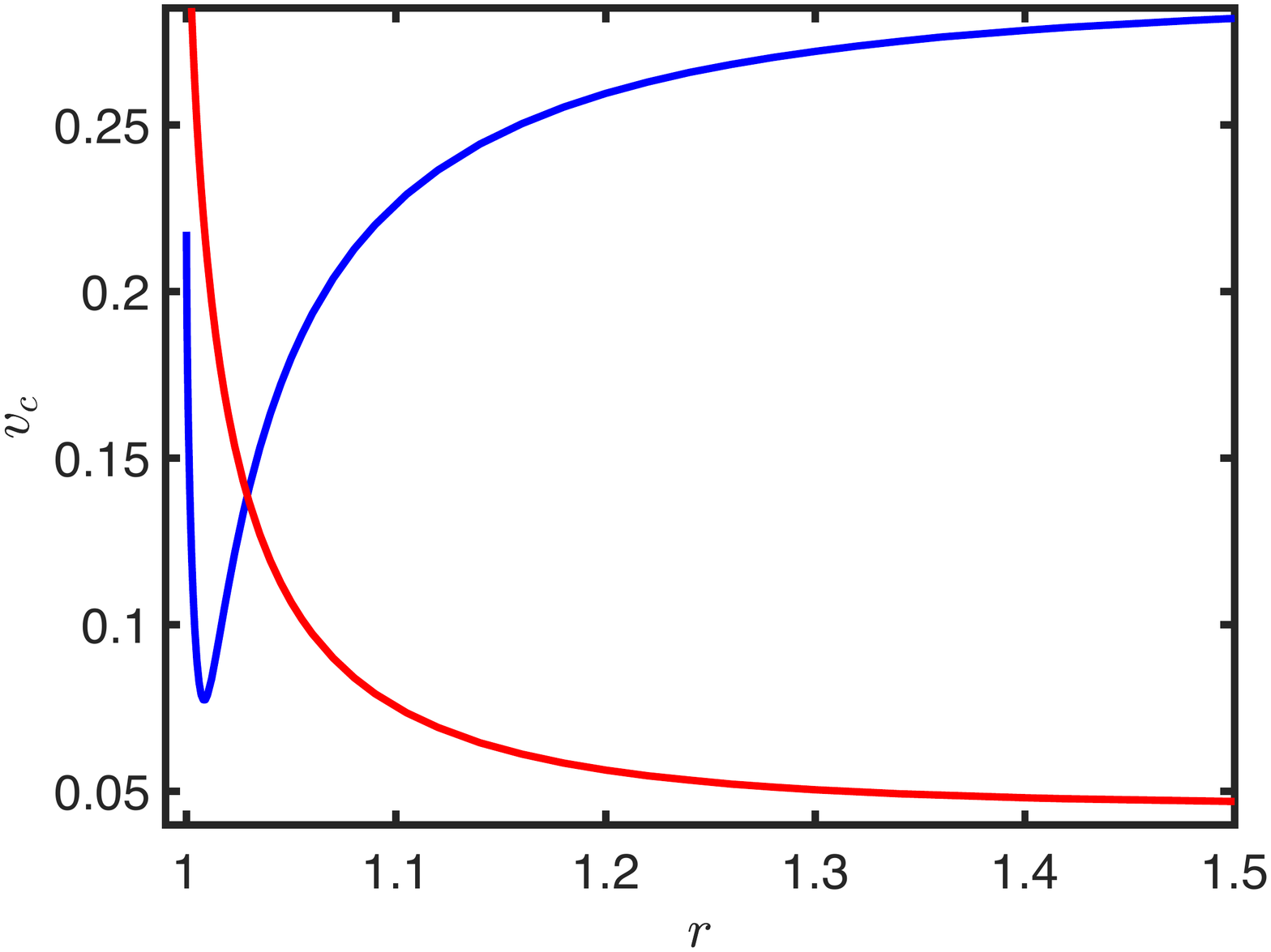}
 	 \caption{Critical velocity in the kink-antikink collision (blue) and antikink-kink collision (red) as function of $r$ of the tanh-deformed $\varphi^6$ model.}
 	\label{fig:hypphi6criticalvelocity}
\end{figure}

We illustrate in the Fig.~\ref{fig:hypphi6criticalvelocity} the critical velocity $v_c$ as a function of $r$ for kink-antikink and antikink-kink collision. The plot for kink-antikink scattering illustrates the establishment of a minimum followed by an increase in critical velocity as $r$ increases. In contrast, as the parameter increases, the antikink-kink scattering displays a diminishing in critical velocity.

It is important to point out some similarities between the tanh-deformed $\varphi^6$ and $\varphi^6$ model. Only the zero mode for a simple kink (or antikink) exists in the region with large $r$ values. However, two-bounces windows are found due to the existence of the vibrational mode for the antikink-kink pair \cite{Dorey.PRL.2011}. Furthermore, the critical velocity in the kink-antikink scattering, which is shown in blue in Fig.~\ref{fig:hypphi6criticalvelocity},  indicates that for high values of $r$, it goes to $v_c=0.289$, comparable to $\varphi^6$ model. However, if $r$ decreases, the critical velocity reaches a minimum at $r=1.008$ and then increases. This behavior is due to the fact that for $r$ below $r=1.008$, there appears a new bound state, not present in the $\varphi^6$ model for the kink-antikink pair. In the antikink-kink collision, on the other hand, the critical velocity diminishes as we increase $r$; see Fig.~\ref{fig:hypphi6criticalvelocity}, red curve. In fact, it goes to $v_c=0.0457$, which is found in the $\varphi^6$ model. In the present model, for $1<r<1.5$, there are several bound states trapped by the antikink-kink pair; in fact, the number of bound states change from five to four as $r$ goes above $r=1.04$, but this is not sufficient to chance the monotonic behavior of the critical velocity in this case.

\begin{figure}
 	\subfigure[]{\includegraphics[width=8cm]{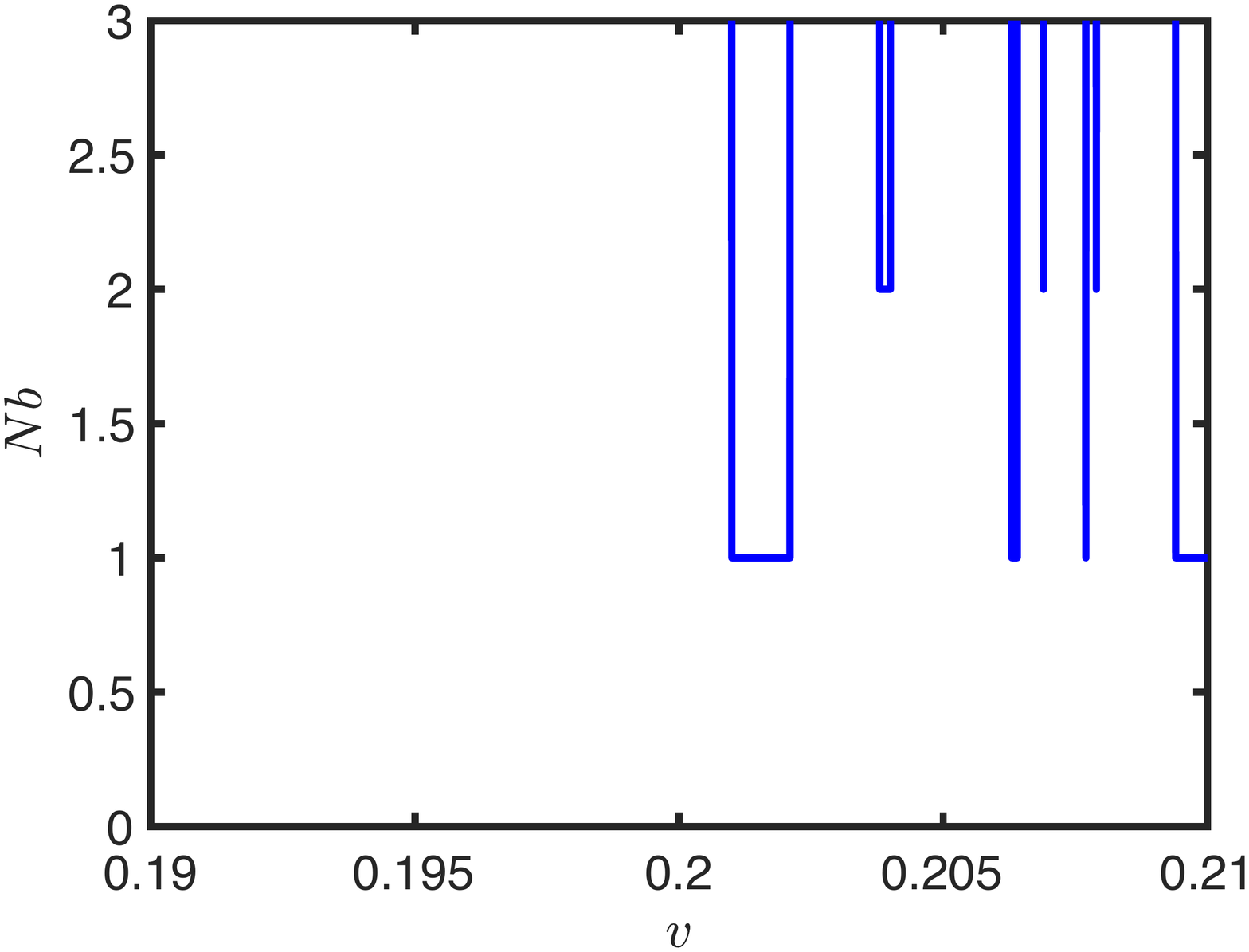}\label{hypphi6NbA}}
 	\subfigure[]{\includegraphics[width=8cm]{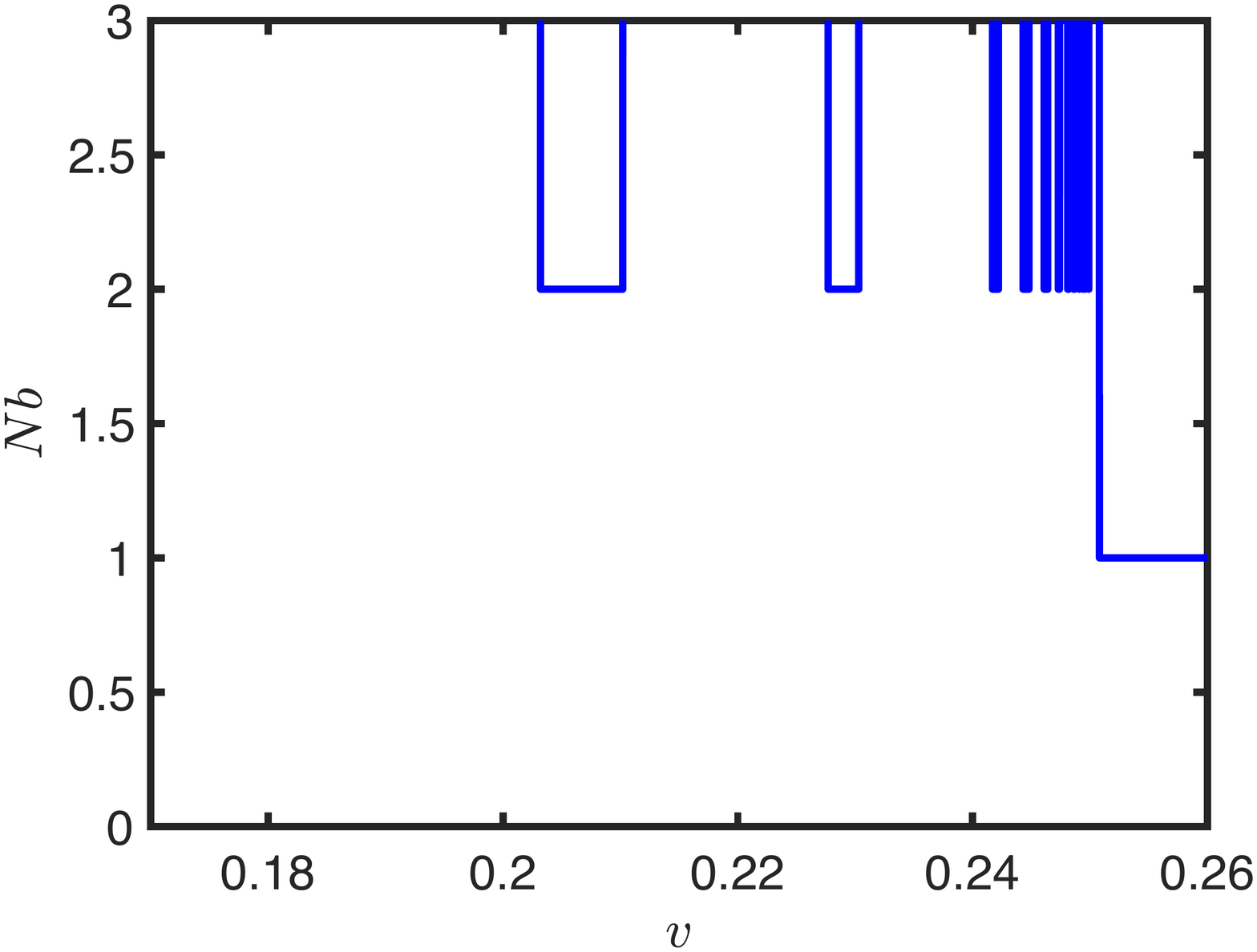}\label{hypphi6NbB}}
 	 \caption{Number of bounces $Nb$ as a function of initial velocity $v$ for (a) kink-antikink collision with $r=1.00005$ and (b) antikink-kink collision with $r=1.005$.}
 	\label{fig:hypphi6Nb}
\end{figure}

The bounce windows structure shown in Fig. \ref{fig:hypphi6Nb} reveals the behavior of the number of bounces $N_b$ as a function of initial velocity. Notice that the thickness of the windows decreases as $v$ increases, for both figures. Fig. \ref{hypphi6NbA} for $r=1.00005$ displays an intriguing resonant structure, which appears for $v<v_c$, with the presence of one and two-bounce windows, hard to see in the related literature. In particular, the Fig. \ref{fig:n100005v020200} and Fig.
\ref{fig:n100005v020380} correspond to the first one-bounce windows and the first two-bounce windows, respectively. The distinction between the other one-bounce windows is defined by the increased number of oscillations as $v$ increases. For higher velocities, we only notice the appearance of one-bounce ($N_b=1$). In this case, the kink-antikink collides only once and produces the antikink-kink pair. This behavior can be seen in Fig. \ref{fig:n100005v020950}. In the Fig. \ref{hypphi6NbB} we show the resonance structure for antikink-kink scattering with $r=1.005$. We may observe the conventional window structure with two-bounce and bion states for low velocities; this behavior changes for collisions close to the critical velocity $v_c=0.2508$.

\section{Conclusion}\label{sec:conclusion}


In this work we have used the deformation procedure to introduce two new models, the tanh-deformed $\varphi^4$ model and the tanh-deformed $\varphi^6$ model. The models contain a real parameter $r>1$, and we have shown that $r$ controls the potentials, the kinklike solutions, their masses, the stability potentials and the internal modes associated to the stability of the systems. 

In the case of the tanh-deformed $\varphi^4$ model, we have studied the kink-antikink collision and shown the appearance of several two-bounce windows. We recognize that the amount of internal modes has an impact on the two-bounce windows suppression process. For the case with three bound states, some windows are suppressed, while false two-bounce windows arise. In contrast, the resonance windows have been fully recovered as the number of bound states has decreased, as is the case for $r=1.86$. This result motivates us to further study the collision of kinks in connection with the presence of fractal structure which was found before in the $\varphi^4$ model. A particularly interesting investigation that could be implemented, is related to the recent study on the fractal structure and collective coordinate and moduli space of kink and antikink collisions in the $\varphi^4$ model \cite{Manton.PRL.2021,Manton.PRD.2021,Adam.PRD.2022}. In the case of the tanh-deformed $\varphi^4$ model introduced in the present work, the deformation requires the parameter $r$, which may perhaps contribute to modify the results included in Refs. \cite{Manton.PRL.2021,Manton.PRD.2021,Adam.PRD.2022}, bringing new information on the fractal behavior of the collisions.

For the tanh-deformed $\varphi^6$ model, we performed the kink-antikink and antikink-kink collision for distinct values of parameter $r$. The stability analysis revealed that lowering $r$ provides a Schr\"odinger-like potential that facilitates the emergence of internal modes for a single kink or antikink. The kink-antikink scattering process revealed a complex structure as a function of the initial velocity and the parameter $r$, which is associated to the appearance of the internal mode. As previously stated, the kink-antikink collision produces oscillating pulses, two-bounce and one-bounce. It is worth pointing out that there are two types of two-bounce windows, as well as the one-bounce type scattering. The distinction between the two is in the final outcome. In one example, the scalar field structures collide twice and scatter in the same initial vacuum. However, for some ranges of velocities, after the two interactions, the kink-antikink pair scatters with a shift to the other vacuum, forming an antikink-kink pair. Our numerical investigation of antikink-kink scattering revealed the presence of a fractal structure. Unlike the case of kink-antikink configuration, in the antikink-kink scenario there is no possibility for the pair to visit the other topological sector after the interaction. Therefore, the two-bounce behavior of the kink-antikink collision differs from the antikink-kink one.

We think the above hyperbolic-deformed models can be used to investigate other issues, in particular, the fractal structure of the collisions, which is directly influenced by the presence of several bound states. This may find connection with the recent study on the scattering for the graphene superlattice equation \cite{chaos} and also, the approach described in \cite{Adam.PRD.2022}, which allows to incorporate relativistic corrections in the scattering, giving excellent description of kink-antikink collisions in the $\varphi^4$ theory, contributing to reproduce the fractal structure in the formation of the final state. Specifically, the graphene superlattice equation was considered to describe propagation of solitary electromagnetic waves in a graphene superlattice and the presence of fractal structure in the kink scattering is interesting characteristic which deserves further investigation, including the use of the deformation to describe generalized models. The hyperbolic-deformed models may also be used to further study issues related to the presence and unbinding of shape modes, as recently investigated in Refs. \cite{turk,other}. These and other related issues are presently under consideration, and we hope to report on them in the near future.

\section*{Acknowledgments}
AMM would like to thank Islamic Azad University Quchan branch for the grant. FCS would like to thank Funda\c c\~ao de Amparo \`a Pesquisa e ao Desenvolvimento Cient\'ifico e Tecnol\'ogico do Maranh\~ao, FAPEMA, Universal Grant 00920/19. DB would like to thank Conselho Nacional de Desenvolvimento Cient\'\i fico e Tecnol\'ogico, CNPq, grant No. 303469/2019-6, and Paraiba State Research Foundation, FAPESQ-PB, grant No. 0015/2019, for partial financial support.

\end{document}